\documentclass[a4paper,fleqn,usenatbib]{mnras}
\usepackage[T1]{fontenc}
\usepackage{ae,aecompl}
 
\usepackage{graphicx}	
\usepackage{amsmath}	
\usepackage{amssymb}	
\usepackage[usenames]{color}

\definecolor{Blue}{rgb}{0,0.5,1}
\definecolor{Purple}{rgb}{1,.38,0}
\definecolor{Orange}{rgb}{1,0.5,0}

\def\New#1{{ #1}}

\title[Zooming in on supermassive black holes]{Zooming in on supermassive black holes: how resolving their gas cloud host renders their accretion episodic}

\author[R. S. Beckmann et al.]{R. S. Beckmann$^{1,2}$\thanks{Email: ricarda.beckmann@physics.ox.ac.uk}, J. Devriendt$^{1,3}$ and A. Slyz$^{1}$
\\
$^{1}$ Sub-department of Astrophysics, University of Oxford, Keble Road, Oxford OX1 3RH, UK\\
$^{2}$ Institut d'Astrophysique de Paris, CNRS \& Sorbonne Universites, UMR7095, 98bis Boulevard Arago, F-75014, Paris, France \\
$^{3}$ Université Lyon 1, ENS de Lyon, CNRS, Centre de Recherche Astrophysique de Lyon UMR5574, F-69230 Saint-Genis-Laval, France\\}

\date{Accepted XXX. Received YYY; in original form ZZZ}

\pubyear{2018}

\begin{document}
\label{firstpage}
\pagerange{\pageref{firstpage}--\pageref{lastpage}}
\maketitle

\begin{abstract}
Born in rapidly evolving mini-halos during the first billion years of the Universe, supermassive black holes (SMBH) feed from gas flows spanning many orders of magnitude, from the cosmic web in which they are embedded to their event horizon. As such, accretion onto SMBHs constitutes a formidable challenge to tackle numerically, and currently requires the use of sub-grid models to handle the flow on small, unresolved scales. In this paper, we study the impact of resolution on the accretion pattern of SMBHs initially inserted at the heart of dense galactic gas clouds, using a custom super-Lagrangian refinement scheme to resolve the black hole (BH) gravitational zone of influence. We find that once the self-gravitating gas cloud host is sufficiently well resolved, accretion onto the BH is driven by the cloud internal structure, independently of the BH seed mass, provided dynamical friction is present during the early stages of cloud collapse. For a pristine gas mix of hydrogen and helium, a slim disc develops around the BH on sub-parsec scales, turning the otherwise chaotic BH accretion duty cycle into an episodic one, with potentially \New{important} consequences for BH feedback. In the presence of such a nuclear disc, BH mass growth predominantly occurs when infalling dense clumps trigger disc instabilities, fuelling intense albeit short-lived gas accretion episodes.
\end{abstract}





%
%

\section{Introduction}
\label{sec:introduction}

First observed to power quasars at redshift $z\geq 6$ \citep{Fan2006}, SMBHs formed during the first billion years of the Universe. This epoch saw not only the formation of SMBHs but also of many other key components of the Universe: the first halos collapsed, the first gas streamed in and the first stars ignited. Given the current paucity of observations from this era, we have to rely almost entirely on analytic considerations and simulations to understand the early evolution of SMBHs.

Three viable channels have been proposed to form the progenitors of SMBHs. Stellar remnants are expected to have masses in the range of $10-500 \rm \ M_\odot$, runaway cluster collapse BHs are expected to form with masses in the range $10^3-10^4 \rm \ M_\odot$ and direct collapse BHs have seed masses in excess of $10^5 \rm \ M_\odot$ (see \citet{Volonteri2010} for a review). Whatever seed mass the BH forms with, it needs to gain many orders of magnitude in mass to reach the observational estimates of $\sim 10^9 \ \rm M_\odot$ at $z=6$ \citep{Fan2006} and gas accretion in gas-rich mini halos is expected to play a vital role particularly in the early mass growth of potential SMBH.

Understanding the early evolution of massive BHs therefore crucially depends on understanding the hydrodynamical evolution of gas in high-redshift mini-halos housing SMBH progenitors. This highly non-linear problem can only be tackled using simulations, but numerical efforts are hampered by the vast range of scales involved. Including all relevant length scales to monitor the co-evolution of a single SMBH and its host galaxy, from the Mpc scales of cosmic filaments through the kpc and pc scales of the galaxy, down to the sub-pc scales on which the BH accretes gas, in a single simulation constitutes a tremendous computational challenge. 

The problem does not lend itself well to idealised simulations because the evolution of a BH depends crucially on the evolution of its host galaxy, and the host galaxy itself is evolving rapidly. Cosmological zoom simulations, where a high resolution region containing all the progenitors of a particular object is embedded in a lower resolution background, is one way to maximise the range of scales resolved in a single simulation. Zoom simulations have given crucial insight into galaxy evolution in the early Universe \citep{DiMatteo2008,Dubois2012a,Ceverino2017}, but even in the most resolved zoom simulations available to date, the BH environment remains under-resolved. BH physics, such as formation, accretion and feedback, therefore have to be included as sub-grid algorithms that include the impact of unresolved scales using analytic or empirical models.

Any sub-grid model necessarily makes assumptions how the input parameters relate to the accretion onto the BH. Arguably the most common way to describe unresolved accretion onto BHs is through the Bondi-Hoyle-Lyttleton (BHL) model \citep{Hoyle1939,Bondi1944,Bondi1952}, discussed in Section \ref{sec:BHL}. In this model, the accretion rate is calculated from the local density and sound speed, as well as the relative velocity between gas and BH. In galaxy evolution simulations, the main difficulty with such sub-grid models lies in correctly estimating these local gas properties in the presence of limited resolution. 

To compensate for unresolved cold, dense gas, some authors advocate using a boost factor in the accretion rate \citep{Booth2009,Vogelsberger2013,DeGraf2014}, while other authors prescribe scaling the accretion rate down to account for angular momentum  \citep{Krumholz2004,Power2011,Dubois2012,Rosas-Guevara2015,Curtis2016}. \citet{Negri2016} study the accuracy of BHL accretion for a BH with feedback, and find that it can lead to both over- and under-estimates of the black hole mass, depending on the resolution and the way variables are calculated. \citet{Gaspari2013} systematically break the assumptions of Bondi accretion by cooling, heating or stirring the gas, and find that the accretion rate is boosted several orders of magnitude above the Bondi value. \citet{MacLeod2015} study accretion onto a point mass in the presence of a pressure gradient at infinity, and find the accretion rate onto the BH to be reduced by up to two orders of magnitude. 

In light of these discrepancies, efforts have been made to develop alternative accretion algorithms, which use large-scale properties of the host galaxy to calculate the accretion rate onto the BH, and naturally account for the angular momentum of accreted gas. \citet{Angles-Alcazar2013}, building on a model by \citet{Hopkins2011}, compute the accretion rate onto the BH based large scale gravitational torques in the galaxy. \citet{DeBuhr2010a} calculate the accretion rate onto the BH from the mean surface density of the galaxy, the angular rotation frequency and a free viscosity parameter, and \citet{Hobbs2011} use a ``ballistic'' accretion model based on the velocity dispersion of the gas in the bulge. 

All these models estimate the mass accreted by the BH from information on scales much larger than the gravitational zone of influence of the BH itself. While this does represent an educated best estimate, and has led to many insights into the evolution of BHs, it can, on its own, only ever constitute an overly simplistic treatment of the problem. This is particularly true for the progenitors of SMBHs, as their host galaxies  are only just assembling during the BH early mass evolution. Embedded in a rapidly changing environment and subject to strong feedback from the first generations of stars, galaxy-wide properties are difficult to define and measure for proto-galaxies, and local gas properties can vary rapidly. High resolution simulations, which have to make fewer a priori assumptions about the state of galactic gas, are therefore the most viable tool to study the origin of potential SMBHs,  judge the importance of gas accretion versus black hole mergers in the formation of SMBHs, and assess the viability of the three proposed seed formation channels as SMBH progenitors.

With such a long term goal in mind, this paper presents a study of BH accretion in collapsing clouds, using a new refinement algorithm for grid codes that allows the BH gravitational zone of influence to be resolved within the full context of its host galaxy at affordable computational cost. Embedded in a highly unsettled, gas rich galaxy intended to mimic a high-redshift mini halo, the simulations in this paper present a pilot study of early accretion onto potential SMBHs candidates in a galactic context {with unprecedented spatial resolution. They demonstrate the lasting impact of resolving the internal structure of the gas cloud feeding the BH, and the vital role played by dynamical friction in allowing the seed BH to accrete efficiently during the early stages of cloud collapse.

Section \ref{sec:simulation} details the simulation setup and Section \ref{sec:bh_zoom} presents the novel BH refinement algorithm. Section \ref{sec:mseed} establishes the key role played by dynamical friction during cloud collapse, and why sub-grid models based on a local relative velocity measure are unreliable at high-resolution (see also Appendix \ref{sec:convergence_levelmax}). Section \ref{sec:resolution} demonstrates the importance of resolving the internal structure of the gas cloud host at all times by presenting a comparative study of resolution impact on BH accretion histories, a discussion of which can be found in Section \ref{sec:discussion}. Conclusions are summarised in Section \ref{sec:conclusions}.

\section{The simulations}
\label{sec:simulation}

All simulations presented in this work use the adaptive mesh refinement code \textsc{ramses} \citep{Teyssier2002}. The Euler equations are solved using a second-order unsplit Godunov scheme, and a HLLC Riemann solver with a MinMod total variation diminishing scheme to reconstruct interpolated variables. The courant safety factor is set to a value of $0.6$.  

A root grid of $64^3$ cells is laid out on the entire simulation volume (a cube of size $L_{\rm box}$ on a side), which is then adaptively refined up to level $l_{\rm glob}$ outside the zoom region around the BH, and to $l_{\rm zoom}$ within this region (see Section \ref{sec:zoom-within-zoom} for details). The size of an individual cell at level $l$ is equal to  $\Delta x_l = \frac{L_{\rm box}}{2^l}$. Refinement is determined using a quasi-Lagrangian criterion: a cell is split into 8 when its total baryonic mass exceeds 8 times the gas mass it initially contained. 
To minimise numerical fragmentation \citep{Truelove1997}, cells are also refined so the local thermal jeans length exceeds the cell size at all times. Gravitational contributions of all baryonic matter, including stars and BH as well as gas, are computed using a multi-grid Poisson solver \citep{Guillet2011}, where the total density of matter is found by assigning particles to the grid using a cloud-in-cell interpolation algorithm. 

\subsection{Initial conditions}
\label{sec:initial_conditions}
In order to test our refinement algorithm in a controlled environment, we set up an isolated cooling halo with a Navarro-Frenk-White (NFW) profile for the dark matter component, similar to the one presented in \citet{Dubois2008}. The dark matter potential of the halo is modelled as a fixed analytic potential, where

\begin{equation}
\rho_{\rm NFW}(r) = \frac{\rho_s}{(r/r_s)(1+r/r_s)^2 }.
\end{equation}
so the halo has a  total integrated mass of 
\begin{equation}
M(<r)=4\pi \rho_s r_s^3 \bigg( \ln(1 + r/r_s) - \frac{ r/r_s}{1+r/r_s} \bigg),
\end{equation}
where $r_s$ and $\rho_s$ are the halo characteristic radius and density respectively. We take $\rho_s = 200 \rho_{\rm c} $ where $\rho_{\rm c} = \frac{3 H_0^2 } {8 \pi G}$ is the current critical density, assuming a Hubble parameter of $H_0 =70 \rm \ km/s^{-1}  Mpc^{-1}$. To define the characteristic radius $r_s$, we set the concentration parameter $c=\frac{r_{200}}{r_{\rm s}}=3.5$ in our simulation, and pick $v_{200}=85  \rm \ km/s^{-1}$. The halo is truncated at two virial radii, with densities outside set to a uniform value of $\rho_{\rm IGM}=3.27 \times 10^{-9} \rm \ H/cm^3$, and embedded in a box of $L_{\rm box}=1.0 \ \rm Mpc$. We impose a universal baryon fraction of $f_{\rm baryon}=15 \%$ within the halo, and distribute gas according to $\rho_{\rm gas} = 0.15 \rho_{\rm NFW}$, with the initial pressure profile set according to hydrostatic equilibrium. The total halo mass is $M_{\rm tot}=5.1 \times 10^{11} \ \rm M_\odot$. 

When adding rotation (Section \ref{sec:mseed} and Section \ref{sec:resolution} only), we distribute angular momentum according to  $ j(r)=j_{\rm max} \frac{M(<r)}{M_{\rm vir}} $, following \citet{Bullock2001}, normalised to a spin parameter $\lambda = J |E|^{1/2} / ( G M_{\rm vir}^{5/2}) =0.04$, where $J$ and $E$ are the total angular momentum and energy respectively. 

\subsection{Cooling and Heating}
\label{sec:cooling}

Above $10^4 \rm \ K$ the gaseous mix of H, He and electrons is assumed to radiatively cool through collisions following the pristine cooling function tabulated by \citet{Sutherland1993}.
Below this temperature radiative cooling occurs via $\rm H_2$ emission following \citet{Grassi2014}. The gas is kept metal-free throughout and follows an ideal equation of state and is assumed to be mono-atomic 
with adiabatic index $\gamma = 5/3$. 

\subsection{Star formation}
\label{sec:star_formation}

Star formation proceeds according to a Kennicutt-Schmidt law $\dot{\rho}_\star = \epsilon_\star \rho / t_{\rm ff}$ \citep{Kennicutt1998,Krumholz2007}, where $\epsilon_\star=0.02$ is the (constant) star formation efficiency, $\rho$ the gas density and $t_{\rm ff}$ is the free fall time of the gas. Star particles are generated following a Poisson random process \citep{Rasera2006,Dubois2008}, in cells that exceed the star formation number density threshold of $\rho_0 / \mu m_H = 4.73 \times 10^3 \rm \ H / cm^3$, where $m_H$ is the mass of a hydrogen atom and $\mu$ the mean molecular weight. As a result, star particles are spawned with masses of $M_\star = 149.2 \ \rm M_\odot $, or integer multiples thereof. No stellar feedback is included in this work.

\subsection{Black hole formation and accretion} 
\label{sec:BHL}

BH formation sites are identified on the fly using the structure-finding algorithm \textsc{phew} \citep{Bleuler2015}, which locates gravitationally bound, collapsing gas clumps in cells above a number density threshold of $10^3 \rm \ H/cm^3 $. After a formation site has been identified, a sink particle \citep{Krumholz2004,Dubois2010} is inserted in the clump densest cell, with a user-defined initial sink mass $M_{\rm seed}$. To conserve mass and momentum, an equivalent amount of gas mass is removed from the host cell, and the new sink particle inherits the host cell velocity vector. Only one BH is allowed to form per simulation.

Accretion proceeds according to the BHL accretion rate interpolation formula \citep{Hoyle1939,Bondi1944}, 
\begin{equation}
	\dot{M}_{\rm BHL} = \frac{4 \pi G^2 M_{\rm BH}^2 \rho_\infty}	{(c_{s,\infty}^2+v_\infty^2)^{3/2}}.
	\label{eq:BHL}
\end{equation}
which describes the amount of matter accreted by a point mass $M_{\rm BH}$ moving at a constant velocity $v_\infty$ through a gas background of uniform density $\rho_\infty$ and sound speed $c_{s,\infty}$. Quantities  ``at infinity'' are measured far from the gravitational zone of influence of the BH. The derivation assumes that the gas has no angular momentum, does not cool, is not self-gravitating and, in the case of supersonic relative velocity, is pressureless.

The BHL problem has two characteristic scale radii. For subsonic motion of the BH, where $\mathcal{M}_\infty = \frac{v_\infty}{c_{s,\infty}} < 1$, the Bondi radius 
\begin{equation}
	R^{\rm B}=\frac{GM_{\rm BH}}{c_{s,\infty}^2}
	\label{eq:Rb}
\end{equation}
is the radius for which pressure forces balance the gravitational acceleration due to the hole. Gas flow outside this radius is subsonic and almost uniform in density, while within this radius, the gas evolves towards a free-fall solution, rapidly becoming supersonic.

For supersonic relative motion, $\mathcal{M}_\infty \geq 1$, the accretion radius 
\begin{equation}
	R^{\rm A}=\frac{2GM_{\rm BH}}{v_\infty^2}.
	\label{eq:Ra}
\end{equation}
defines a sphere which contains all accreted streamlines. $R^{\rm A}$ and $R^{\rm B}$ can be understood as the scale radius of the BH gravitational potential at a given relative velocity with respect to the background medium, $v_\infty$.
 
The accretion rate onto the sink particle therefore depends on the properties of the gas from which it accretes. ``At infinity'' is a concept that is difficult to define for a BH in a galactic context, but the properties of gas in the cells immediately surrounding the sink particle should provide a good approximation if the BH gravitational zone of influence is unresolved. To sample the local environment, \textsc{ramses} uses so-called ``cloud particles'', which are distributed with constant spacing $\Delta x_{\rm min}/2$ around the sink particle, filling a sphere with radius $r_{\rm cloud} = 4 \Delta x_{\rm min}$, \New{where $\Delta x_{\rm min}$ is the size of the smallest cell in the simulation}. Each cloud particle samples the gas properties of the cell that contains it.

Mass-weighted gas properties for use in Equation \ref{eq:BHL}, $\rho_\bullet$, $c_{s,\bullet}$ and $v_\bullet$, are calculated by summing over the cloud particles, weighted using the cell mass and a gaussian kernel 
\begin{equation}
\omega \propto \exp(-r^2/r^2_K),
\end{equation}
where $r$ is the distance of the cloud particle to the sink. The same kernel is also used to weigh the amount of mass removed per cell during accretion. $r_K$ is a scale radius, set according to
\begin{equation}
r_K= \begin{cases} 
\Delta x_{\rm min} / 4 & \text{if  }  r_{\rm BHL} < \Delta x_{\rm min} / 4 \\ 
r_{\rm BHL} &  \text{if  }  \Delta x_{\rm min} /4 \leq r_{\rm BHL} \leq 2 \Delta x_{\rm min}\\ 
2 \Delta x_{\rm min} & \text{otherwise}
 \end{cases}
\end{equation}
depending on the size of the interpolated BHL radius 
\begin{equation}
r_{\rm BHL} = \frac{G M_{\rm BH}}{ c_{s,{\rm cell}}^2 + v_{\rm cell}^2}
\end{equation}
relative to the local resolution $\Delta x_{\rm min}$. $r_{\rm BHL}$ 
is calculated on the fly using the BH mass, $M_{\rm BH}$, the sound speed and relative velocity of the gas in the cell containing the BH, $c_{s,{\rm cell}}$ and $v_{\rm cell}$. At each timestep, only up to 75 \% of a given cell mass is allowed to be removed to avoid numerical issues arising from too important instantaneous gas removal. For more details on the sink particle algorithm, see \citet{Krumholz2004,Dubois2010}.

Contrary to many works on BH accretion, we do not employ a boost factor in the Bondi accretion rate, as suggested in \citet{Booth2009}, as it is intended to compensate for the unresolved gravitational attraction of the BH. Nor do we reduce the accretion rate as advocated by \citet{Curtis2016} to account for the angular momentum of the accreted gas.  We also do not limit accretion to the Eddington rate,  
\begin{equation}
	\dot{M}_{\rm Edd}= \frac{4 \pi G M_{\rm BH} m_{\rm p}}{ \epsilon_{\rm r} \sigma_{\rm T} c}
	\label{eq:eddington}
\end{equation}
 where \( \epsilon_{\rm r} \) is the radiative efficiency, \( \sigma_{\rm T} \) the Thompson cross section, \(m_{ \rm p} \) the proton mass, and \(c \) the speed of light in vacuum, but plot its value for comparison.  \New{The accretion rate is however de facto limited by the available local gas mass. Indeed, by virtue of mass conservation, the maximum gas mass accreted per timestep is always capped by the total amount of gas contained within the cells located at a distance smaller than $r_{\rm cloud}$ from the BH, regardless of whether the BHL accretion rate (Equation \ref{eq:BHL}) requires more mass be removed}.

\subsection{Dynamical friction}
 \label{sec:fdrag}

\begin{figure}
	\centering
	\includegraphics[width=\columnwidth]{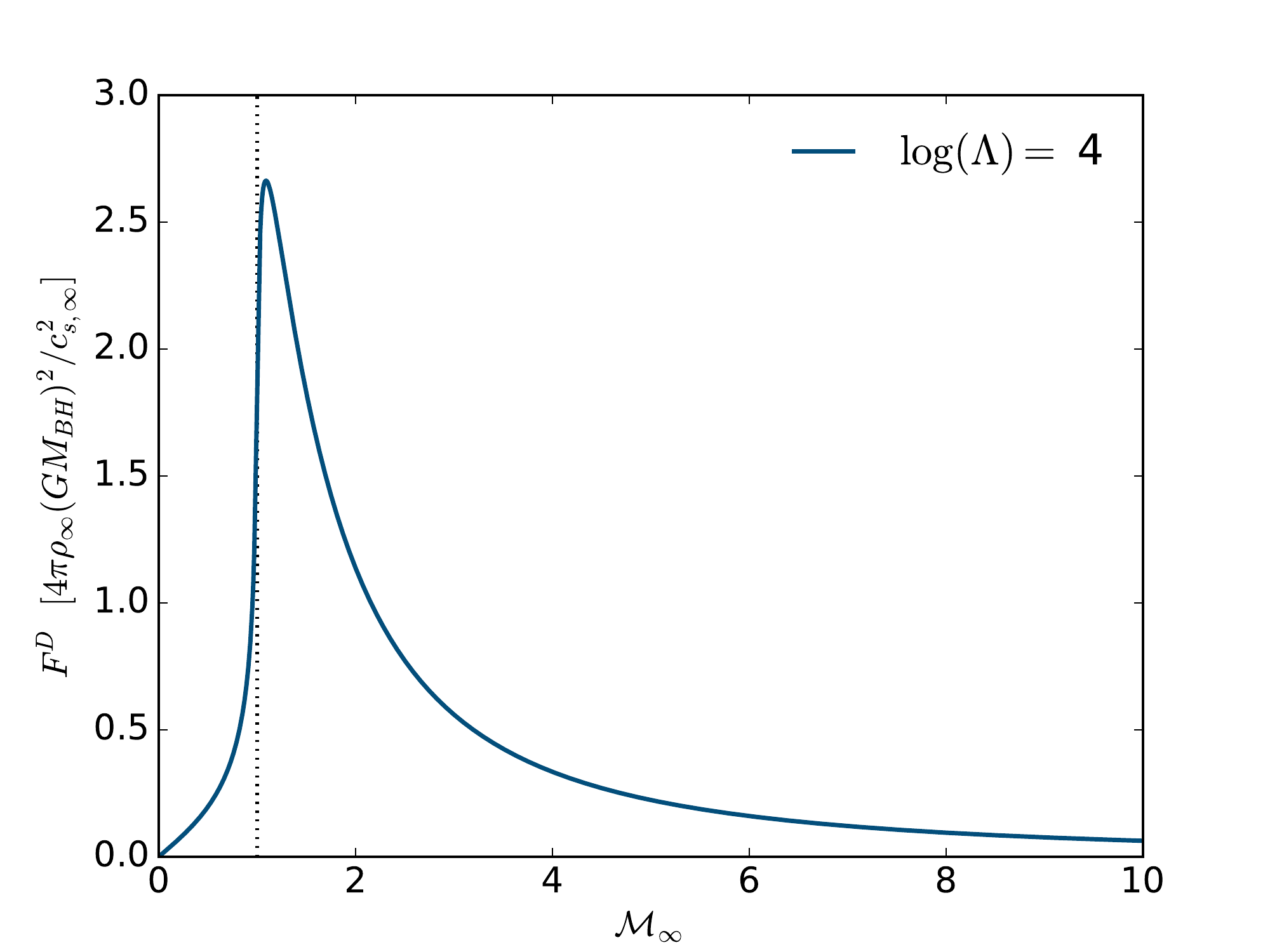}
	\caption{Magnitude of the drag force on the sink particle due to dynamical friction, which depends on the mach number $\mathcal{M}$. The dotted line denotes the sonic point. This figure is based on the analytic formula by \citet{Ostriker1998}.}
	\label{fig:Idrag}
\end{figure}

In Section \ref{sec:mseed}, we explore the option of adding a sub-grid drag force on the BH, to compensate for unresolved dynamical friction caused by the gaseous background. It follows the analytic solution of the problem given by \citet{Ostriker1998}:
\begin{equation}
	\bold{F}^{D} = - I \frac{4 \pi G^2 M_{\rm BH}^2 \rho_\infty}{v_\infty^2} \hat{\bold{v}}_\infty =  F^{D} \hat{\bold{v}}_\infty
	\label{eq:fdrag}
\end{equation}
where $ \hat{\bold{v}}_\infty $ is the unit vector of the relative velocity between gas ``at infinity'' and the BH, and 
\begin{equation}
	I = \begin{cases}
		\frac{1}{2} \ln \Big( \frac{ 1+ \mathcal{M}_\infty}{1-\mathcal{M}_\infty} \Big) - \mathcal{M}_\infty  &  { \rm if }\ \ \mathcal{M}_\infty < 1\\
		\frac{1}{2} \ln \Big( 1 - \frac{1}{\mathcal{M}_\infty^2} \Big) + \ln (\Lambda)& { \rm if }\ \ \mathcal{M}_\infty > 1 \\
		\end{cases}
		\label{eq:Idrag}
\end{equation}	
The magnitude of the drag force depends strongly on the Mach number, as can be seen in Figure \ref{fig:Idrag}. The sub-grid algorithm in \textsc{ramses} evaluates the analytic formula in Equations \ref{eq:fdrag} using the same mass-weighted quantities also used in the BHL accretion rate, $\rho_\bullet$, $v_\bullet$ and $c_{s,\bullet}$ as proxies for the same quantities at infinity. $\ln(\Lambda)=4.0$ is the Coulomb logarithm, with the value chosen based on work in \citet{Beckmann2018a}. To circumvent the discontinuity at $I(\mathcal{M}) = 1$ and numerical instabilities as $\mathcal{M} \rightarrow 0$, we extrapolate linearly between $ I(\mathcal{M}=0.99)$ and $ I(\mathcal{M}^3=1.01)$, and use a linear Taylor expansion $I \approx \frac{\mathcal{M}}{3}$ for $\mathcal{M} < 0.01$.

\section{Zooming in on the black hole}
\label{sec:bh_zoom}

\subsection{Algorithm description}
\label{sec:zoom-within-zoom}

\begin{figure}
	\centering
		\includegraphics[width=\columnwidth]{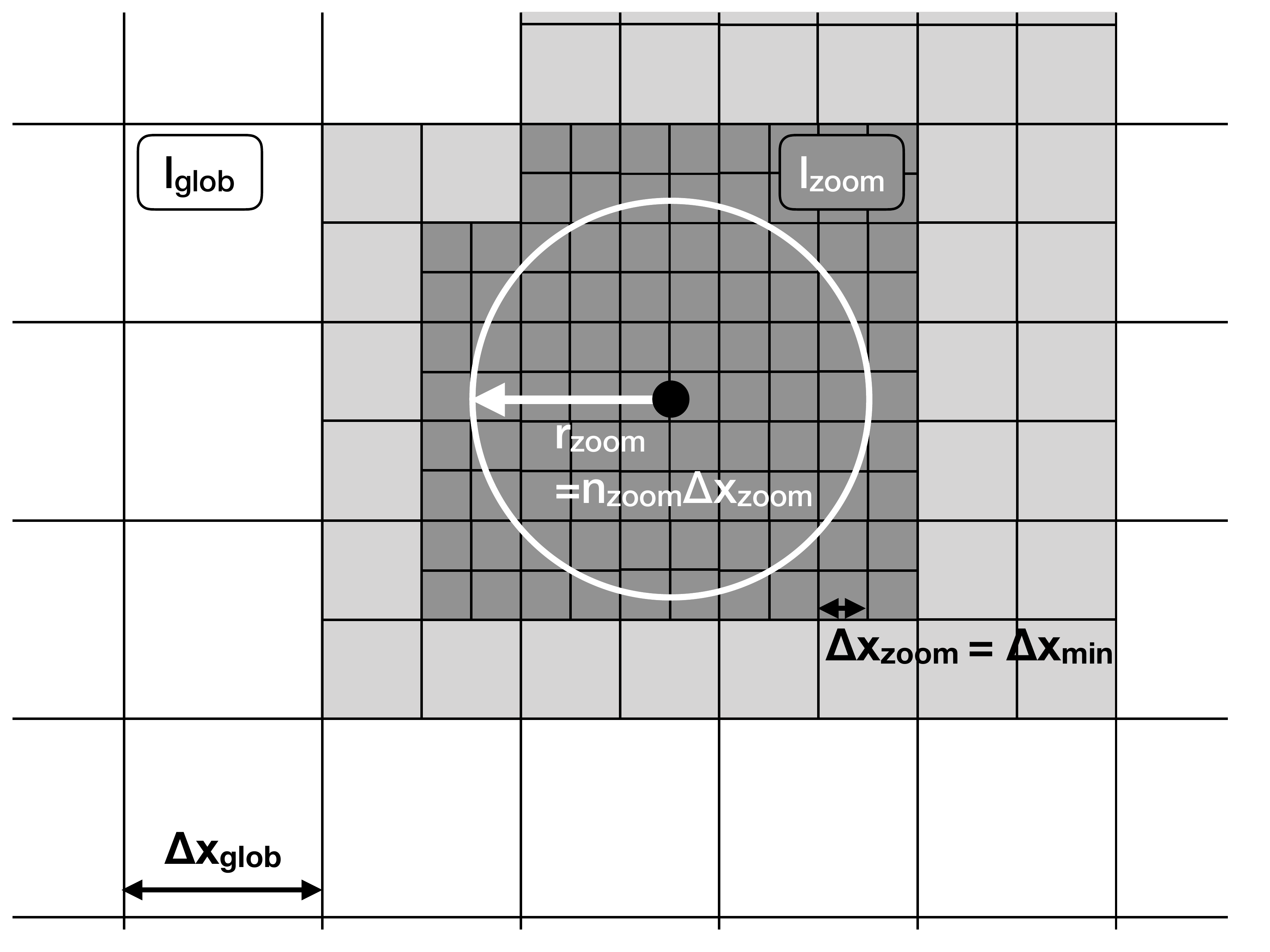}	
	\caption{Schematic grid structure generated by the BH zoom refinement algorithm, for the case with two extra refinement levels ($l_{\rm zoom} = l_{\rm glob+2}$) and $n_{\rm refine}=4$. All cells at level $l_{\rm zoom}$ are highlighted in dark grey, cells at level $l_{\rm glob}$ are shown in white and intermediate cells are highlighted in light grey. The BH is annotated as a filled black circle and solid black lines outline the grid structure. The zoom refinement radius, $r_{\rm zoom}$ is represented as a white circle.}
	\label{fig:schematic}
\end{figure}

\begin{figure}
	\centering
		\includegraphics[width=\columnwidth]{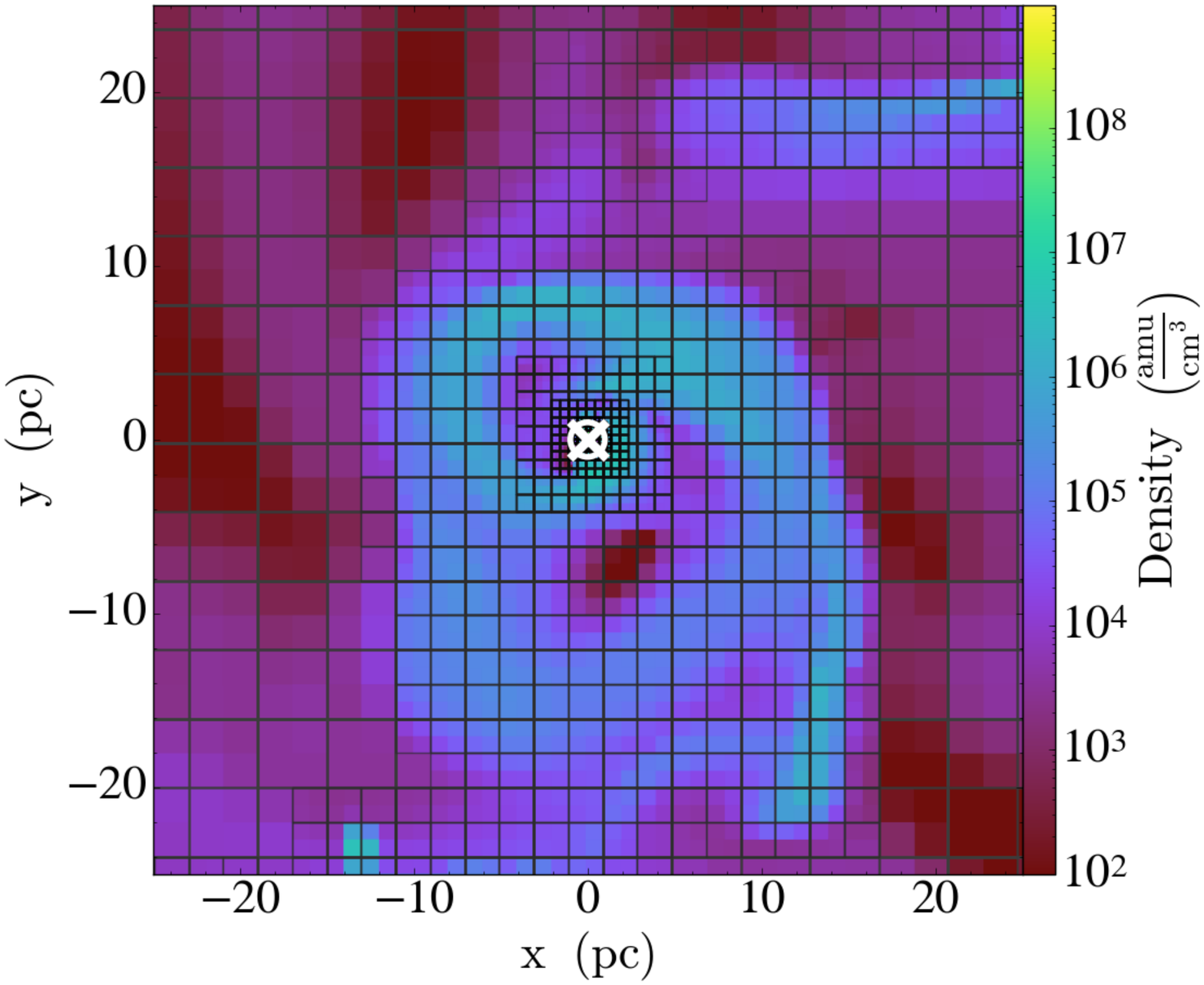}	
		\includegraphics[width=\columnwidth]{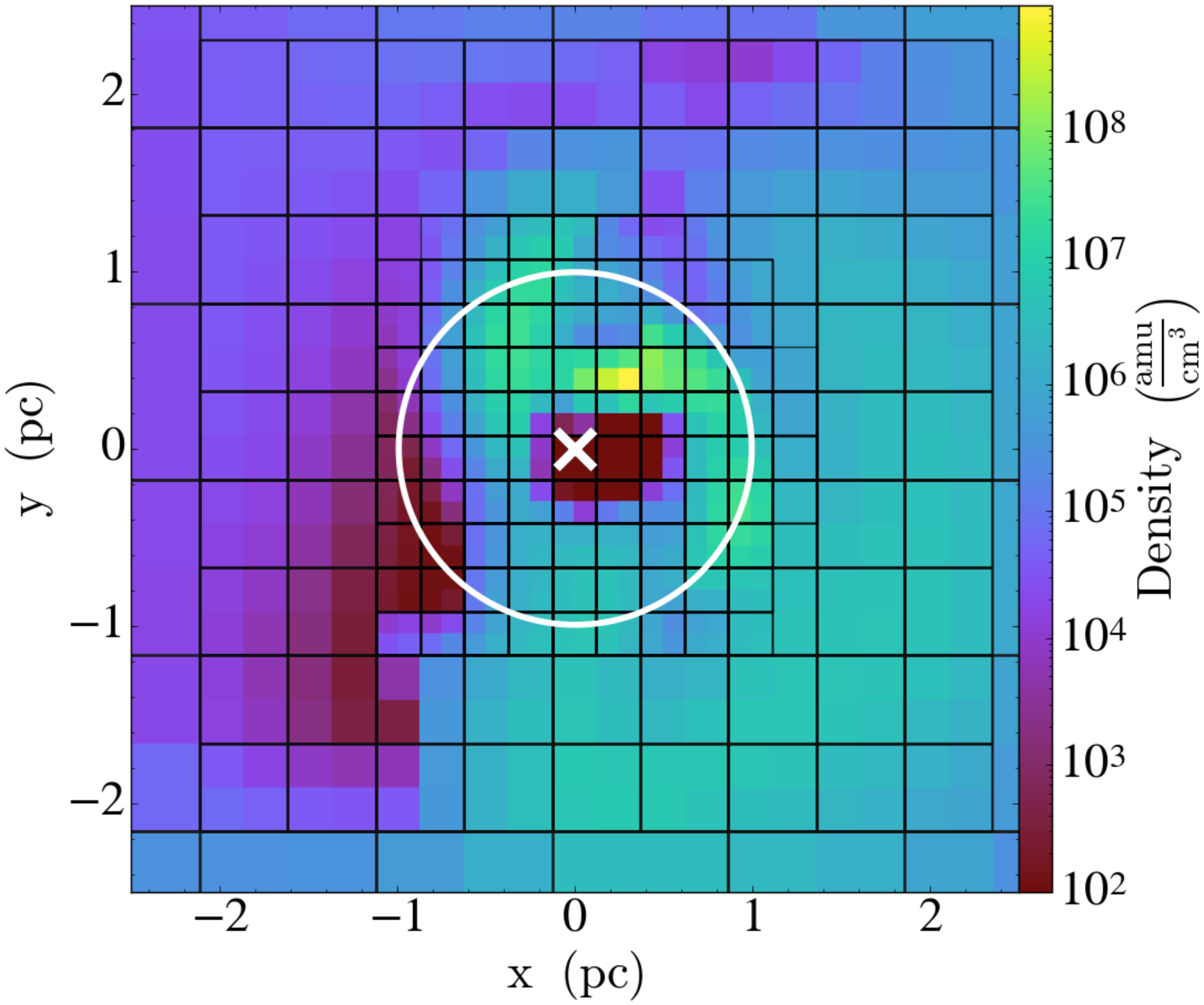}	
	\caption{Grid structure generated by the BH zoom refinement algorithm, at different length scales, plotted over a gas density projection. The example shown here has three extra levels, so $l_{\rm zoom} = l_{\rm glob}+3$. The BH position is indicated by the white cross, and the radius of the BH zoom region,  $r_{\rm zoom}$, by the white circle. Outside the BH zoom region, the galaxy is adaptively refined following a quasi-lagrangian criterion.}
	\label{fig:grids}
\end{figure}

In order to reach high resolution in the vicinity of the BH, but keep the simulations numerically affordable, we developed a novel BH zoom refinement algorithm\New{, which is shown schematically in Figure \ref{fig:schematic}}. This super-Lagrangian refinement scheme surrounds the sink particle with a spherical region at fixed, user-defined resolution $\Delta x_{\rm zoom} = {L_{\rm box}}/{2^{l_{\rm zoom}}} $, embedded in an adaptively refined galaxy at lower resolution, $ \Delta x_{\rm glob} ={L_{\rm box}}/{2^{l_{\rm glob}}}$, where $l$ describes a refinement level. \New{For simulations in which the BH zoom is activated, $\Delta x_{\rm zoom} = \Delta x_{\rm min}$}. The highly refined spherical region around the BH has a radius of $r_{\rm zoom}=n_{\rm zoom} \times \Delta x_{\rm zoom}$  where $n_{\rm zoom}$ is a free parameter kept fixed throughout the simulation (see Section \ref{sec:convergence_nrefine} for a convergence study). In order to minimise edge effects at grid boundaries, the high refinement region is surrounded by concentric shells of progressively lower refinement until $\Delta x_{\rm glob}$ is reached, as can be seen in Figure \ref{fig:grids}. This ensures that neighbouring cells differ at most by a single refinement level. The refinement region is centred on the sink particle position at all times, tracking its movement through the simulation box, with cells (de-)refining as appropriate. \New{As the timestep in RAMSES is limited in such a way that no particle can cross more than a single cell in any given timestep, the grid around the BH is refined and de-refined sufficiently frequently to always be quite accurately centred on the BH.}

Once a BH particle has been created, the BH zoom algorithm is activated. To allow newly created cells to be distributed over several processors, and to minimise numerical artefacts due to the number of levels being added, cells are only allowed to be refined by two extra levels per coarse timestep of the simulation. The simulation is load-balanced after each coarse timestep during the level triggering phase. Accretion onto the BH is only activated once the target level $l_{\rm zoom}$ has been reached, to allow the structure of the gas flow to emerge and avoid contaminating the early accretion history. As explored in detail in Section \ref{sec:mseed}, a maximum drag force is applied to the sink during the initial level triggering phase, i.e. we set the relative velocity between the sink particle and the gas $v_\bullet =0 $ at each timestep.  

The mass of newly created star particles depends on the gas mass contained by the cells in which they formed, so that at a given gas density smaller cells form lighter star particles. To avoid overwhelming the simulation with a large number of star particles that carry only a small amount of mass, star formation is prevented in any cell with $\Delta x < \Delta x_{\rm glob}$. This has negligible influence on the total stellar mass of the galaxy (see Section \ref{sec:resolution}).

\textsc{ramses} uses a cloud-in-cell method to calculate the total density distribution  to solve the Poisson equation, where the mass of each particle is distributed over local grid cells according to
\begin{equation}
	W(x-x_p) = \begin{cases}
		1 - | x - x_p | / \Delta x & | x - x_p| \leq \Delta x \\
		0 & {\rm otherwise. }
		\end{cases}
\end{equation}
where $x$ and $x_p$ are the position of the cell centre and particle, and $\Delta x$ is the cell size. In simulations with a large range of refinement levels, this can lead to spurious local maxima in the density field if a massive particle is deposited into an extremely small cell. To avoid this issue with the BH zoom, we deposit the mass of star particles at a maximum level $l_{\rm glob}$, and use the OctTree structure of RAMSES to distribute mass into child cells at higher refinement levels (constant density extrapolation). This is a commonly used technique to deal with massive dark matter particles in cosmological zoom simulations (e.g. \citet{Powell2011}). 

\subsection{{Convergence of black hole zoom}}
\label{sec:convergence_nrefine}

To quantify the the impact of varying $n_{\rm zoom}$ in BH zoom simulations, a BH of mass $M_{\rm BH} = 260 \ \rm M_\odot$ is placed at the centre of a non-rotating version of the halo described in Section \ref{sec:initial_conditions} and the gas is allowed to cool. Simulations with  $2 \leq n_{\rm zoom} \leq 16$ are compared to two fiducial simulations using only the standard quasi-Lagrangian refinement, called C\_halo16 and C\_halo20 respectively (see Table \ref{tab:convergence} for details). When not using the BH zoom, the BH host cell is forced to remain at $l_{\rm glob}$ to avoid spurious local de-refinement. For simulations using BH zoom,  $l_{\rm glob} = 16$ and $l_{\rm zoom} = 20$, leading to a resolution of $\Delta x_{\rm zoom} = 0.99 \rm \ pc$ within the BH zoom region, and a maximum of $ \Delta x_{\rm glob} = 15.8 \rm \ pc$ outside. 

The initial density distribution is refined up to $l_{\rm glob}$, so a higher $l_{\rm glob}$ leads to a more peaked initial density profile. The BH is inserted at the centre of the simulation volume at $t=0 \times \ t_{\rm ff}/10^3$, where $t_{\rm ff}$ is the free fall time of the halo. The BH zoom levels are added during the first coarse timesteps of the simulation until $\Delta x_{\rm min} = \Delta x_{\rm zoom}$, where $\Delta x_{\rm min}$ is the size of the smallest cell in the simulation.

\begin{table*}
\setlength{\tabcolsep}{15pt}
\centering
	\begin{tabular}{  l | c | c | c | c  |  c | r}
		\hline
		\multicolumn{6}{| c |}{\bf{Convergence test simulations}} \\
		\hline
 		name & $l_{\rm glob}$ & $ l_{\rm zoom}$ & $\Delta x_{\rm glob}$ $[\rm pc]$ & $\Delta x_{\rm zoom}$ $[\rm pc]$ & $n_{\rm zoom}$ & CPU hours ** \\ 
		\hline
		C\_l20n2 & 16 & 20 & $15.8 \rm \ pc$ &  $ 0.99 \rm \ pc$ & 2 & 3.35 \\
		C\_l20n4 & 16 & 20 & $15.8 \rm \ pc$ &  $ 0.99 \rm \ pc$ & 4 & 3.21\\
		C\_l20n8 & 16 & 20 & $15.8 \rm \ pc$ &  $ 0.99 \rm \ pc$ & 8 & 3.33\\
		C\_l20n16 & 16 & 20 & $15.8 \rm \ pc$ &  $ 0.99 \rm \ pc$ & 16 & 3.54\\
		\hline
		C\_halo16$^*$ & 16 & - & $15.8 \rm \ pc$ &  $  - $ & - & 0.47 \\
		C\_halo20$^*$ & 20 & - & $0.99 \rm \ pc$ &  $ - $ & - & 4.2 \\
		\hline
		C\_l18n8 & 16 & 18 & $15.8 \rm \ pc$ &  $ 3.9 \rm \ pc$ & 8 & 1.61\\
		C\_l20n8 & 16 & 20 & $15.8 \rm \ pc$ &  $ 0.99 \rm \ pc$ & 8 & 3.33\\
		C\_l22n8 & 16 & 22 & $15.8 \rm \ pc$ &  $ 0.25 \rm \ pc$ & 8 & 6.83\\
		C\_l24n8 & 16 & 24 & $15.8 \rm \ pc$ &  $ 0.06 \rm \ pc$ & 8 & 9.37\\
		\hline
		\multicolumn{6}{l}{* equivalent to setting $n_{\rm zoom}=0$} \\
		\multicolumn{6}{l}{** per $t_{\rm ff}$ of evolution on 12 cores}
	\end{tabular}
	\caption{Parameters for simulations used in the convergence tests of Sections \ref{sec:convergence_nrefine} and  \ref{sec:convergence_levelmax}.}
	\label{tab:convergence}
\end{table*}

\begin{figure}
	\centering
	\includegraphics[width=\columnwidth]{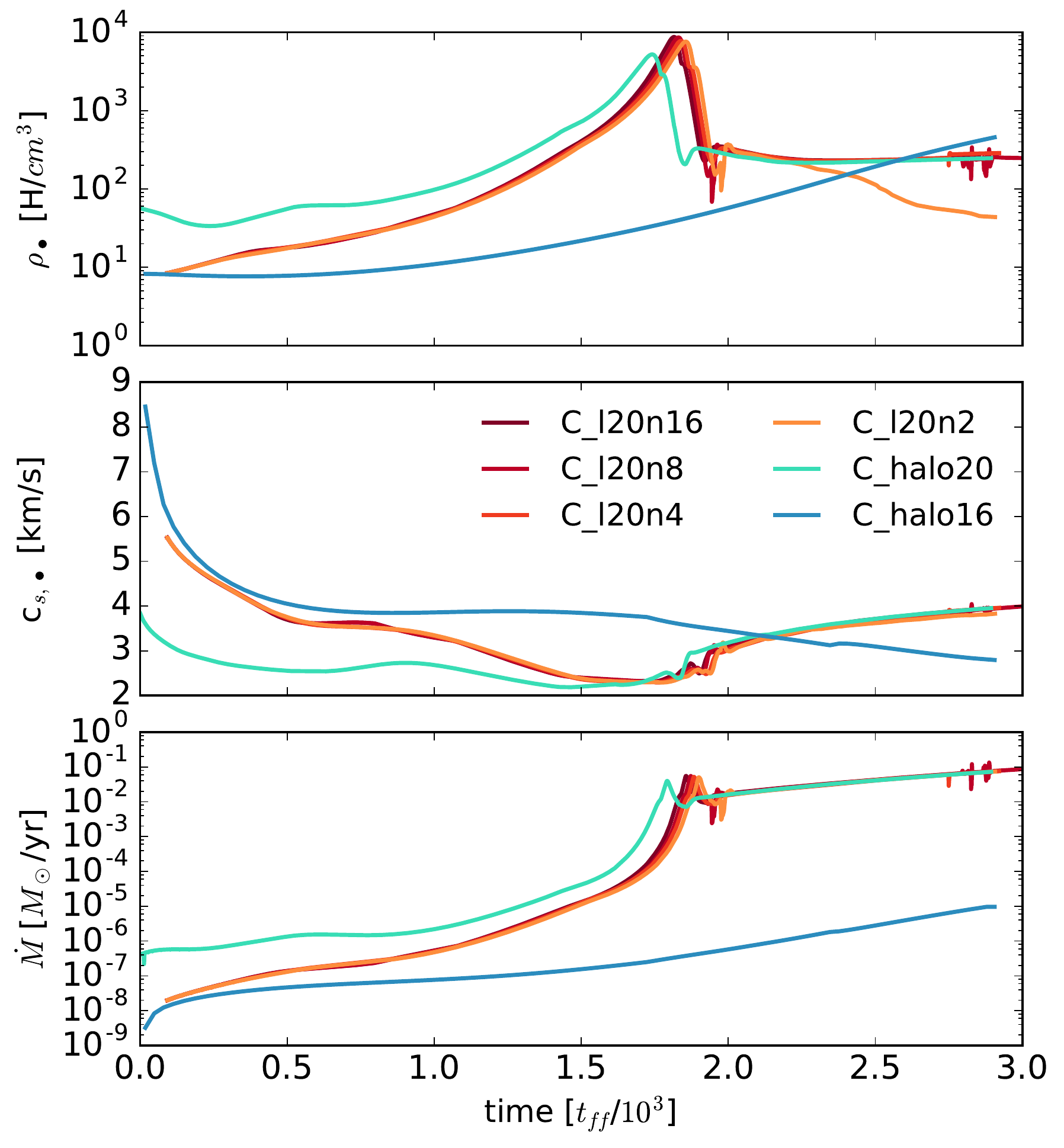}
	\caption[Density, sound speed and accretion rate, as measured by the BH, for a range of different radii of the BH zoom region]{Density (top panel), sound speed (middle panel) and accretion rate (bottom panel), as measured by the BH, for different radii of the BH zoom region,  \mbox{$r_{\rm zoom} = n_{\rm zoom} \Delta x_{\rm zoom}$} (see Table \ref{tab:convergence}). Convergence is achieved for $n_{\rm zoom} \geq 4 $.}
	\label{fig:nsink_convergence}
\end{figure}

Figure \ref{fig:nsink_convergence} shows that BH zoom refinement allows the gas in the vicinity of the BH to evolve from the initial conditions of a low resolution simulation, C\_halo16, to the solution of the high resolution simulation, C\_halo20. Convergence behaviour of mass weighted density, $\rho_\bullet$, and sound speed, $c_{s,\bullet}$, is very good for \mbox{$n_{\rm zoom} \geq 4 $}. When $n_{\rm zoom}<4$, gas properties fail to converge as the maximum extent of the cloud particles, set at $r_{\rm cloud} = 4  \Delta x_{\rm zoom}$, is larger than the size of the smallest BH zoom grid, $r_{\rm zoom}=n_{\rm zoom}  \Delta x_{\rm zoom}$. When this happens, the outermost cloud particles probe regions outside $r_{\rm zoom}$ where gas properties reflect the density profile a factor of two further away than if the cell were refined an extra level. 

In conclusion, \New{in the accretion only case}, the BH zoom scheme is not sensitive to the choice of $n_{\rm zoom}$ as long as  $n_{\rm zoom} \geq 4 $, so that the highest refinement region is larger than the accretion region, i.e. $r_{\rm zoom} \geq r_{\rm cloud}$.

\subsection{{Supply-limited accretion}}

One notable feature of Figure \ref{fig:nsink_convergence} is the discontinuity in the evolution of the gas near $t=1.8 \ t_{\rm ff}/10^3$. 
At this point, the amount of mass scheduled for accretion (calculated using \mbox{Equation \ref{eq:BHL}}), \mbox{$\dot{M}_{\rm BHL} \times \Delta t  = G^2 M_{\rm BH}^2 \rho_\bullet \Delta t / (v^2_\bullet + c^2_{s,\bullet}) ^{3/2} $}, exceeds the total gas mass available within the accretion region, $\rho_\bullet V_\bullet$, where  $V_\bullet$ is the constant volume of the accretion region and $\Delta t$ is the timestep. \New{Assuming BHL accretion,} this occurs for a minimum BH mass of:
\begin{equation}
	M_{\rm SLA} = \sqrt{\frac{V_\bullet  }{G^2 \Delta t  } (v^2_\bullet + c^2_{s,\bullet}) ^{3/2}}.
	\label{eq:Mflux}
\end{equation}
As mass is conserved, $M_{\rm BH} > M_{\rm SLA}$ results in the maximum available mass being removed at each timestep, transitioning the accretion rate from being based on the BHL formula to being set by the mass flux into the accretion region, so called supply limited accretion (SLA). This occurs approximately when the BH accretion scale radius becomes equal to the resolution.

The BH remains in SLA as long as $ M_{\rm BH}  > M_{\rm SLA} $. At fixed $l_{\rm zoom}$, $V_\bullet$ and $\Delta t$ are approximately constant. In the spherically symmetric case studied here, $v_\bullet \approx 0$, and Figure \ref{fig:nsink_convergence} shows that $c_{s,\bullet}$ varies by less than a factor of two over the course of these simulations. For all intents and purposes, the right-hand side of Equation \ref{eq:Mflux} can therefore be considered constant, and the BH remains in SLA as $M_{\rm BH} > M_{\rm SLA}$ continues to hold. If there exists a non-negligible relative velocity $v_\bullet > 0$, conservation of momentum during accretion decreases the relative velocity, lowering $M_{\rm SLA}$ while increasing $M_{\rm BH}$ and thus driving the system towards the SLA regime. In less idealised simulations, where $c_{s,\bullet}$ and  $v_\bullet$ can vary strongly, it is possible for the BH to transition back to BHL accretion, but again, gas cooling and accretion will tend to reduce sound speed and relative velocity respectively, thus steering towards SLA. \New{If the BH accretion rate is further restricted, e.g. capped at the Eddington accretion rate given in Equation \ref{eq:eddington}, the transition mass will be unchanged but the transition will be postponed as the BH will take (much) longer to reach the SLA  threshold mass.}

\begin{figure*}
	\centering
	\setlength{\tabcolsep}{0pt}
	\begin{tabular}{lcr }
		\includegraphics[width=0.32\textwidth]{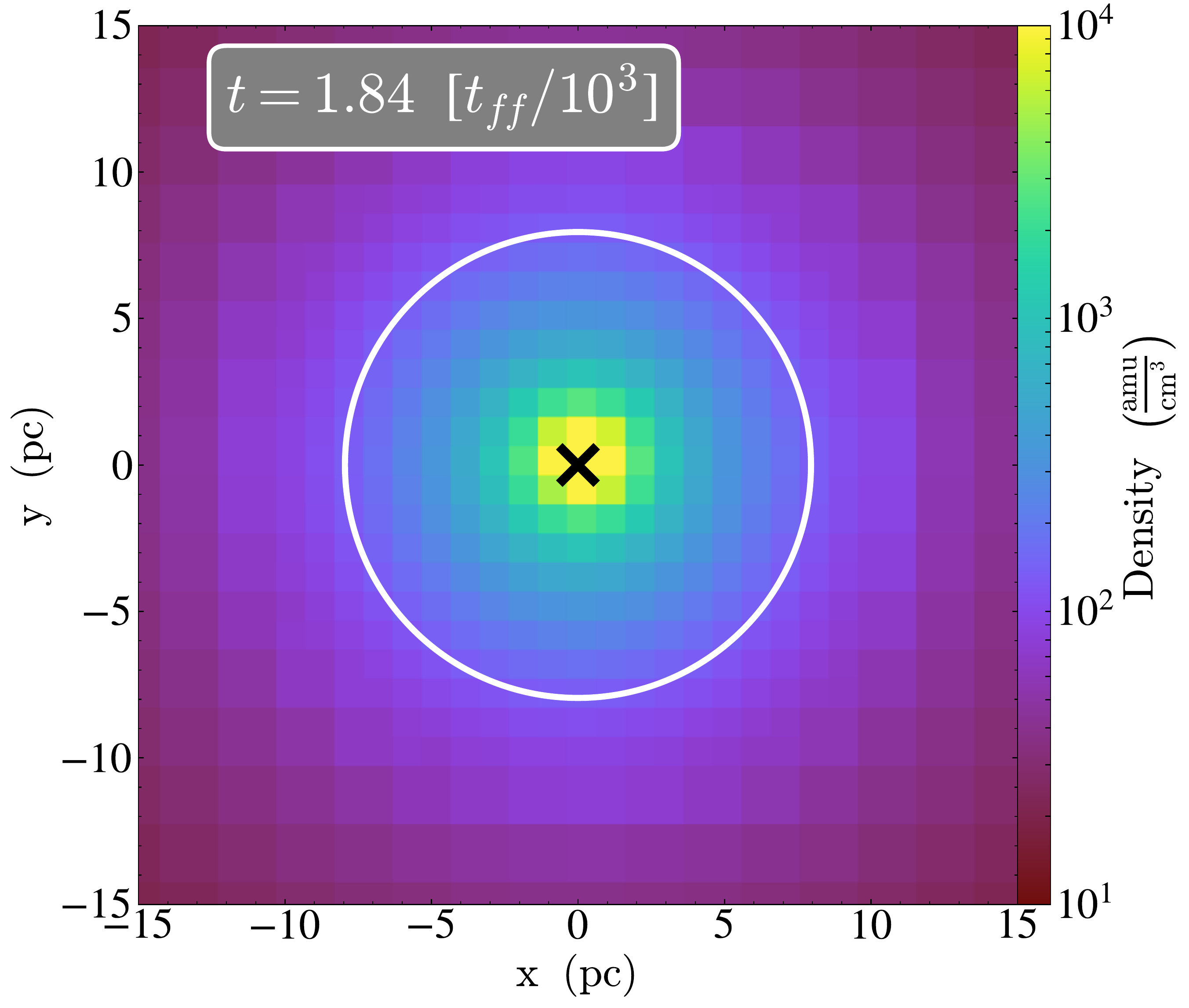} &
		\includegraphics[width=0.32\textwidth]{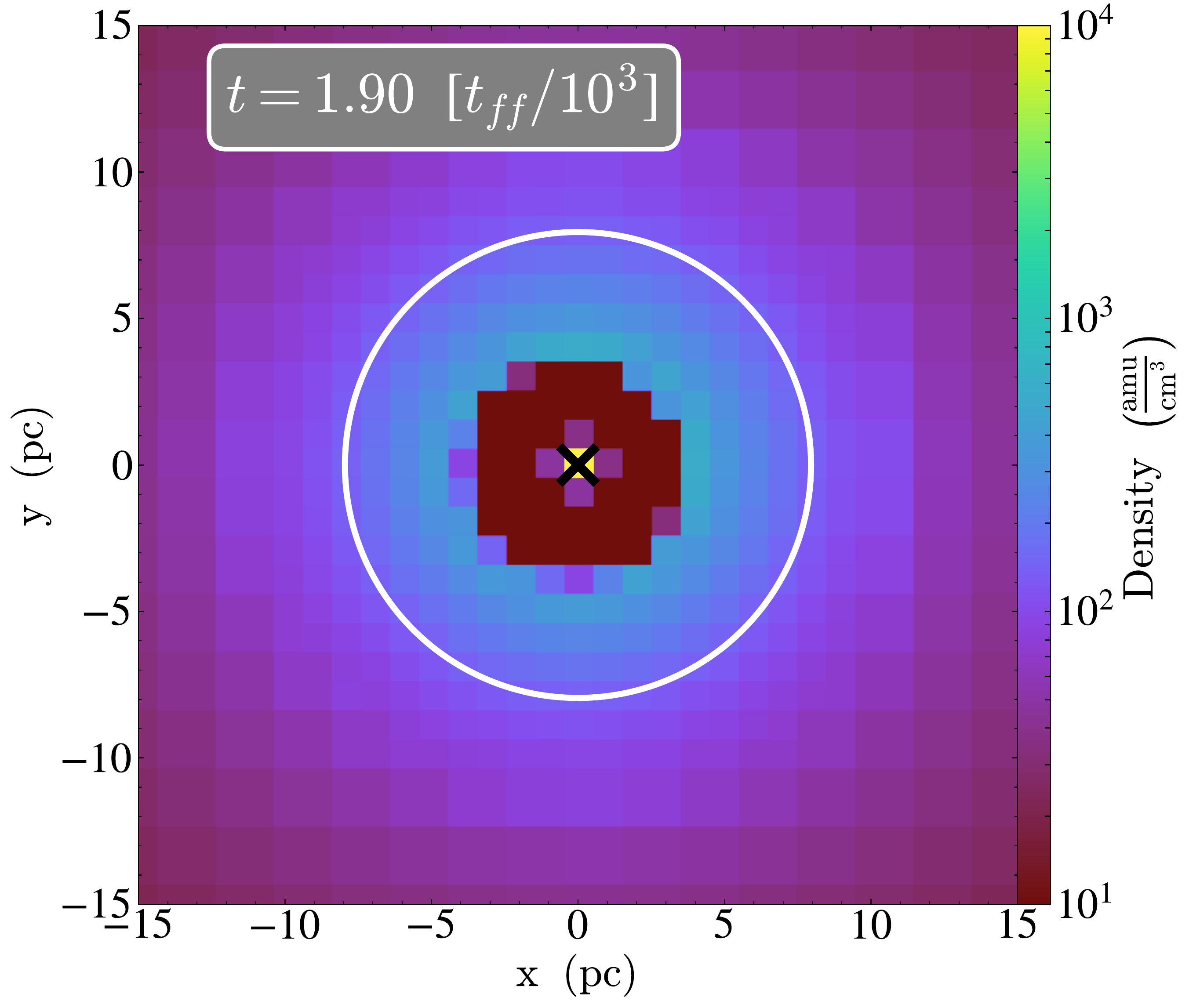} &
		\includegraphics[width=0.32\textwidth]{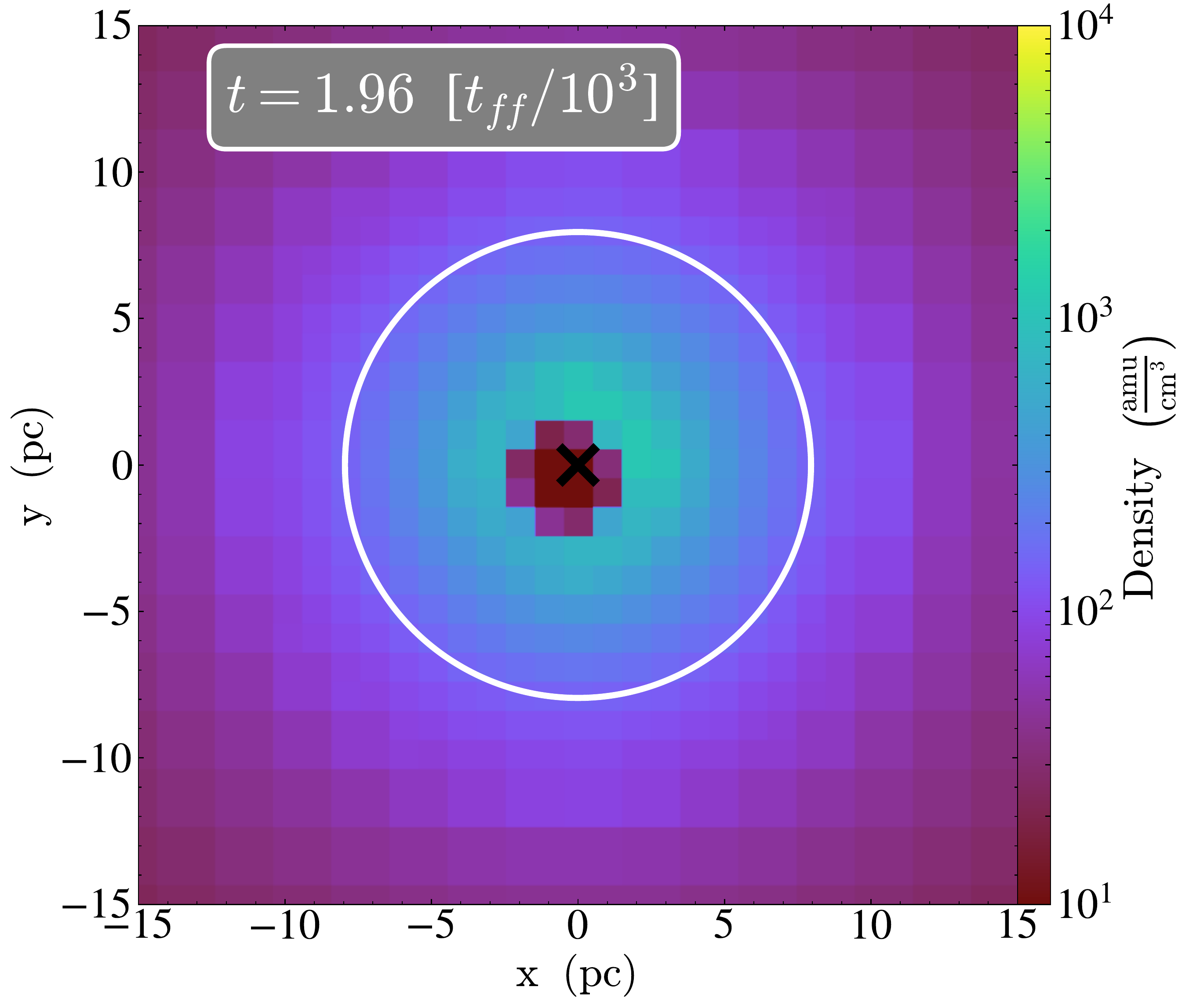} \\
	\end{tabular}
	\caption[Density slices of the gas in the vicinity of the BH during the transition from BHL to SLA]{Density slices of the gas in the vicinity of the BH in C\_l20n8 at various epochs during the transition from BHL accretion to SLA. The black cross denotes the position of the Bh, and $r_{\rm zoom}$ is marked by the white circle.}
	\label{fig:flux}
\end{figure*}

During the transition to SLA, the central density structure is accreted by the BH and replaced with a low density accretion region (see Figure \ref{fig:flux}), decreasing the kernel weighted average $\rho_\bullet$. For reasons of numerical stability, only $75 \% $ of a cell mass can be removed in a single accretion step, so densest cells are emptied more slowly (middle panel of Figure \ref{fig:flux}). As evident in Figure \ref{fig:nsink_convergence}, the transition to SLA is therefore not an instantaneous process. 

\New{The transition mass $M_{\rm SLA}$ does show an explicit resolution dependence. For BH zoom, the BH is always located in a cell at maximum resolution so $V_\bullet = \Delta x_{\rm zoom}^3$. We can approximate the minimum timestep of the simulation as the sound crossing time of the smallest cell, $\Delta t = \Delta x_{\rm zoom}/ c_s$. Equation \ref{eq:Mflux} then shows that $M_{\rm SLA}$ depends linearly on resolution, as $M_{\rm SLA} \propto \sqrt{\frac{V_{\bullet}}{\Delta t}} = \sqrt{\frac{\Delta x_{\rm zoom}^3 }{ \Delta x_{\rm zoom} / c_s}} \propto \Delta x_{\rm zoom}$. Simulations with smaller $\Delta x_{\rm zoom}$ but otherwise identical setup therefore transition earlier to SLA, as shown and discussed in more detail in Appendix \ref{sec:convergence_levelmax}. However, for the simple test case of a spherically collapsing cloud, the delay in transition has no influence on the longterm mass evolution of the BH, which converges for simulations spanning over two orders of magnitude in $\Delta x_{\rm zoom}$. As both the BH and the gas are bound at the bottom of the gravitational well, infalling gas not yet accreted by the BH simply collects in its vicinity to be accreted during the transition to SLA. In a more complex environment, the finite lifetime of the cloud feeding the BH could become a concern, as gas not yet accreted could become gravitationally unbound from the BH before the transition to SLA occurs.}

In conclusion, for sufficiently high resolution, accretion onto the BH automatically transitions from being based on the analytic BHL model to being determined by the mass flux into the accretion region of the BH, i.e. to the so called supply limited accretion (SLA) mode.


\section{Black hole accretion in collapsing clouds}
\label{sec:mseed}

As laid out in the introduction to this paper, three different formation channels have been proposed for SMBH progenitors, producing seed BHs in the range $10-10^5 \rm \ M_\odot$ \New{\citep[see][for a review]{Volonteri2010}}. In literature, it is frequently reported that stellar mass seeds fail to grow in comparison to more massive seed masses, which has been taken as tentative evidence in favour of the direct collapse BH seed model \New{\citep{Pelupessy2007,Pacucci2015a}}. The failure of light seeds to grow is commonly attributed to their low masses that limit their ability to gravitationally capture gas, and to feedback blowing away their gas supply. In this section, we demonstrate that light seeds can fail to grow simply because their dynamics are under-resolved in a given simulation. We then present an algorithm that produces a converged dynamical evolution history for a BH seeded at a given time in a given gas cloud, taking advantage of the extra information revealed by the BH zoom algorithm.

\subsection{The impact of dynamical friction on early black hole accretion}
\label{sec:dynamical_friction}

\begin{table}
	\setlength{\tabcolsep}{10pt}
	\centering
	\begin{tabular}{lccr}
		\hline
		\multicolumn{4}{|c|}{\bf{Disc galaxy simulations}} \\
		\hline
		\bf{name} & $l_{\rm zoom}$  & $M_{\rm seed}$ [$\rm M_\odot$]   & Drag force \\
		\hline
		D\_l26\_tiny & 26 & 260  & None\\
		D\_l26\_small & 26 & $5\times10^3$ & None\\
		D\_l26\_medium & 26 & $2.6\times10^4$ & None\\
		D\_l26\_big & 26 & $1\times10^5$ & None\\
		D\_l26\_huge & 26 & $2.6\times10^5$ & None\\
		\hline
		O\_l26\ & 26 & 260  & Ostriker1999 \\
		\hline
		F\_l26\_a & 26 & $10^{-2}$ & Max \\
		F\_l26\_b & 26 & $10^{-1}$  & Max \\
		F\_l26\_c & 26 & $1$ & Max \\
		F\_l26\_d & 26 & $260$ & Max \\
		F\_l26\_e & 26 & $10^{4}$ & Max \\
		F\_l26\_f & 26 & $10^{5}$ & Max \\
		\hline
	\end{tabular}
	\caption{Parameters for simulations in Section \ref{sec:mseed}. All simulations have $n_{\rm zoom}=8$, $l_{\rm glob}  = 20$  and $l_{\rm zoom} = 26$. This leads to a resolution if $\Delta x_{\rm glob}=0.99 \rm \ pc$ within the galaxy and $\Delta x_{\rm zoom} = 0.015 \rm \ pc$ in the black hole vicinity.}
	\label{tab:disc_parameters}
\end{table}

To study the formation and evolution of BHs in collapsing clouds in a galactic context, rotation is added to the halo described in Section \ref{sec:initial_conditions} and it is evolved for $100 \rm \ Myr$ at a resolution of $\Delta x_{\rm glob} = 0.99 \rm \ pc$ to allow the gas to cool and fragment. A collapsing cloud is identified using the {\sc phew} algorithm, and a sink particle of a given mass $M_{\rm seed}$, which inherits the position and velocity of its parent cell, is inserted. Once the BH has been seeded, the BH zoom algorithm refines the black hole environment to $\Delta x_{\rm zoom}=0.01\rm \ pc$, using $n_{\rm zoom} = 8$. The parameters for simulations in this section are summarised in Table \ref{tab:disc_parameters}.

\begin{figure}
	\centering
	\includegraphics[width=0.9\columnwidth]{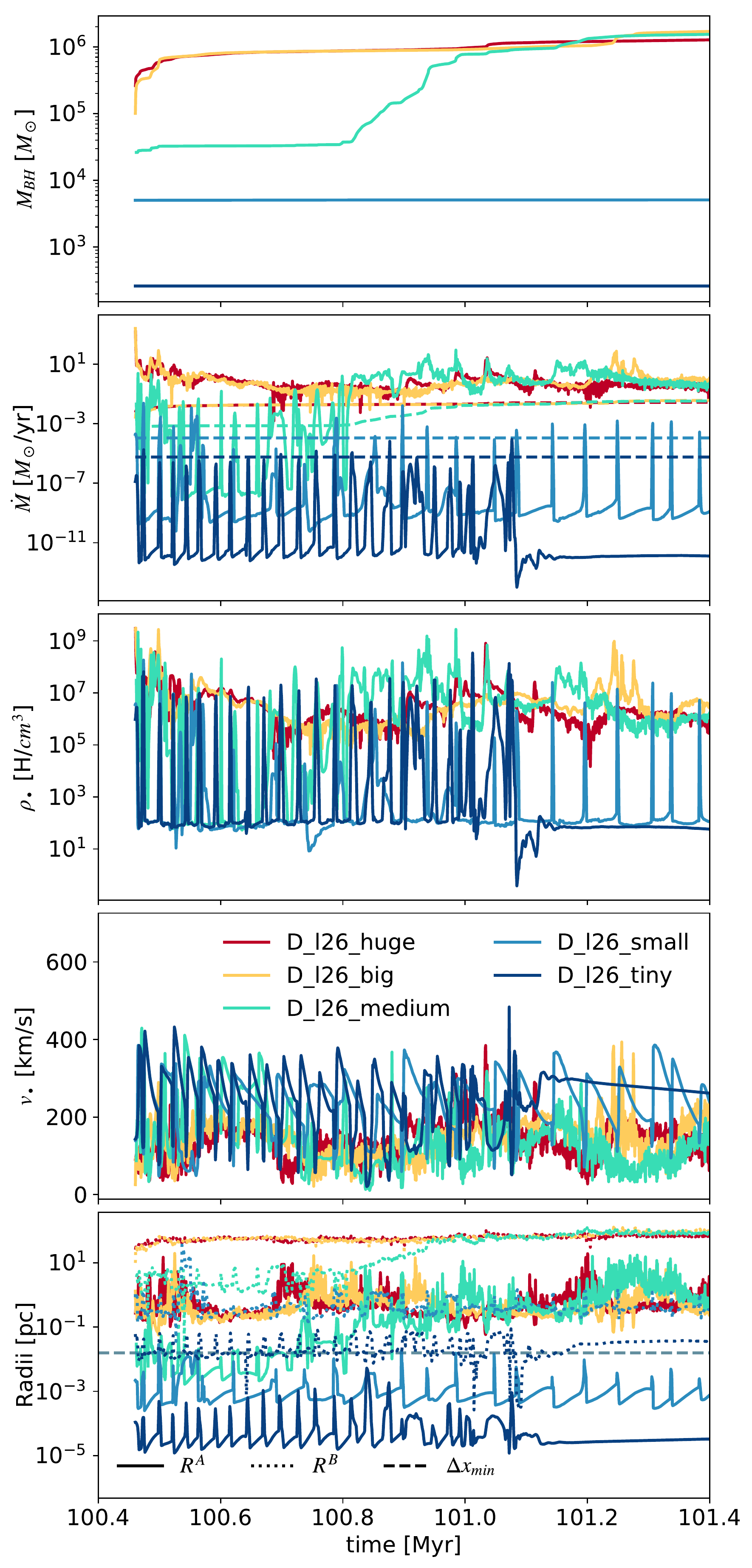}
	\caption[Time evolution of mass averaged quantities in the accretion region of BHs seeded with various masses]{Time evolution of mass averaged quantities in the accretion region of BHs seeded with various BH seed masses in identical collapsing clouds (see Table \ref{tab:disc_parameters} for details). The dashed lines on the second panel, $\dot{M}$, represent the Eddington accretion rates. The bottom panel shows the scale radii for each simulation, $R^{\rm A}$ and $R^{\rm B}$ (see Equations \ref{eq:Ra} and \ref{eq:Rb}), in comparison to the minimum resolution of the simulation, $\Delta x_{\rm min} = \Delta x_{\rm zoom}$, which is plotted as a grey dashed line. }
	\label{fig:bondiplot_lvl26_nodrag}
\end{figure}

Figure \ref{fig:bondiplot_lvl26_nodrag} shows that the evolution of a BH depends crucially on the choice of $M_{\rm seed}$, even in the absence of feedback. The two simulations with the most massive seeds, D\_l26\_big and D\_l26\_huge, using  $M_{\rm seed} = 1\times 10^5 \ \rm M_\odot $ and $M_{\rm seed} = 2.6 \times 10^5 \ \rm M_\odot$ respectively, rapidly converge. 
For the remainder of the simulations, while not identical due to the chaotic nature of the problem, their accretion rates, density $\rho_\bullet$ and relative velocity, $v_\bullet$, show the same trends (see central three panels of Figure \ref{fig:bondiplot_lvl26_nodrag}). After a delay of $0.6 \rm \ Myr$, and a prolonged growth phase, the mass of D\_l26\_medium also converges to the value reached by the two larger seed masses. By contrast BHs with smaller seed masses, D\_l26\_tiny and D\_l26\_small do not grow appreciably in mass.

\begin{figure*}
	\centering
	\includegraphics[width=0.95\textwidth]{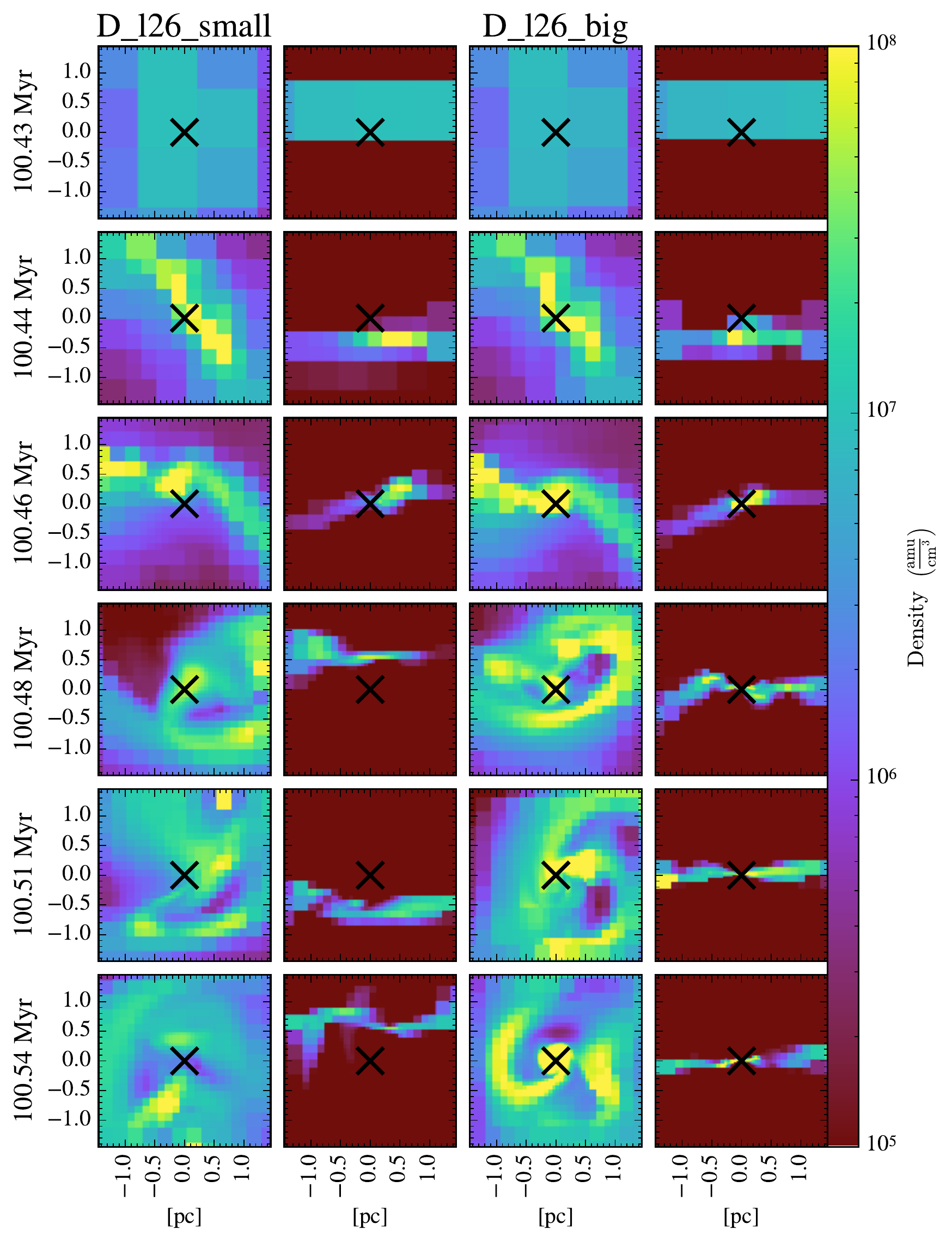}
	\caption[Density projections face-on and density slices edge-on of the collapsing cloud in D\_l26\_small and D\_l26\_huge]{Density projections face-on (1st and 3rd column) and density slices edge-on (2nd and 4th column) of the collapsing cloud in D\_l26\_small (left columns) and D\_l26\_huge (right columns) at 5 consecutive points in time. The black hole location is marked by a black cross in each panel.}
	\label{fig:slices_D_l26_masses}
\end{figure*}

The accretion histories of light and heavy seeds are driven by their diverging dynamics. As can be seen in Figure \ref{fig:slices_D_l26_masses}, the cloud studied here has sufficient angular momentum to collapse to a disc. Black holes with $M_{\rm BH} \geq 10^5 \ \rm M_\odot$ remain within the plane of the disc during collapse, (right hand columns of Figure \ref{fig:slices_D_l26_masses}). Lighter black holes (left hand columns of Figure \ref{fig:slices_D_l26_masses}) oscillate perpendicularly to the disc, with a vertical amplitude that significantly exceeds the scale height of the disc. Therefore, massive seeds with $M_{\rm BH} \geq 10^5 \ \rm M_\odot$ remain attached to dense gas features and accrete efficiently, while lighter seeds spend the majority of their time in low density regions (see third panel of Figure \ref{fig:bondiplot_lvl26_nodrag}) above and below the disc and therefore accrete less. While it is possible that the choice of sub-grid accretion algorithm has some influence on the results, the more stringent limitation clearly is the inability of the BH to remain attached to the cloud during the collapse phase.

Keeping sink particles attached to gas features is a common challenge in hydrodynamical simulations \citep{Sijacki2007,Volonteri2016,Biernacki2017}. Physically, a black hole has two mechanism to exchange momentum with the gas: (i) accretion, which transfers momentum from the gas to the BH and (ii) gravitational focusing of the gas into an overdense wake downstream of the BH, whose gravitational attraction acts as a drag force \citep[see][for an analytic derivation]{Just1990}. Resolving this wake on the grid requires the local resolution to be comparable to the gravitational scale radius of the BH, i.e. $R^{\rm A} = \frac{ 2 G M_{\rm BH}}{v_{\infty}^2} \sim \Delta x_{\rm zoom}$ for a supersonically moving BH (as explored in detail in \mbox{\citet{Beckmann2018a}}), where $v_{\infty}$ is the relative velocity between the BH and the bulk of the gas. Resolving the drag force therefore depends on the local cell size (identical for all D\_l26 simulations) and the mass of the BH. Unfortunately, the drag force is not numerically self-correcting in the same way that accretion is. Indeed, if $R^{\rm A} > \Delta  x$ initially, the BH reduces its relative velocity, $v_{\rm BH}$, by transferring momentum to the gas, increasing $R^{\rm A}$ further. By contrast, if $R^{\rm A} < \Delta x$ originally, the relative velocity remains high and $R^{\rm A}$ remains unresolved. 

The analytic work used to calculate $R^{\rm A}$ assumes that gas on scales larger than $R^{\rm A}$ is homogeneous, uniform, not self-gravitating, and not subject to a gravitational potential except that of the BH, none of which are reasonable assumptions in the context presented here. Despite these limitations, the final panel of Figure \ref{fig:bondiplot_lvl26_nodrag} shows that the accretion radius $R^{\rm A}_\bullet$, evaluated using local mass-weighted quantities, remains a reliable test of drag force resolution. For the two heaviest seeds, D\_l26\_big and D\_l26\_huge, $R^{\rm A}_\bullet > 10 \Delta x_{\rm zoom}$ at all times. The two smallest simulations, D\_l26\_small and D\_l26\_tiny, by contrast have $R^{\rm A}_\bullet << \Delta x_{\rm zoom}$. D\_l26\_medium has an accretion radius $R^{\rm A}_\bullet$ that is marginally resolved. The brief intervals when $R^{\rm A}_\bullet > \Delta x_{\rm zoom}$ produce sufficient drag on the BH for the oscillations around the disc plane to decay, allowing the BH to settle into the disc at $t=100.7 \rm \ Myr$ where it accretes the disc dense core and converges to the same mass as D\_l26\_big and D\_l26\_huge. 

Dynamical friction therefore clearly plays a decisive role in the early evolution of BH accretion in collapsing clouds. Under-resolving this force by choosing too light a seed mass has a lasting impact on the BH dynamics and mass accretion history, making the choice of seed mass just as crucial as the choice of accretion or feedback algorithm. Too light, and the BH never grows. Too massive, and too much mass is removed on pre-collapse scales, smoothing out the internal structure of the cloud and polluting the early evolution of the BH mass, the very process to be investigated. Just right, and the BH mass evolution reflects the mass evolution of the cloud core  (see mass convergence in the top panel of Figure \ref{fig:bondiplot_lvl26_nodrag}). For the cloud studied in this section, at a resolution of $\Delta x_{\rm glob} = 0.99 \rm \ pc$ and $\Delta x_{\rm zoom} = 0.01 \rm \ pc$, this yields a seed mass range $5 \times 10^4 \ \rm M_\odot < M_{\rm seed} < 5 \times 10^5 \ \rm M_\odot$, which is quite narrow. Choosing $M_{\rm seed}$ therefore requires detailed prior knowledge about the cloud in question at the chosen resolution, and the seed mass will inevitably vary from cloud to cloud.

\subsection{The choice of black hole seed mass}
\label{sec:fixes}

 \begin{figure}
 	\centering
	\includegraphics[width=0.9\columnwidth]{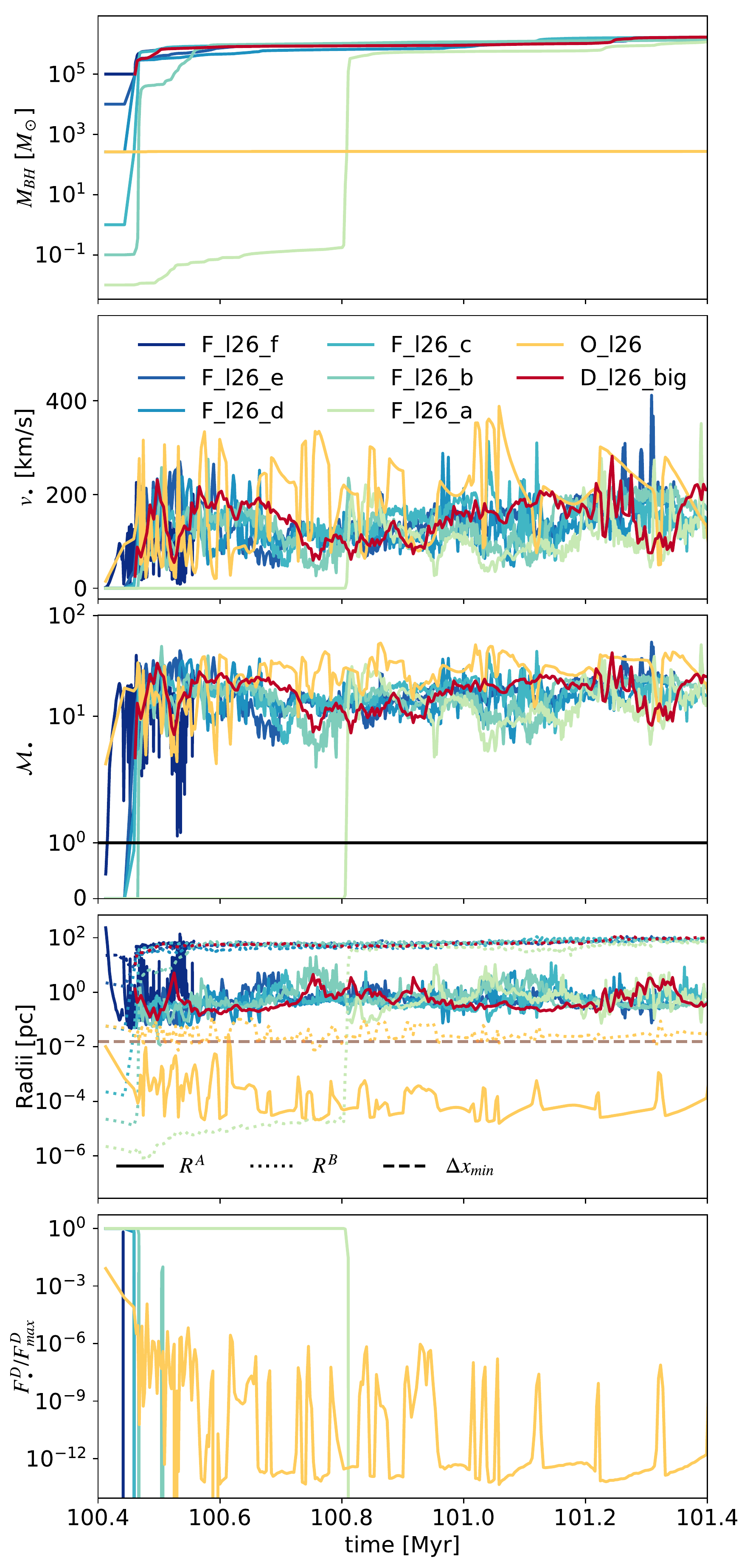}
	\caption[Time evolution of BH related quantities for a range of different seed masses and drag force algorithms]{Time evolution of BH related quantities for a range of different seed masses and drag force algorithms (see Table \ref{tab:disc_parameters} for details), including the magnitude of the drag force, $F^{\rm D}$, and the Mach number, $\mathcal{M}_\bullet$ . All curves are time averaged over 200 data-points for clarity. The second to last panel also includes the minimum cell size of the simulation as a dashed grey line, $\Delta x_{\rm min} = \Delta x_{\rm zoom}$, for comparison. Parameters for all simulations used in this section can be found in Table \ref{tab:disc_parameters}}
	\label{fig:bondiplot_fixes}
\end{figure}

\begin{figure}
	\centering
	\includegraphics[width=\columnwidth]{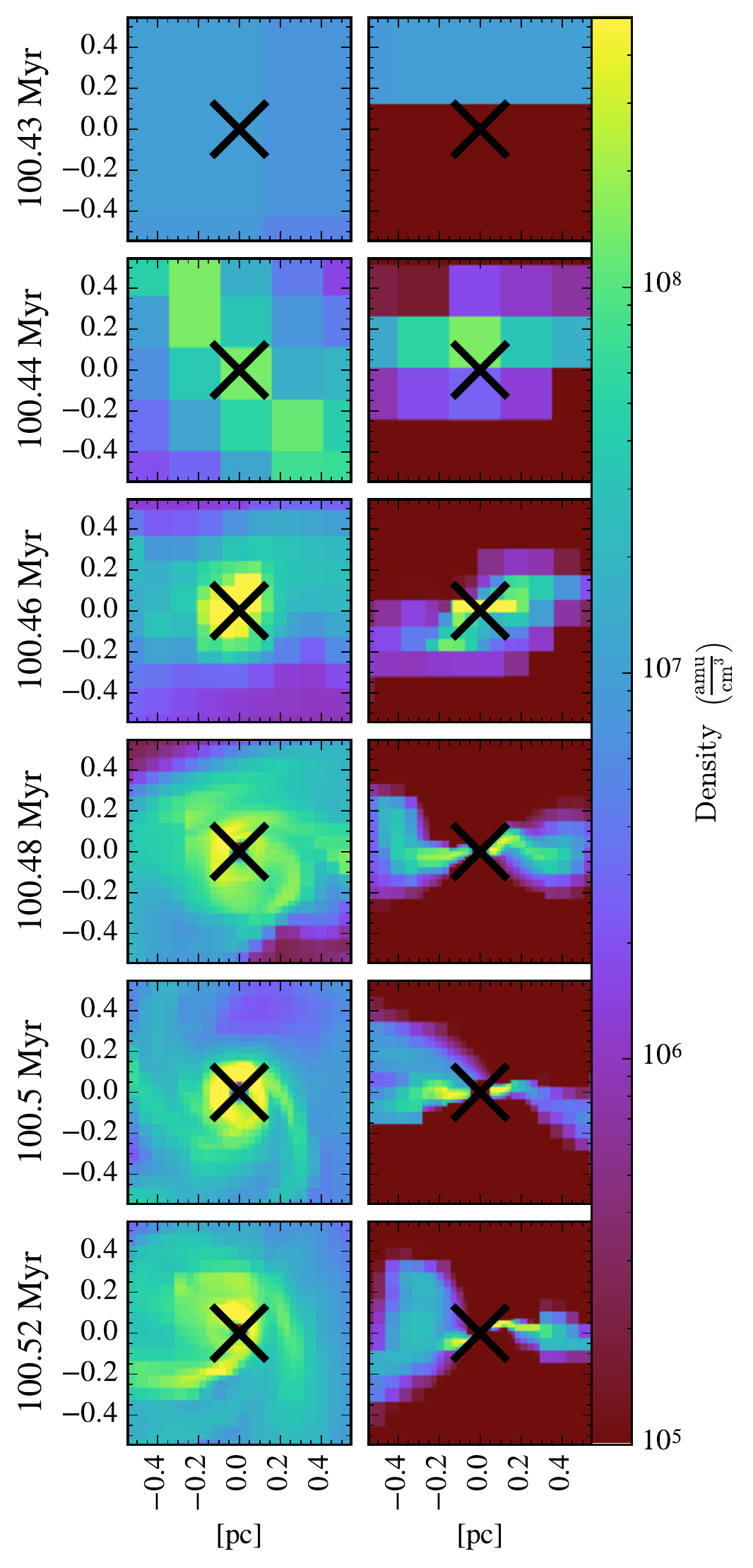} 
	\caption[Edge-on and face-on density slices for F\_l26\_c when  applying the maximum drag force]{Edge-on and face-on density slices for F\_l26\_c ($M_{\rm seed} = 1 \ \rm M_\odot$) at various epochs, when  applying the maximum drag force while $R^{\rm A}_\bullet < 0.2 \Delta x_{\rm loc}$. The BH position is marked by the black cross.}
	\label{fig:multiplot_fixes}
\end{figure}

As shown in the previous section, the mass accretion onto a BH embedded in a cloud collapsing under its own gravity is determined by the host cloud, not by the seed BH mass, if all relevant physical processes are included. A sub-grid model for dynamical friction therefore remains essential in order to reliably recover the converged mass accretion history for a given cloud. This section presents simulations with different drag force models in order to highlight problems with the common approach, and present a model that reliably returns the converged mass accretion history. Parameters for all simulations used in this section can be found in Table \ref{tab:disc_parameters}, where the converged solution for massive seeds is represented by  D\_l26\_big. All simulations with a sub-grid drag force only apply the force while $R^{\rm A}_\bullet < 0.2 \Delta x_{\rm loc}$, as work presented in \mbox{\citet{Beckmann2018a}} shows that applying an analytic force based on local quantities when the accretion radius is resolved can un-physically accelerate the BH. $\Delta x_{\rm loc}$ is the size of the BH host cell at any given point in time.

Dynamical friction can be added as a sub-grid drag force on the BH, for example using the analytic model by \mbox{\citet{Ostriker1998}} briefly introduced in \mbox{Section \ref{sec:fdrag}}. The simulation called O\_l26 combines a seed mass of $M_{\rm seed} = 260 \rm \ M_\odot$ with such a sub-grid model for dynamical friction. As can be seen in the top panel of \mbox{Figure \ref{fig:bondiplot_fixes}}, the model based on this linear analytic dynamical friction estimate is unable to solve the problem: the BH in O\_l26 does not grow. Comparing the Mach number for O\_l26 to the fiducial D\_l26\_big shows that the drag force does not significantly influence the motion of the BH. Physically, the magnitude of the drag force due to the wake is strongly dependent on the Mach number (see \mbox{Figure \ref{fig:Idrag}}), with a pronounced peak in force magnitude near $\mathcal{M}=1$, as increasing the relative velocity of the perturber decreases the opening angle of the gravitational wake, which reduces the wake mass and decreases its gravitational attraction. As discussed in detail in Appendix \ref{sec:convergence_levelmax}, the measure of the relative velocity between the sink particle and the gas, $v_\bullet$ becomes very unreliable in highly resolved, collapsing clouds. \New{This occurs because we are modelling a spherically symmetric gas flow using a cartesian grid. As fluxes at the bottom of the potential well cancel imperfectly, an apparently significant residual velocity between the BH and the local gas is measured even when the global relative velocity between BH and host cloud is negligible. This is one striking example of the phenomenon that local measures of gas properties can become less reliable with increasing resolution, a topic discussed in more detail in Appendix \ref{sec:convergence_levelmax}. As sub-grid models crucially depend on the quality of their input parameters, this seemingly benign numerical error turns out to have dramatic consequences.}

In the example studied here, $\mathcal{M}_{\bullet,init} \approx 7$  even at the point when the BH is seeded (see O\_l26  in the third panel of Figure \ref{fig:bondiplot_fixes}), as its relative velocity is measured using cloud particles spread across several cells, the measure inevitably picks up some of the local velocity dispersion. The magnitude of dynamical friction falls off rapidly for $\mathcal{M} > 5$ (see Figure \ref{fig:Idrag}), and its impact therefore remains very low throughout the simulation (see bottom panel) so the BH fails to settle in the disc. As clearly demonstrated by this example, any analytic sub-grid model is only as good as its input parameters, so that unreliable measures of the gas properties it requires can completely undermine its usefulness.

\begin{figure*}
	\centering
	\setlength{\tabcolsep}{0pt}
	\begin{tabular}{lcr}
		\includegraphics[width=0.32\textwidth]{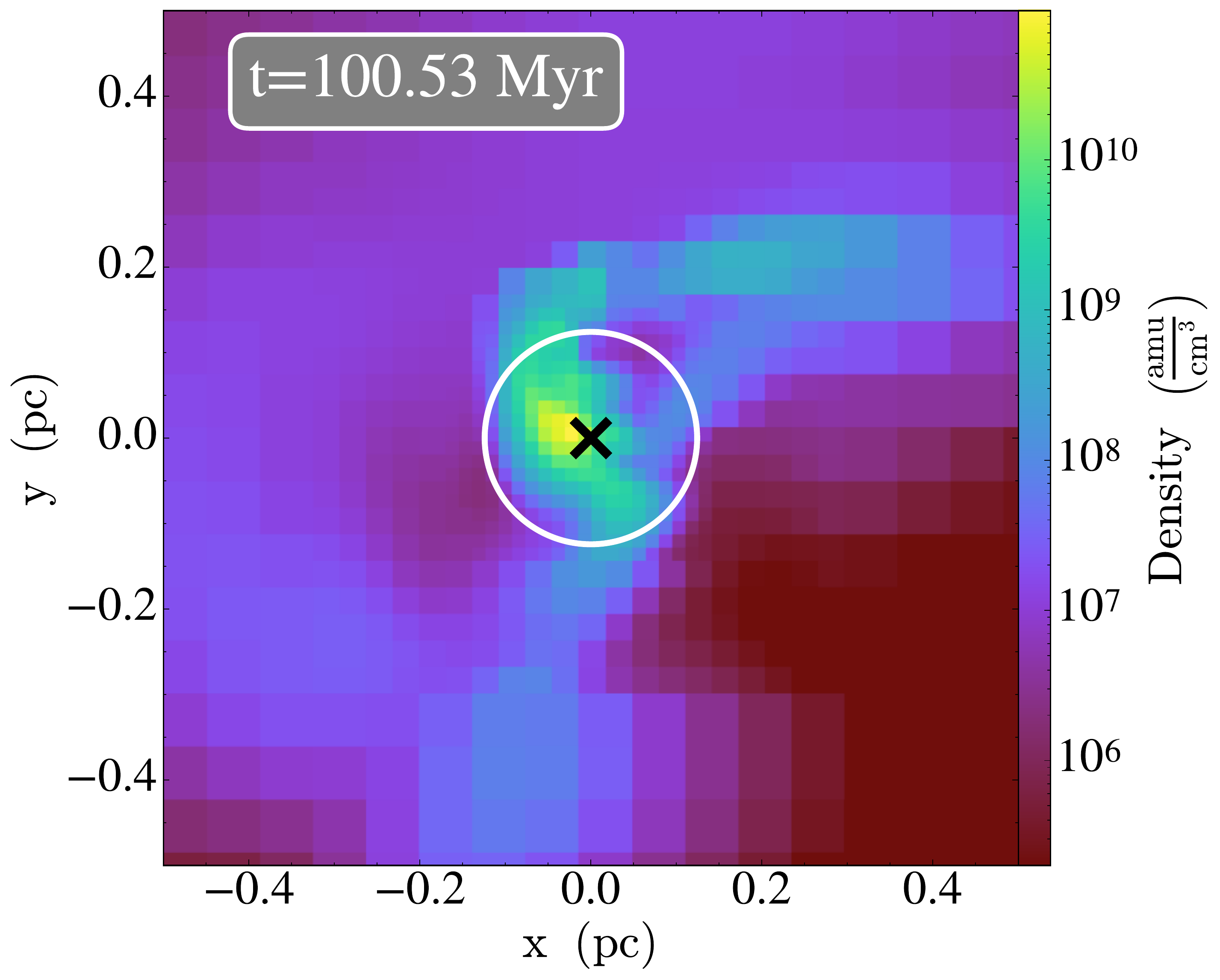} &
		\includegraphics[width=0.32\textwidth]{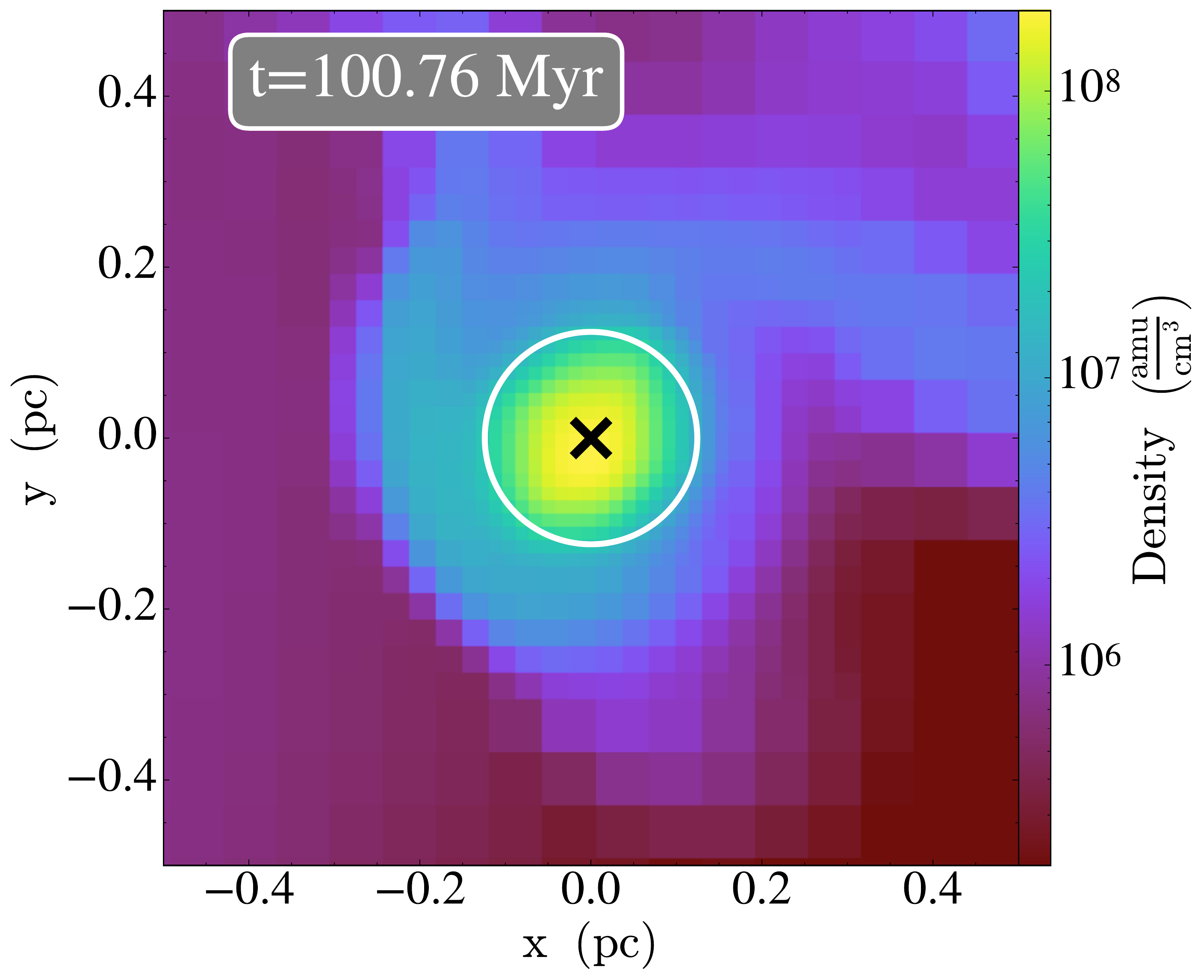}&
		\includegraphics[width=0.32\textwidth]{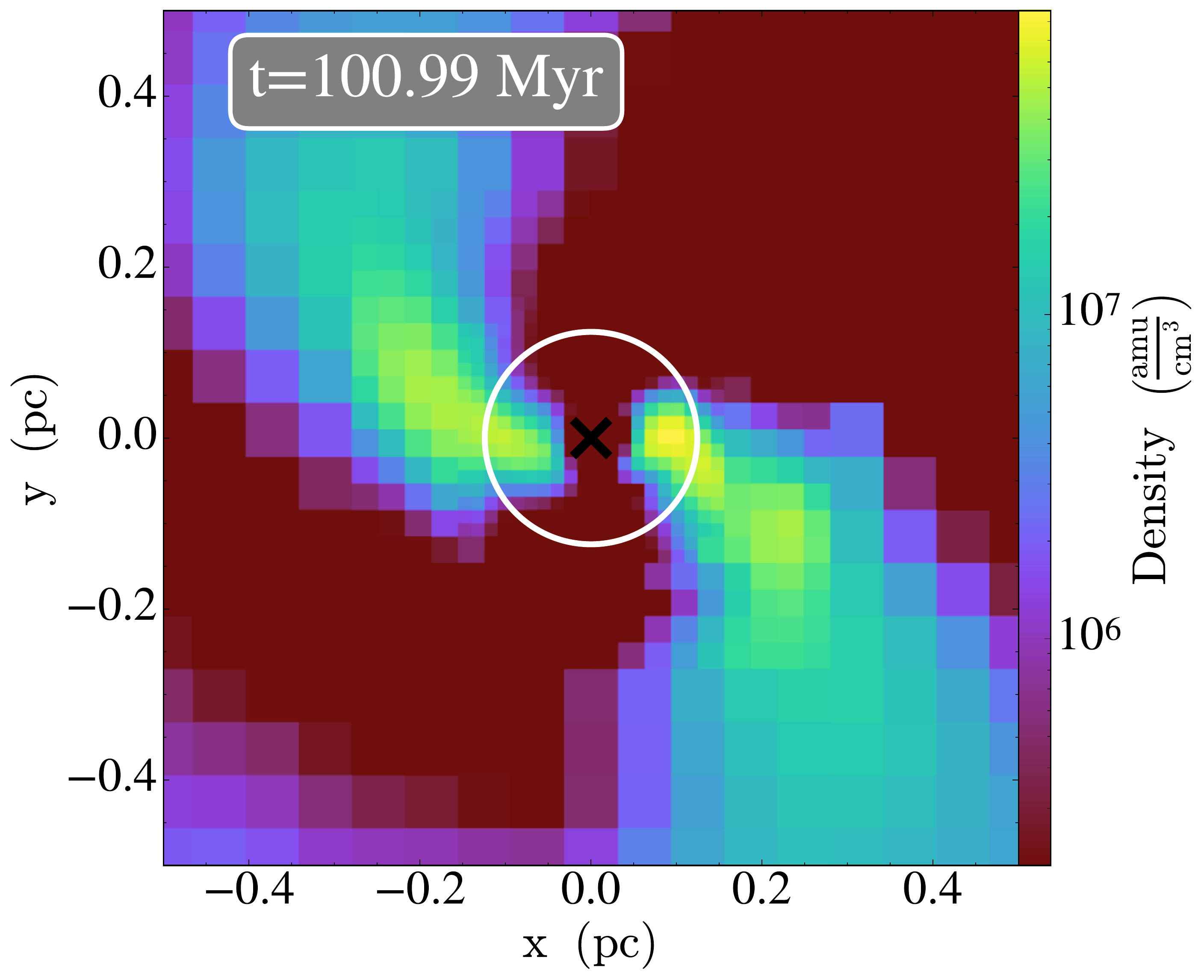}\\
	\end{tabular}
	\caption[Density slices of F\_l26\_a at three points in time]{Density slices of F\_l26\_a at three points in time. The location of the BH is marked by a black cross, and the size of $r_{\rm zoom}$ by a white circle.}
	\label{fig:slices_tiny_seed}
\end{figure*}

To circumvent the issue, we propose updating the BH velocity along with the evolving velocity of the gas in its host cell during cloud collapse, by simply subtracting the relative velocity (which is equivalent to applying the maximum possible drag force). As can be seen in Figure \ref{fig:bondiplot_fixes} (see simulations F\_l26\_a - F\_l26\_f), forcing $v_\bullet = 0$ at each fine time step of the simulation allows BHs to remain attached to the host cloud core independently of seed mass. It produces converged mass evolutions for all seed masses tested, allowing even BHs with very small seed masses (F\_l26\_c has $M_{\rm seed} = 1 \ \rm M_\odot$) to settle smoothly into the emerging gas disc (see the density slices of F\_l26\_c in Figure \ref{fig:multiplot_fixes}). Once a BH is sufficiently massive to accrete the cloud core and transition to SLA, all mass evolutions converge. At this point, $R^{\rm A}_\bullet > 0.2 \Delta x_{\rm zoom}$, and the sub-grid dynamical friction algorithm becomes inactive. 

In the most extreme case,  F\_l26\_a, with $M_{\rm seed} = 10^{-2} \ \rm M_\odot$, growth is delayed as the BH seed mass is initially well below the SLA transition mass (Equation \ref{eq:Mflux}) in spite of the 0.01 pc resolution, and so it initially grows via BHL accretion. While in the BHL regime, the drag force remains active (see bottom panel in Figure \ref{fig:bondiplot_fixes}), and the BH remains attached to the cloud core (see density slices in Figure \ref{fig:slices_tiny_seed}). Like in the spherical collapse case presented in Section \ref{sec:convergence_levelmax}, gas and BH are gravitationally bound so the delay in accretion has no influence over the long-term evolution of the BH mass. However, this could change in the presence of feedback which could disrupt the cloud before the BH is sufficiently massive. \New{If the accretion rate had been capped at the Eddington accretion rate (see Equation \ref{eq:eddington}), this period of low accretion before convergence with the larger seeds would have been much longer yet, stifling mass growth and delaying the point in time when the transition mass $M_{\rm SLA}$ is reached.}

As previously stated, the difficulty stems from the fact that applying any analytic drag force model inherently requires a measure of the relative velocity
of the BH, which can be defined in several ways. As shown in Appendix \ref{sec:convergence_levelmax}, $v_\bullet$ is the better choice for a spherically symmetric collapse, where the relative velocity of the BH with respect to its host cell, $v_{\rm cell}$, is even more overestimated than with respect to the bulk relative velocity. In non-spherically symmetric cases, like the ones discussed in this section, where the scale height of the disc is smaller than the radius of the accretion region and cloud particles probe both sides of the disc simultaneously, the local velocity dispersion will increase $v_\bullet$  relative to $v_{\rm cell}$. Setting $v_\bullet = 0$ allows the BH to remain sufficiently attached to the cloud to continue accreting from its core until its dynamical friction is self-consistently resolved (compare right two columns of Figure \ref{fig:slices_D_l26_masses} to Figure \ref{fig:multiplot_fixes}) and an analytic drag force becomes irrelevant. This 'maximum drag force' prescription is more robust than an imprecise relative velocity estimate due to its self-correcting nature for a force linearly proportional to relative velocity. The reader is however cautioned that it does not necessarily guarantee that the BH remains perfectly attached to the cloud core at all times. In fact, to minimise the need for any drag force prescription, the seed mass should be chosen such that $R^{\rm B}  = \frac{GM}{c_{s,{\rm cell}}^2} \approx \Delta x_{\rm zoom}$, where $c_{s,{\rm cell}}$ is the sound speed of the host cell at formation, so that the accretion radius is marginally resolved at the end of cloud collapse. Applying a maximum drag force while dynamical friction is unresolved should only be considered as a way to ensure that the outcome of the simulation will not sensitively depend on this choice. 

In summary, a small BH seed mass, in combination with forcing $v_\bullet = 0$ until the dynamical friction of the gas on the BH is resolved, is a robust way of producing a BH mass evolution that reflects the emerging structure of the collapsing cloud from which it forms. The resulting accretion history becomes independent of the somewhat arbitrary BH seed mass choice if no feedback is included, as accretion onto the BH is driven by the self-gravity of the host cloud rather than by the gravitational potential of the BH itself. In the absence of any kind of feedback process to alter the cloud collapse, the three competing BH formation mechanisms previously mentioned cannot be distinguished from one another: the BH initial mass is quickly washed out by efficient gas accretion. Bearing this caveat in mind, we investigate in the next section the impact on BH accretion of better resolving the internal structure of the cloud.

\section{The impact of resolution on black hole accretion}
\label{sec:resolution}

\begin{table}
	\centering
	\setlength{\tabcolsep}{6pt}
	\begin{tabular}{lcccr}
		\hline
		\multicolumn{5}{|c|}{\bf{Disc galaxy simulations}} \\
		\hline
		\bf{name} & $l_{\rm zoom}$ & $\Delta x_{\rm min} $ $[\rm pc]$ & $M_{\rm seed}$ [$\rm M_\odot$]   & CPU hours* \\
		\hline
		R\_l20 & 20 & $0.992$& $5 \times 10^3$ & 10 \\
		R\_l23 & 23 & $0.124$ &  $ 700$ & 62 \\
		R\_l26 & 26 & $0.016$  &$40 $ &  404\\
		R\_l28 & 28 & $0.004$  &$10 $ &  2,122\\
		\hline
		\multicolumn{5}{l}{*per Myr of evolution on 36 Intel X5650 (2.67 GHz) cores}\\
	\end{tabular}
	\caption{Parameters for simulations in Section \ref{sec:resolution}. All simulations have $l_{\rm glob}  = 20$ and the three zoom simulations have $n_{\rm zoom}=8$.}
	\label{tab:fixed_simulations}
\end{table}

 \begin{figure}
 	\centering
	\includegraphics[width=\columnwidth]{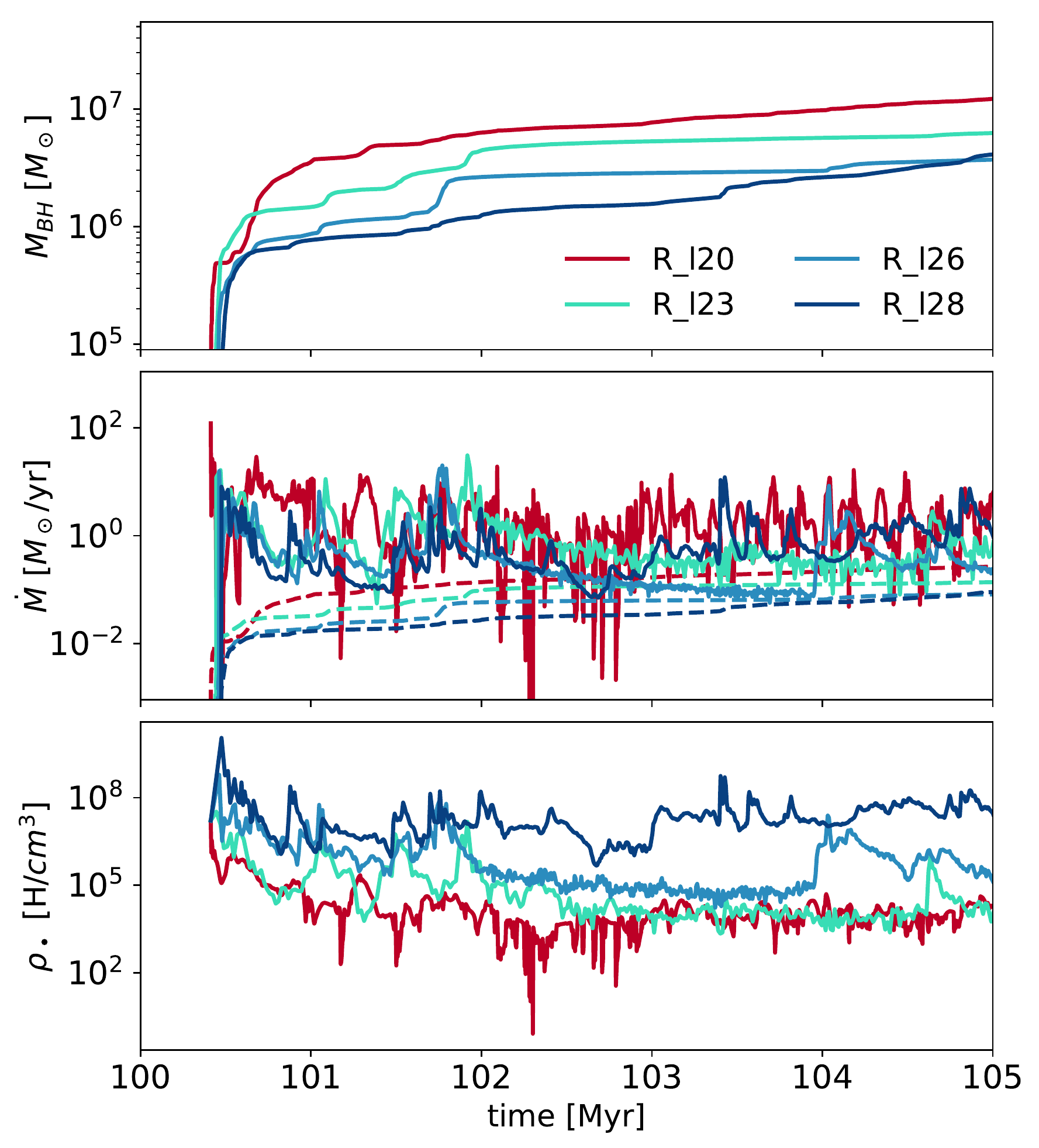}
	\caption[Time evolution of BH and gas properties at different resolutions]{Time evolution of BH and local gas properties at different resolutions (see Table \ref{tab:fixed_simulations} for details). The dashed line on the middle panel represents the Eddington accretion rate. All data is time averaged over 100 data-points for clarity.}
	\label{fig:bondiplot_fixed}
\end{figure}

If accretion onto the BH is driven by the internal structure of the host cloud, then resolving the cloud further may have a decisive impact on the BH mass evolution. To investigate this issue, three simulations with different $l_{\rm zoom}$, R\_l23, R\_l26 and R\_l28 are compared to one simulation without BH zoom, R\_l20 (see Table \ref{tab:fixed_simulations} for details). \New{The seed masses for all four simulations were chosen according to the criterion $R^{\rm B}  \approx \Delta x_{\rm zoom}$ proposed at the end of Section \ref{sec:fixes}.} All four simulations have $l_{\rm glob} = 20$ and $n_{\rm refine} = 8$. The cooling halo and seed formation location are identical to the simulations presented in Section \ref{sec:mseed}. 

 \begin{figure*}
 	\centering
	\setlength{\tabcolsep}{0pt}
	\begin{tabular}{ccc}
		\includegraphics[width=0.33\textwidth]{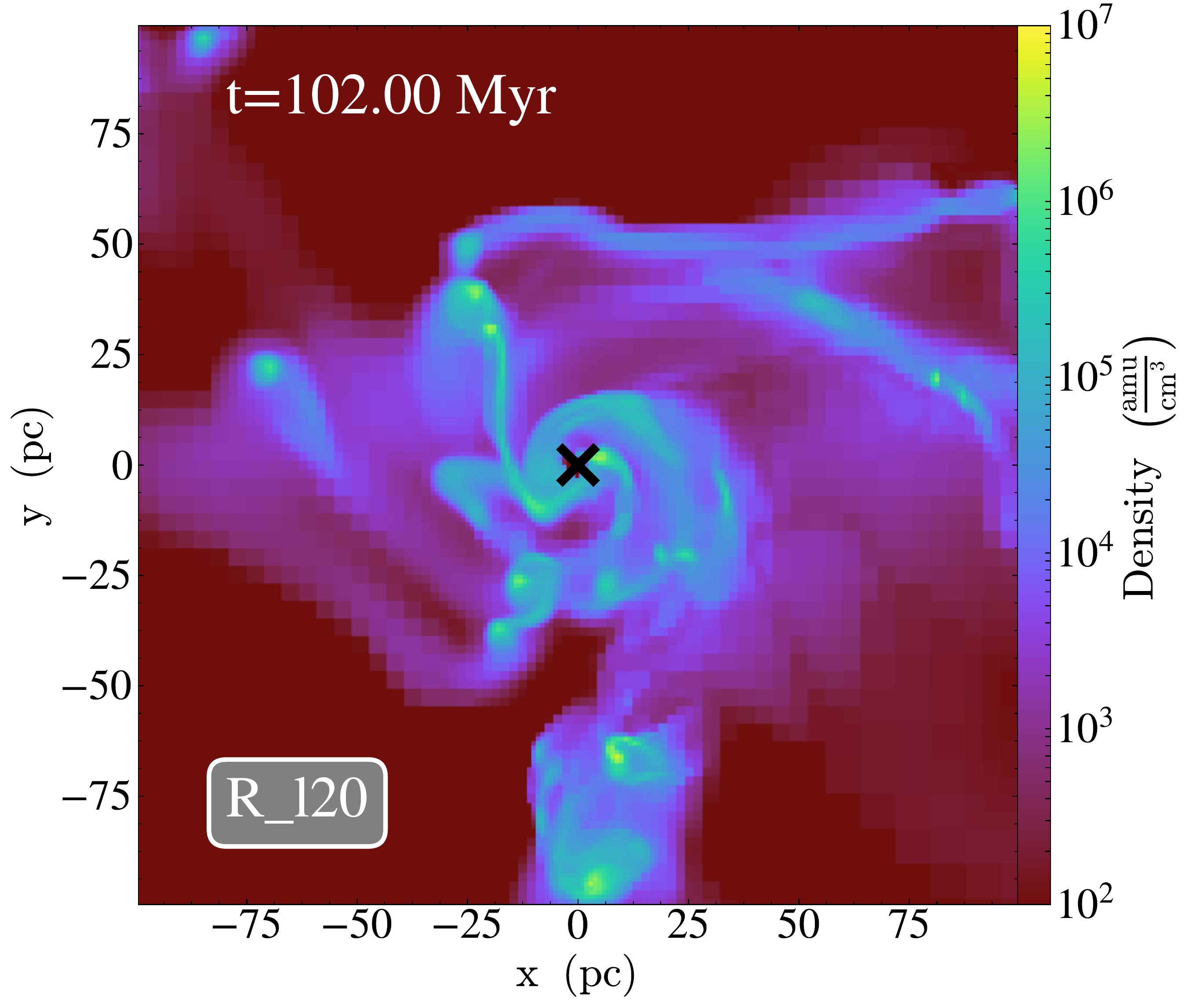}&
		\includegraphics[width=0.33\textwidth]{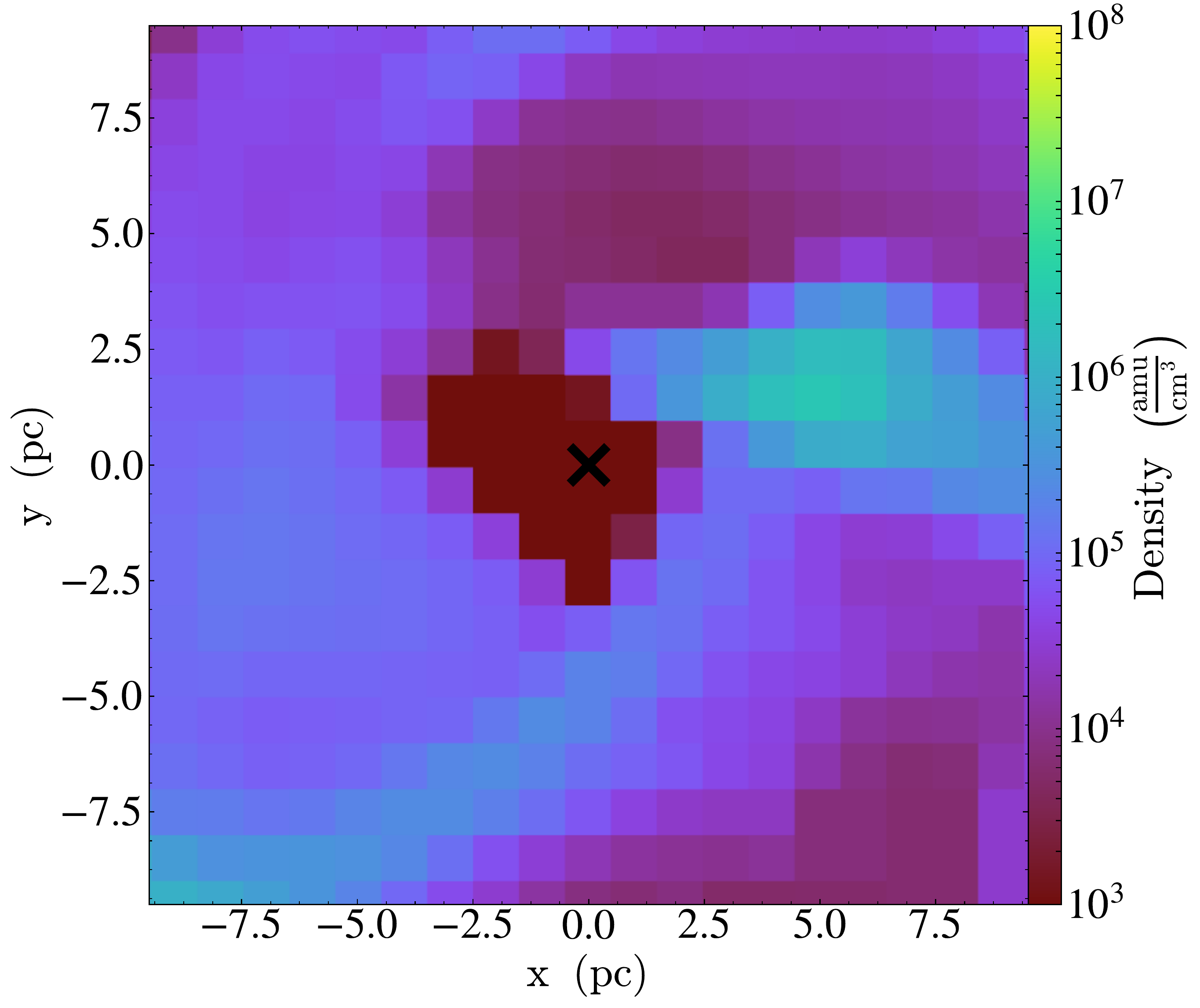}&
		\includegraphics[width=0.33\textwidth]{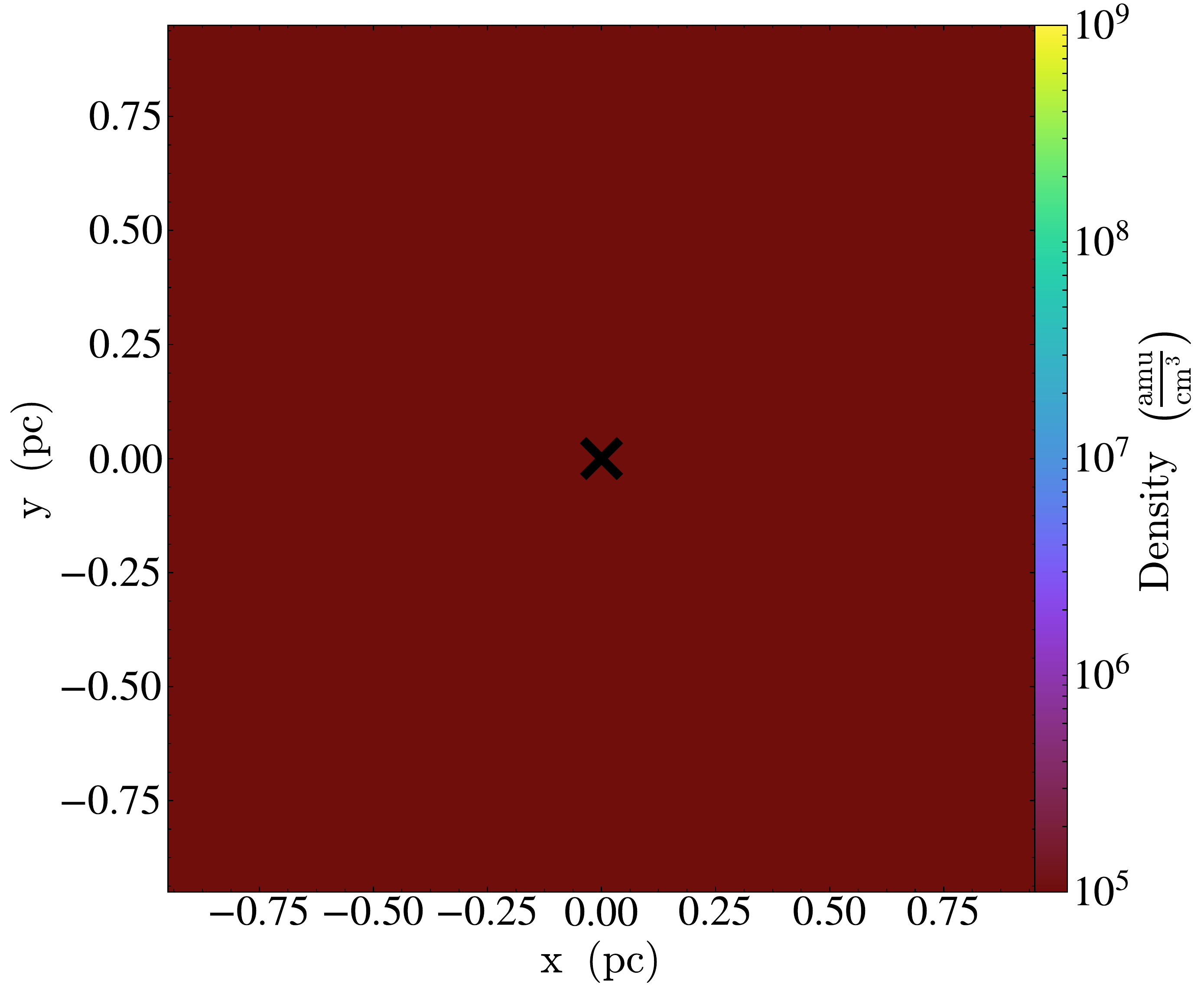}\\
		\includegraphics[width=0.33\textwidth]{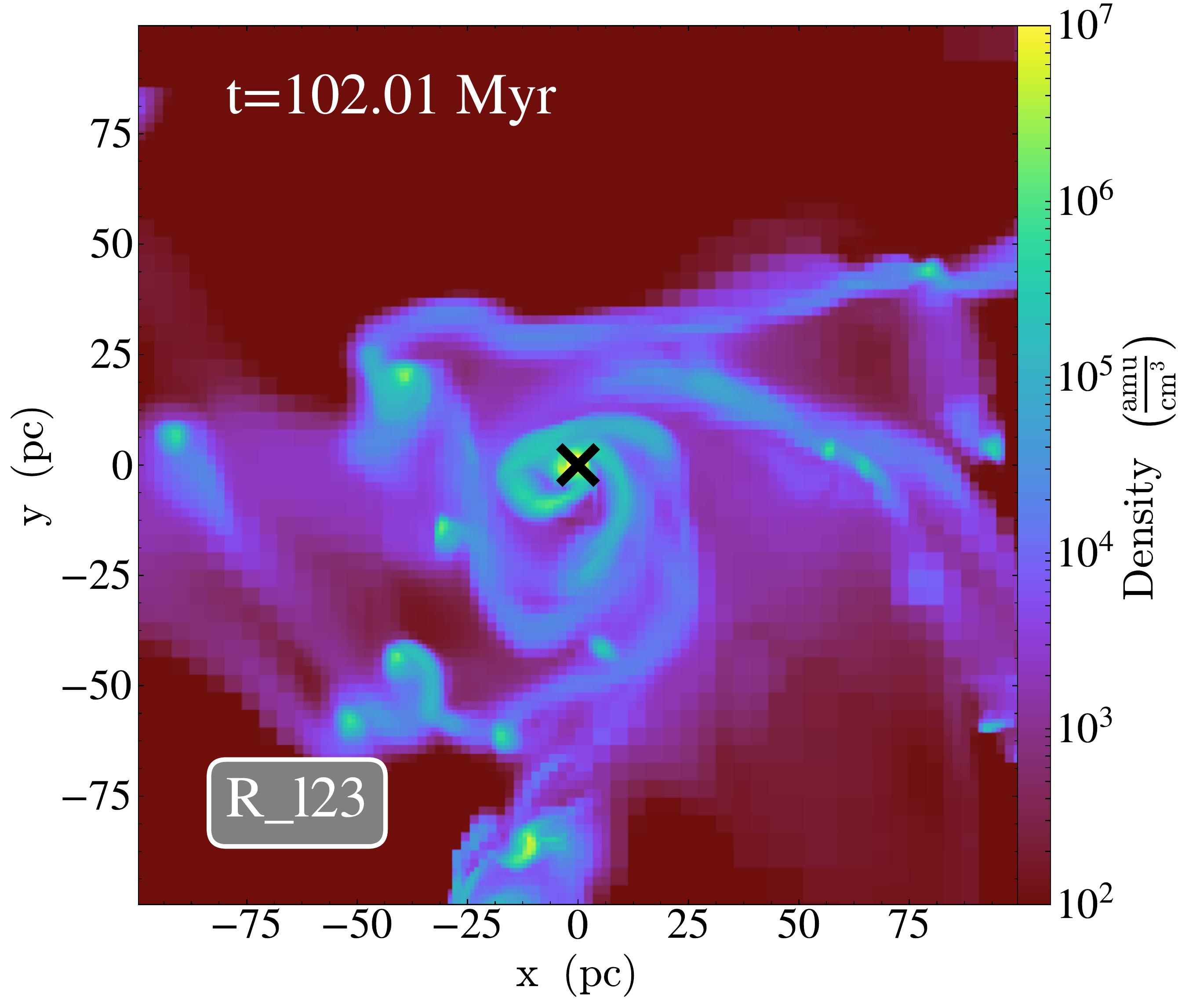}&
		\includegraphics[width=0.33\textwidth]{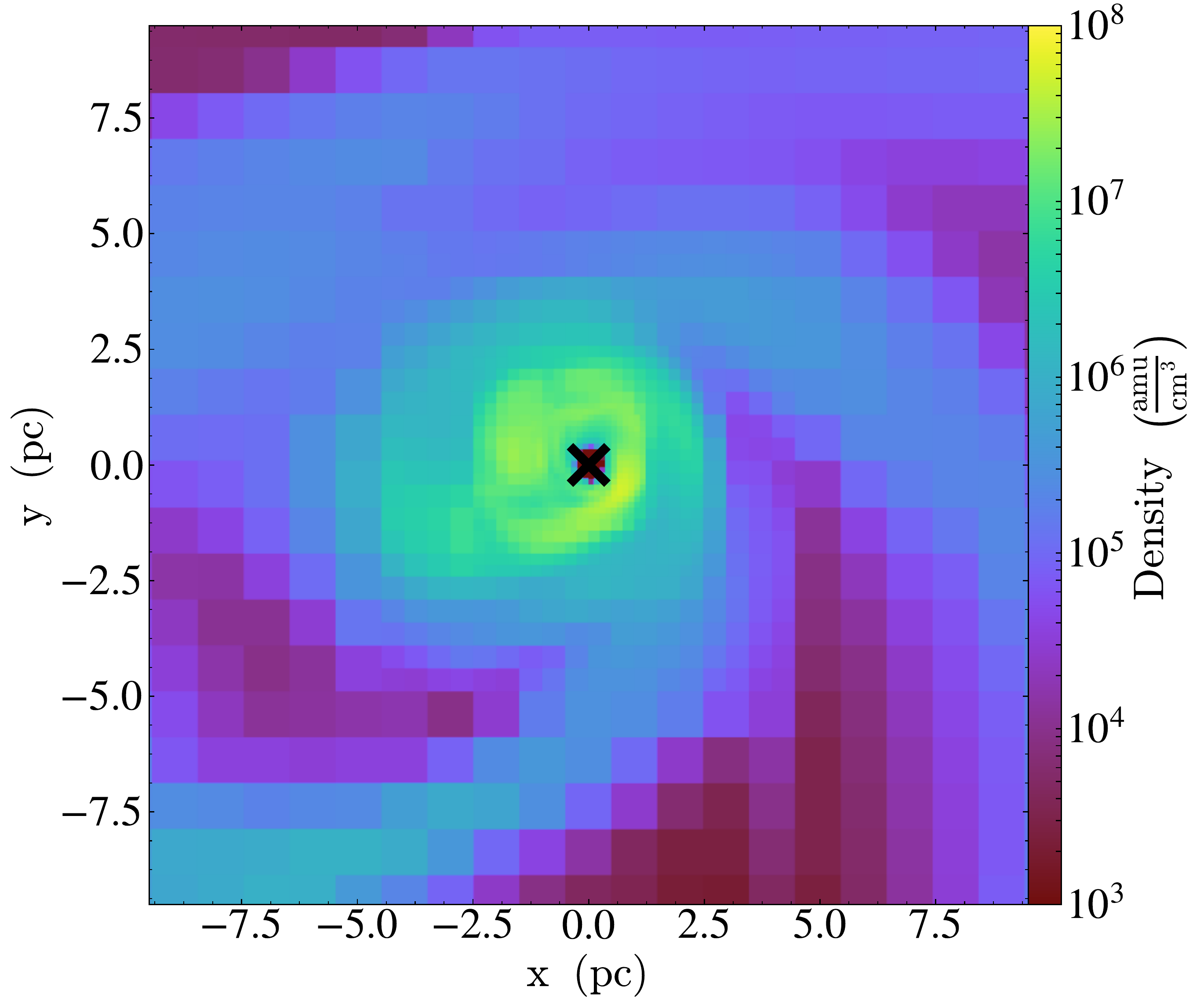}&
		\includegraphics[width=0.33\textwidth]{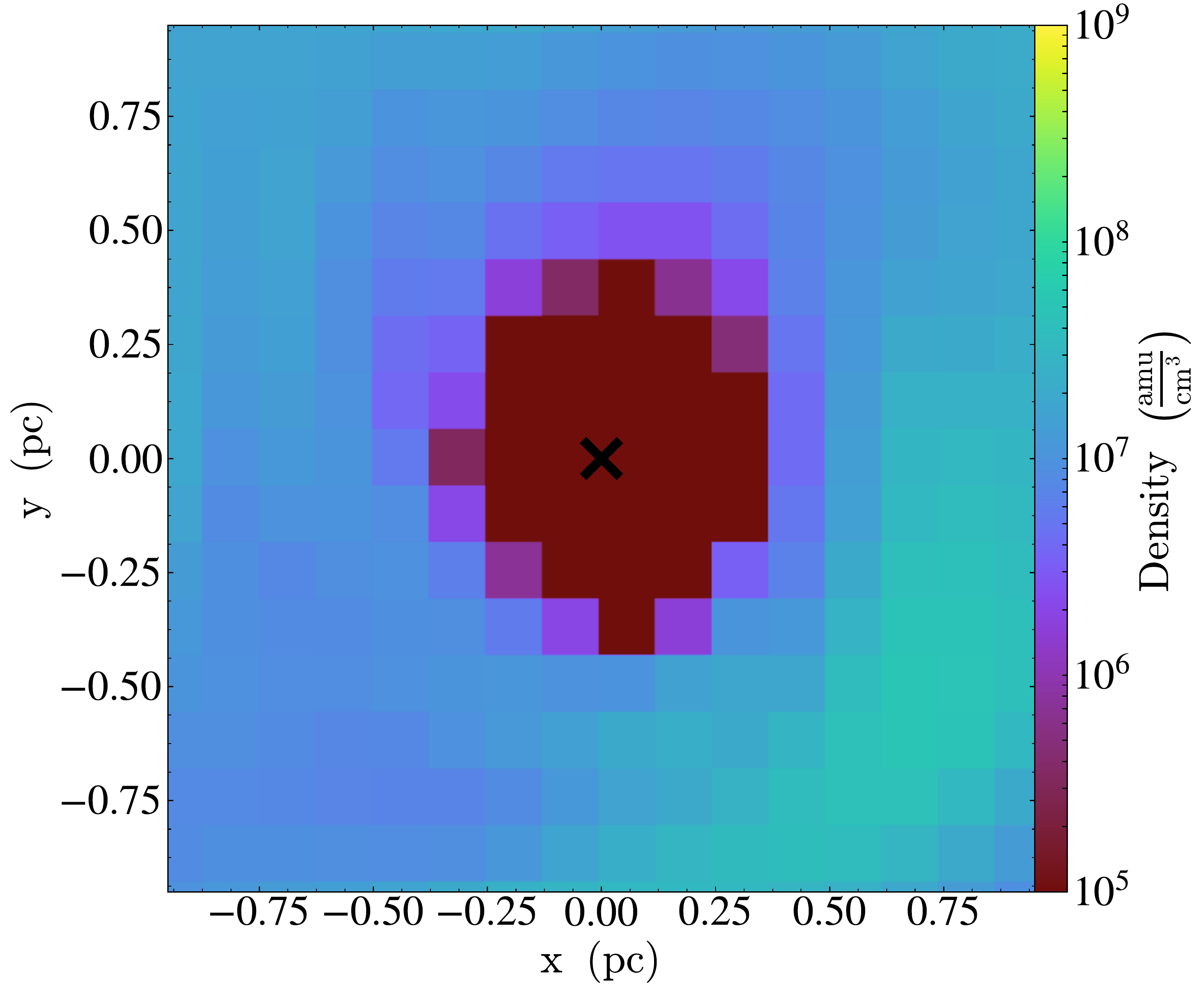}\\
		\includegraphics[width=0.33\textwidth]{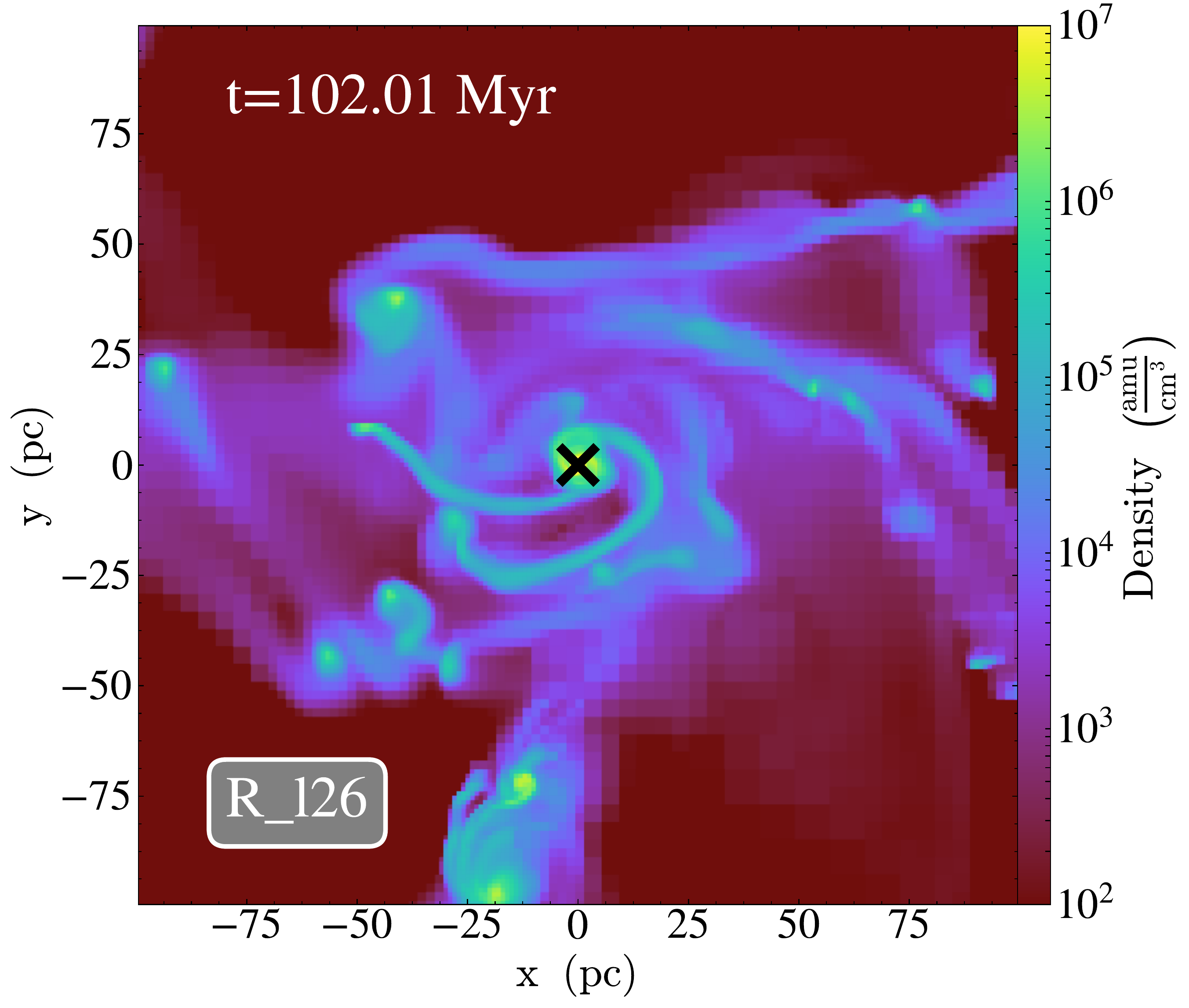}&
		\includegraphics[width=0.33\textwidth]{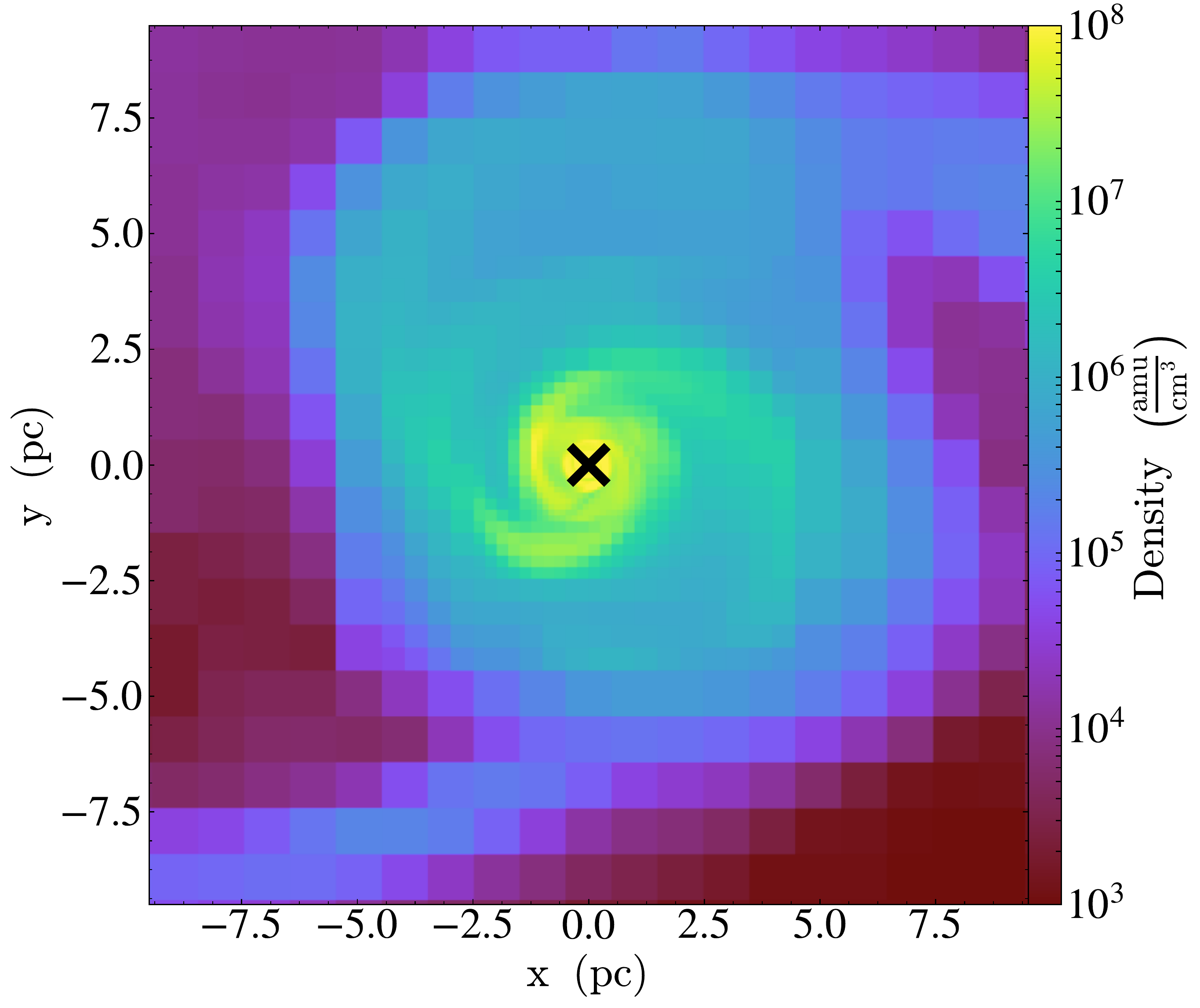}&
		\includegraphics[width=0.33\textwidth]{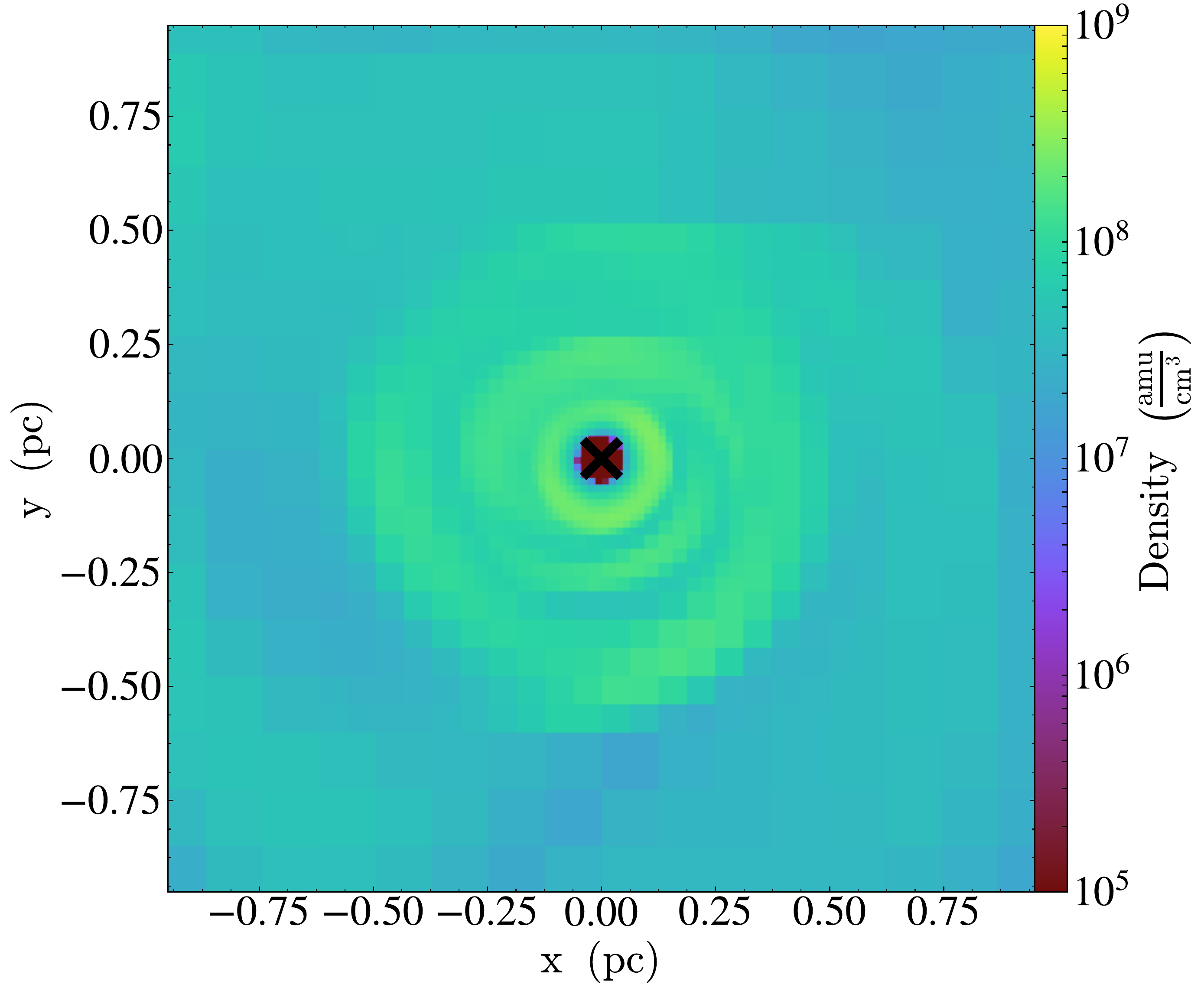}\\
		\includegraphics[width=0.33\textwidth]{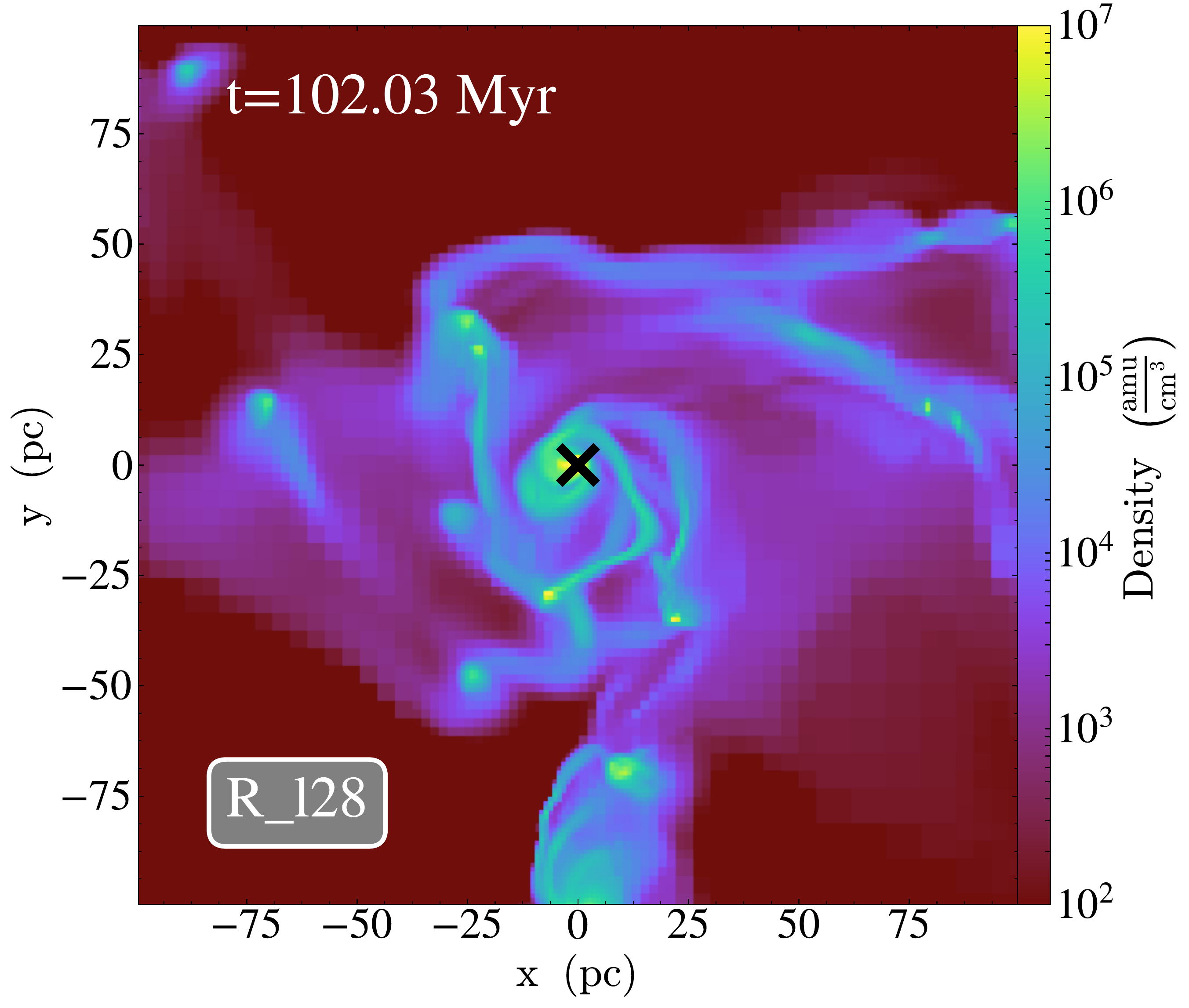}&
		\includegraphics[width=0.33\textwidth]{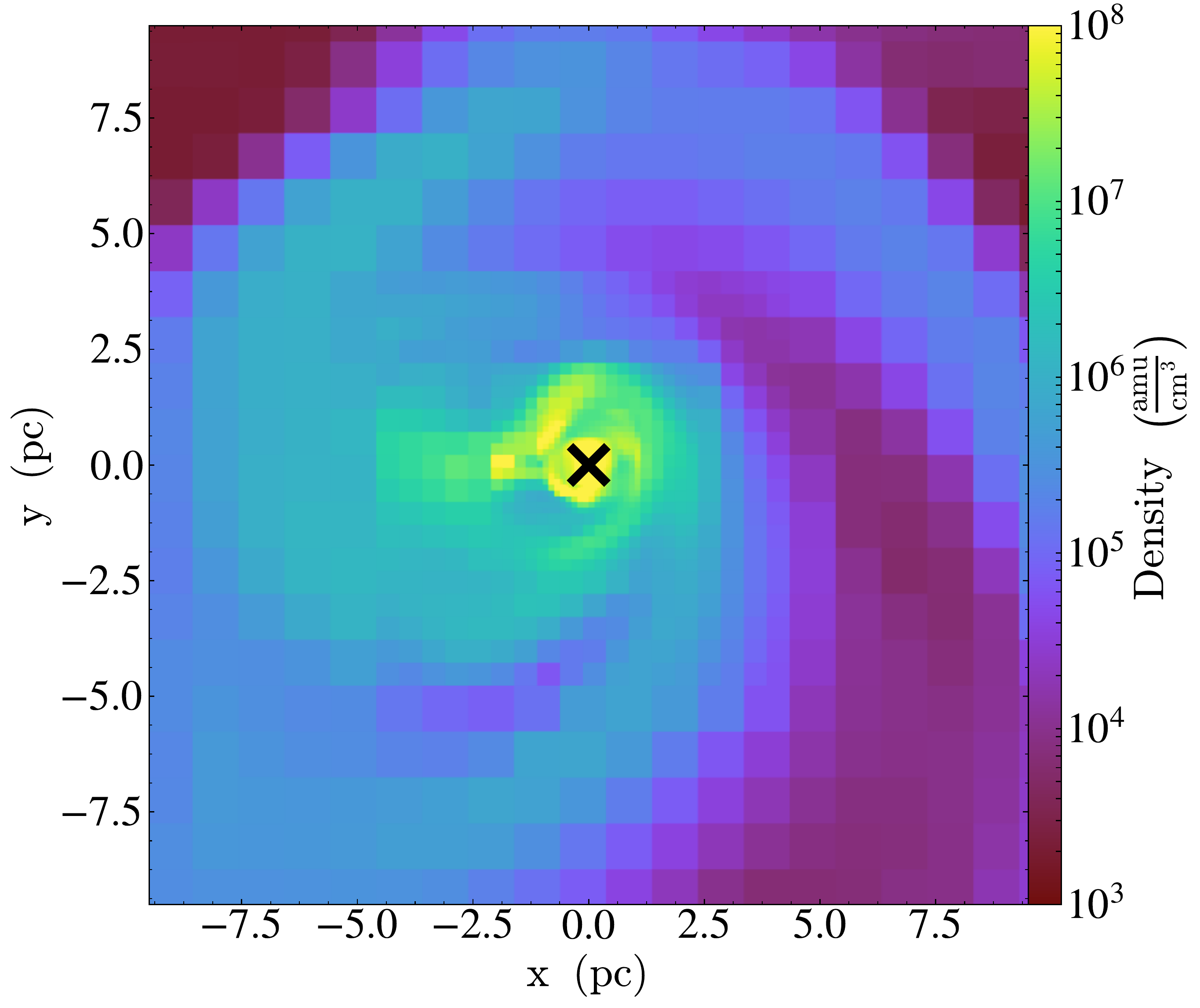}&
		\includegraphics[width=0.33\textwidth]{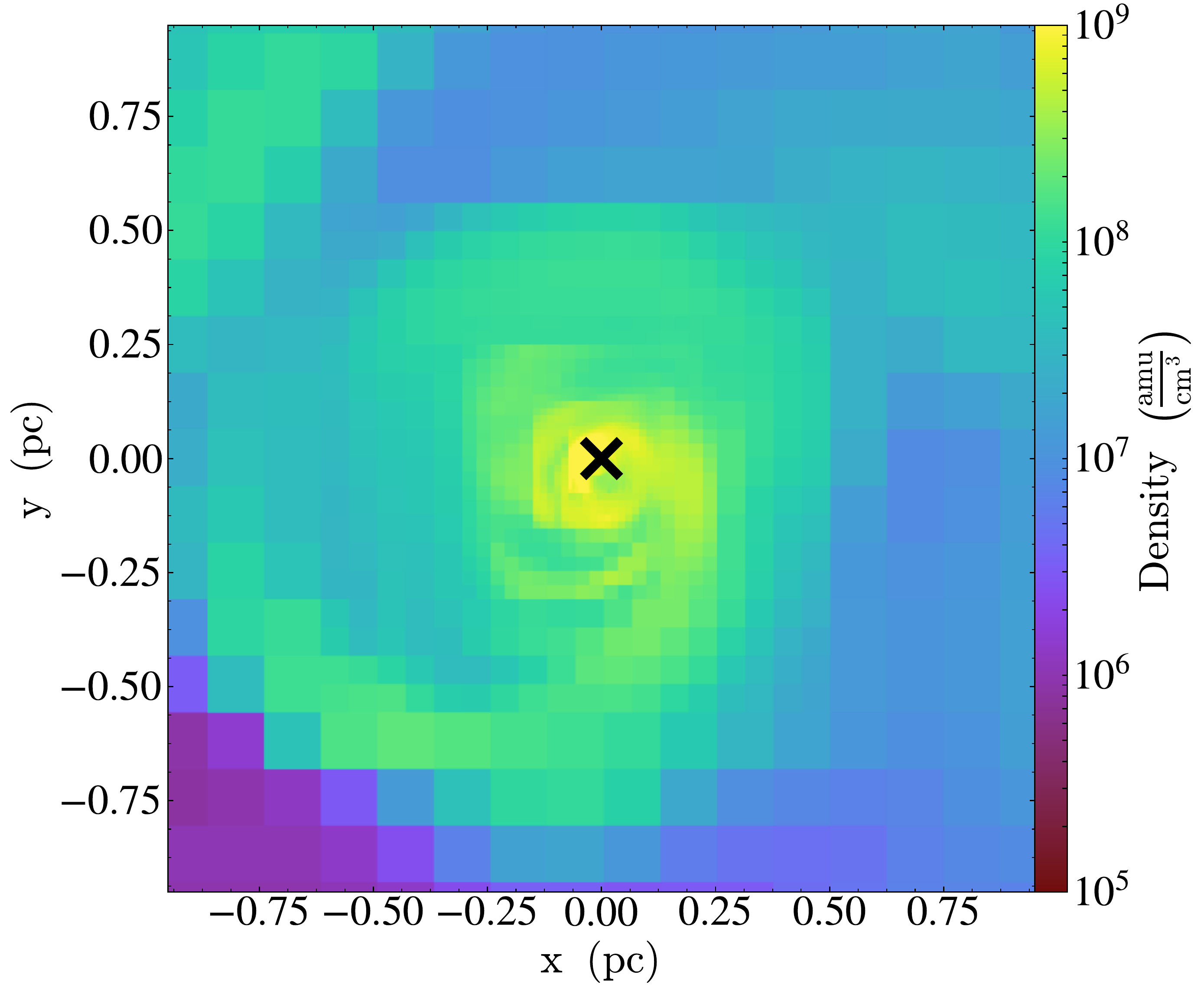}\\
	\end{tabular}
	\caption[Gas density projections of disc galaxy simulations at three different resolutions]{Gas density projections for (from top to bottom) R\_l20, R\_l23, R\_l26 and R\_l28  at $t=102 \rm \ Myr$, at a range of different magnifications. The BH location is marked by a black cross.}
	\label{fig:slices_fixed}
\end{figure*}

Improving resolution in the vicinity of the BH has an immediate and significant impact on its evolution, as can be seen in Figure \ref{fig:bondiplot_fixed}. While it may seem to only mildly affect the BH's rapid initial mass growth (all four simulations produce a BH in the mass range $3-6 \times 10^5$ M$_\odot$ in less than 0.1 Myr, despite starting from orders of magnitude different seed masses), the later accretion patterns and gas properties in the immediate vicinity of the BH differ so notably that the simulations never converge. 

\New{ The similar magnitudes of BH masses after only 0.1 Myr of evolution, despite seed masses spanning over two orders of magnitude, strengthen the conclusion from Section \ref{sec:fixes} that the value of the BH seed mass is inconsequential, as long as it is below the mass acquired during the initial mass boost and above  the SLA transition mass for a given resolution. Both conditions are met for all four simulations presented in this section.}

\begin{figure}
	\centering
	\includegraphics[width=1.0\columnwidth]{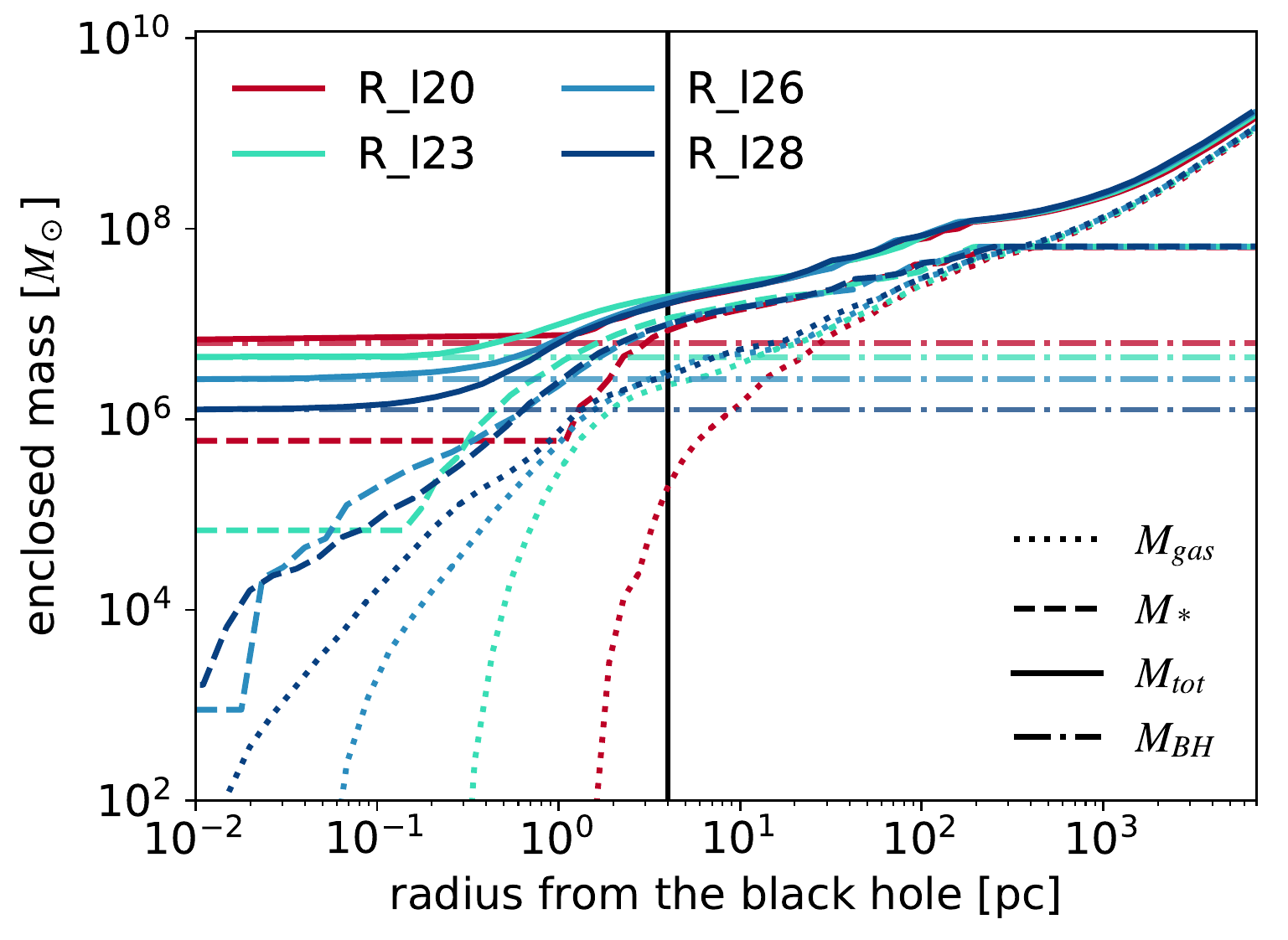}
	\caption{Cumulative spherical mass profiles for four different $l_{\rm glob}$, including gas mass, stellar mass, BH mass and total mass. The vertical line marks the size of the accretion region in the lowest resolution simulation, R\_l20. These profiles were produced at $t=102 \rm \ Myr$.}
	\label{fig:mass_profiles_fixed}
\end{figure}

Structures on pc scale and larger remain similar in all four simulations, as can be seen by looking at the density projections presented in Figure \ref{fig:slices_fixed}. This is confirmed by the mass profiles plotted in Figure \ref{fig:mass_profiles_fixed}, which show that $M_{\rm tot}$ and $M_{*}$  converge for all four simulations on scales above $4 \rm \ pc$, the size of the accretion region in R\_l20. 

By contrast, on smaller scales, the BH in R\_l20 accretes the entire dense core of the cloud (Figure \ref{fig:slices_fixed}, middle column), emptying an irregularly shaped accretion region whose size is of the same order as in-falling clumps. Any cloud that subsequently falls into the centre is disrupted and accreted immediately by the BH, without re-forming a dense core. On the other hand, the accretion region of BHs in the BH zoom simulations, R\_l23, R\_l26 and R\_l28, is significantly smaller than the physical extent of the core. Instead of being accreted, the core collapses into a nuclear gas disc whose rotationally supported structure is captured by the simulation. 

\subsection{The nuclear disc}
\label{sec:nuclear_disc}

\begin{figure}
	\centering
	\includegraphics[width=1.0\columnwidth]{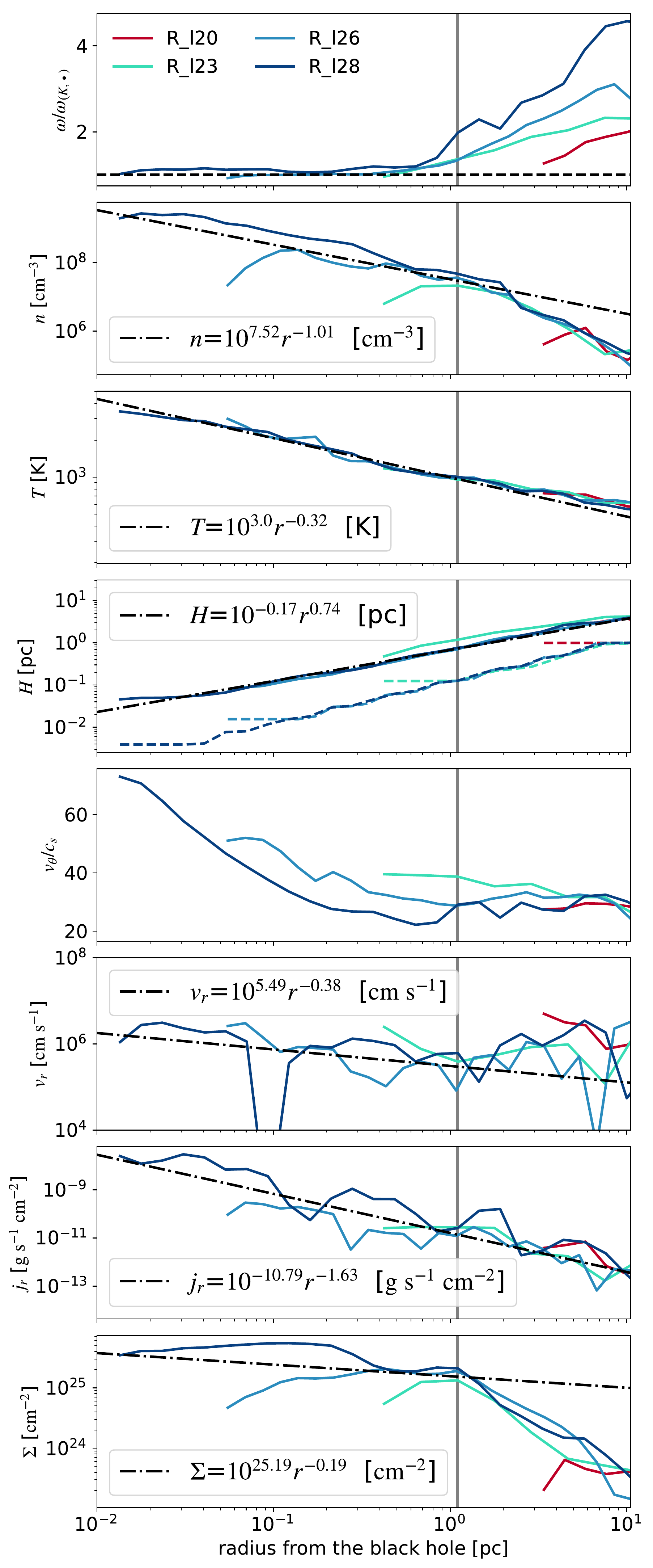}
	\caption{Radial profiles of the nuclear disc at $t=102$  Myr. From top to bottom: epicyclic frequency $\omega$ normalised to the BH Keplerian frequency $\omega_{(K,\bullet)}$, number density $n$, temperature $T$, disc height $H$, ratio of tangential velocity to sound speed $v_{\theta}/c_s$, radial velocity $v_r$, radial mass flux $j_r$ and surface number density $\Sigma$. Each profile is plotted from the edge of each simulation accretion region at $4 \Delta x_{\rm min}$ to a distance of 10pc from the BH. Power law fits, shown as dashed-dotted lines, are calculated for the Keplerian part of the disk only, i.e. at $r \leq 1.1 $ pc (marked by the vertical line).}
	\label{fig:dens_profiles_fixed}
\end{figure}

The density projections in Figure \ref{fig:slices_fixed} show that the physical extent of the nuclear disc is the same in R\_l26 and R\_l28. With a factor 4 higher resolution, R\_l28 captures more of the internal structure of the core, in the form of spiral features, than R\_l26 in which the core appears smoother (compare R\_l26 and R\_l28 in the right column of Figure \ref{fig:slices_fixed}). Despite the internal structure, the gas density, temperature and disk height profiles in Figure \ref{fig:dens_profiles_fixed} show that the Keplerian disk is self-similar, with single power law profiles extending smoothly up to a radius of about 1 pc. In-between accretion events, the disc profiles vary little over the 5 Myr studied here, far longer than the dynamical time scale of the disc, which we estimate at $ 2 \pi r / v_\theta \approx 0.1$ Myr for $r =1 \rm \ pc$. 

The nuclear disk has a maximum radius of $r=1$ pc, where it self-gravitates and fragments. This can be seen visually in the density projections in Figure \ref{fig:slices_fixed}, and is confirmed by the value of the Toomre parameter $Q=c_s \kappa / (\pi G \Sigma)$, which falls below one at $r=1.1$ pc, when calculated using the power law fits from Figure \ref{fig:dens_profiles_fixed}. The inner edge of the disc is determined by the extend of the BH accretion region at $4 \Delta x_{\rm min}$. Between the inner and outer edge, the disk is Keplerian and entirely dominated by the gravitational potential of the BH (see top panel of Figure \ref{fig:dens_profiles_fixed}). The disc is rotationally supported, as can be seen by the fact that it rotates at the Keplerian velocity of the potential, and the ratio of tangential velocity to sound speed, $v_{\theta} / c_{s} > 20 $ at all radii.

\begin{figure}
	\centering
	\includegraphics[width=1.0\columnwidth]{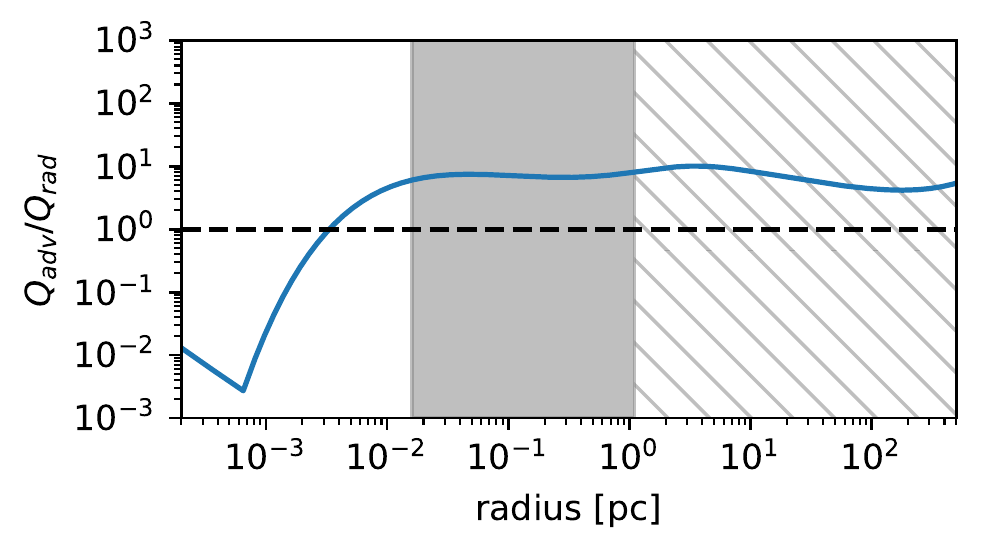}
	\caption{Ratio of the advective, $Q_{adv}$, and the radiative, $Q_{rad}$, cooling rates within the nuclear disc. Above the dashed line, cooling in the disc is advection dominated, whereas below the line, radiative cooling dominates. The range of radii covered by the nuclear disc shown with a solid grey background, and the hashed region denotes the remainder of the computational domain. A white background shows an extrapolation to unresolved scales.}
	\label{fig:disc_cooling}
\end{figure}

With a disc height profile of $H \propto r^{0.74}$, the nuclear disc is too thick to be a classical Shakura-Sunyaev disc \cite{Shakura1973}, and instead resembles the slim discs discussed in \citet{Abramowicz1988,Chen1995} and \citet{Sadowski2013}. Such a large disc scale height is a consequence of inefficient radiative cooling within the disc. In the simulations presented here, the gas is assumed to be primordial and therefore metal-free throughout, so that for temperatures below a few $10^3$ K, (corresponding to the bottom of the cooling curve for atomic hydrogen), radiative cooling can only proceeds via molecular hydrogen transitions. As discussed in  detail in \citet{Grassi2014}, molecular hydrogen cooling is inefficient for gas densities in the range $10^4 -10^{10}  {\rm cm^{-3}}$ due to the low fraction of molecular hydrogen present. Indeed, for a total gas number density of $n=10^{7}-10^{9} {\rm cm^{-3}}$ this fraction is on the order of $10^{-3}$. Unable to cool efficiently with a radiative cooling rate 
\begin{equation}
	Q_{\rm rad} = \Lambda_{\rm H2} n \frac{\Sigma}{m_p} f_{\rm H2}^2 \ \ \ [{\rm erg \ s^{-1} cm^{-2}}]  ,
\end{equation} 
the temperature profile of the disc is set by the advective cooling rate 
\begin{equation}
	Q_{\rm adv}= \frac{\Sigma v_r}{r} \frac{T k_{\rm B}}{ m_p} \xi \ \ \ [{\rm erg \ s^{-1} cm^{-2}]} ,
\end{equation}
which exceeds $Q_{\rm rad}$ 
by an order of magnitude within the entire nuclear disc (see Figure \ref{fig:disc_cooling}). In these equations for $Q_{\rm rad}$ and $Q_{\rm adv}$,
$\Sigma$, $n$, $T$,  and $v_r$ are the surface density, number density, temperature and radial velocity profile fits from Figure \ref{fig:dens_profiles_fixed},  and $\xi=-0.65$ is a dimensionless parameter following \citet{Chen1995}. $\Lambda_{\rm H2}$ is the molecular gas cooling function in $[\rm erg \ s^{-1} cm^{3}]$, and $f_{\rm H2}$ the molecular fraction.

Note however that such a nuclear disc differs from traditional optically thick slim discs in that the gas remains optically thin. Indeed, primordial gas only becomes optically thick around $n = 10^{10} \ { \rm cm}^{-3}$, which is above the maximum density of $n_{\rm max} = 10^{9} \ { \rm cm}^{-3}$ reached in our simulations. At densities above $n=10^{10}  \ { \rm cm}^{-3}$, $f_{\rm H2}$ also drastically increases due to 3-body interactions \citet{Grassi2014}, invalidating the assumption that the gas is mono-atomic, and that $\gamma=5/3$. We therefore conclude that to extend our simulations to scales smaller than an inner disc radius of $r_{\rm in} \approx 5 \times 10^{-3}$ pc, which is on the order of $10^5$ Schwarzschild radii (R$_{\rm SS}$) for the BH masses studied here, one would need to account for such a change of regime. This caveat notwithstanding, we expect the disc morphology to change significantly on scales below $10^{-3} \rm \ pc$. As can be seen in Figure \ref{fig:disc_cooling}, the expected larger $f_{\rm H2}$ allows $Q_{\rm rad}$ to dominate over $Q_{\rm adv}$ at such small radii, which we speculate will turn the disc from slim to thin.

\subsection{Consequences for black hole accretion}
\label{sec:BHaccretion}

The difference between accreting the entire core directly or embedding the BH at the centre of a resolved core results in a factor 5 difference in BH mass after just $0.5 \rm \ Myr$ of evolution and has a lasting influence on BH accretion. 

Comparing the duty cycle of R\_l20 to R\_l23-l28 in the second panel of Figure \ref{fig:bondiplot_fixed} shows that resolving the nuclear disc around the BH markedly changes the accretion pattern. The BH in R\_l20 grows chaotically, i.e. with an accretion rate that fluctuates by orders of magnitude over time periods of less than $0.1 \rm \ Myr$. By contrast, BHs in the three other simulations grow in an episodic fashion, where long periods of smooth accretion are interspersed with accretion bursts that can last up to $0.5 \rm \ Myr$. 

While R\_l26 and R\_l23 show similar accretion patterns, the factor 8 difference in resolution between these two runs affects the gas properties measured in the vicinity of the BH. First of all, the initial BH mass is smaller in R\_l26 as the volume covered by the accretion region, and therefore the mass it contains when transitioning to SLA, is resolution dependent. Secondly, the BH is fed by gas that loses sufficient angular momentum to enter the accretion region at the centre of the nuclear disc. BHs in higher resolution simulations have smaller accretion regions, so the gas needs to lose more angular momentum before being accreted, which results in a lower accretion rate during the smooth accretion phases (compare R\_l23 to R\_l26 around $t=103 \rm \ Myr$ in Figure \ref{fig:bondiplot_fixed}). Finally, as the gas is removed deeper in the BH potential well, higher local densities than in R\_l23 are reached. 

 \begin{figure*}
 	\centering
	\setlength{\tabcolsep}{0pt}
	\begin{tabular}{ccc}
		\includegraphics[width=0.32\textwidth]{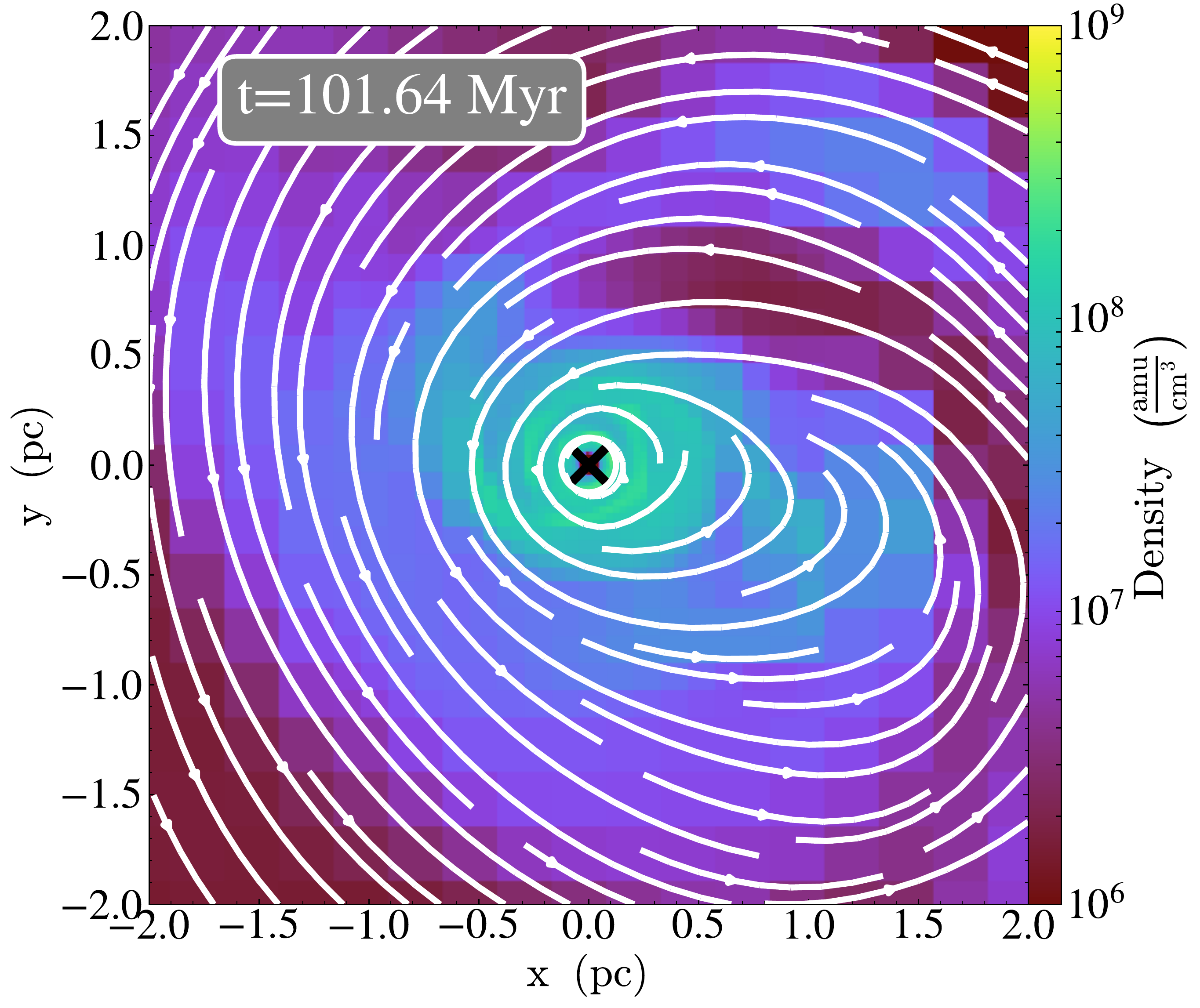}&
		\includegraphics[width=0.32\textwidth]{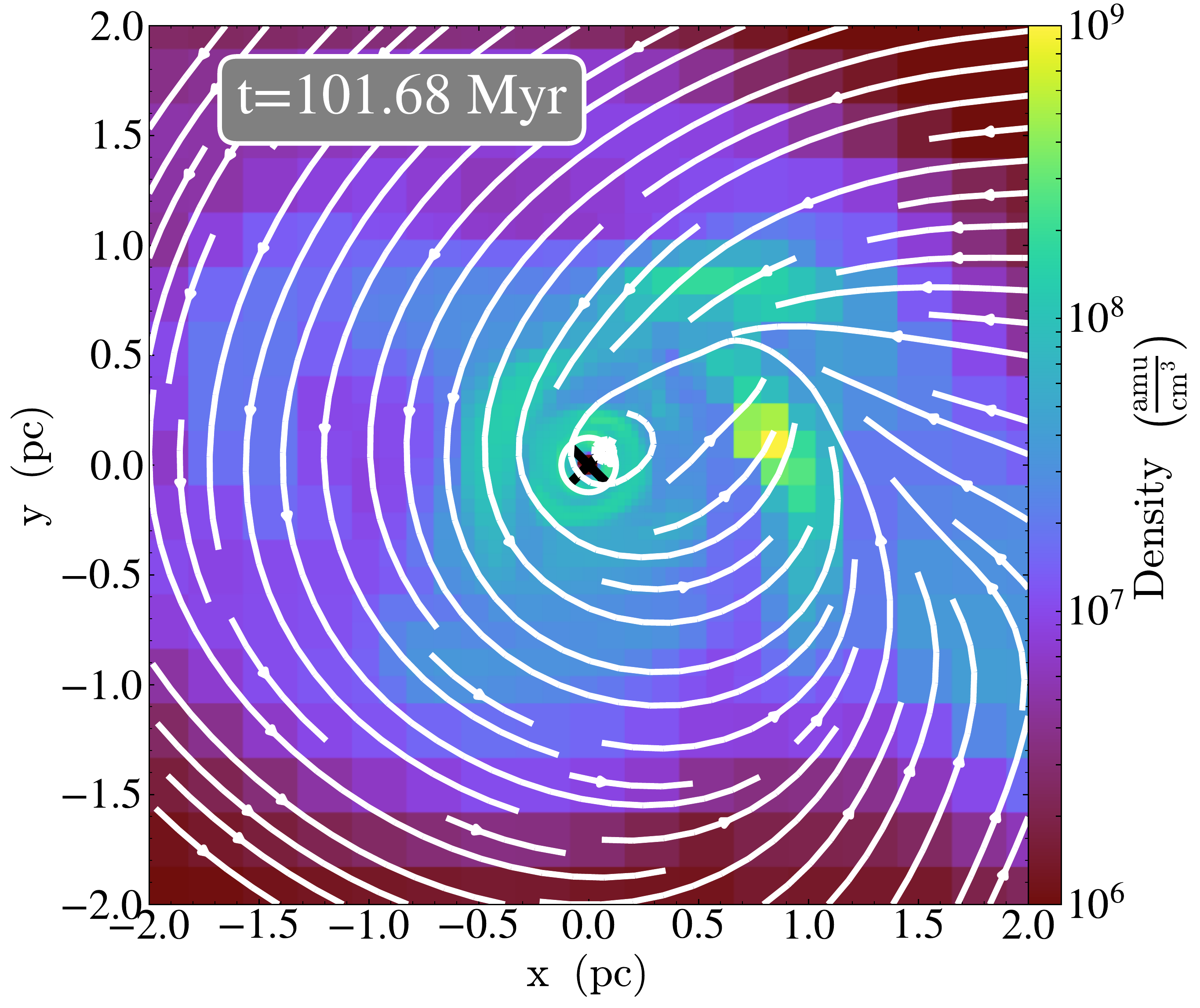}&
		\includegraphics[width=0.32\textwidth]{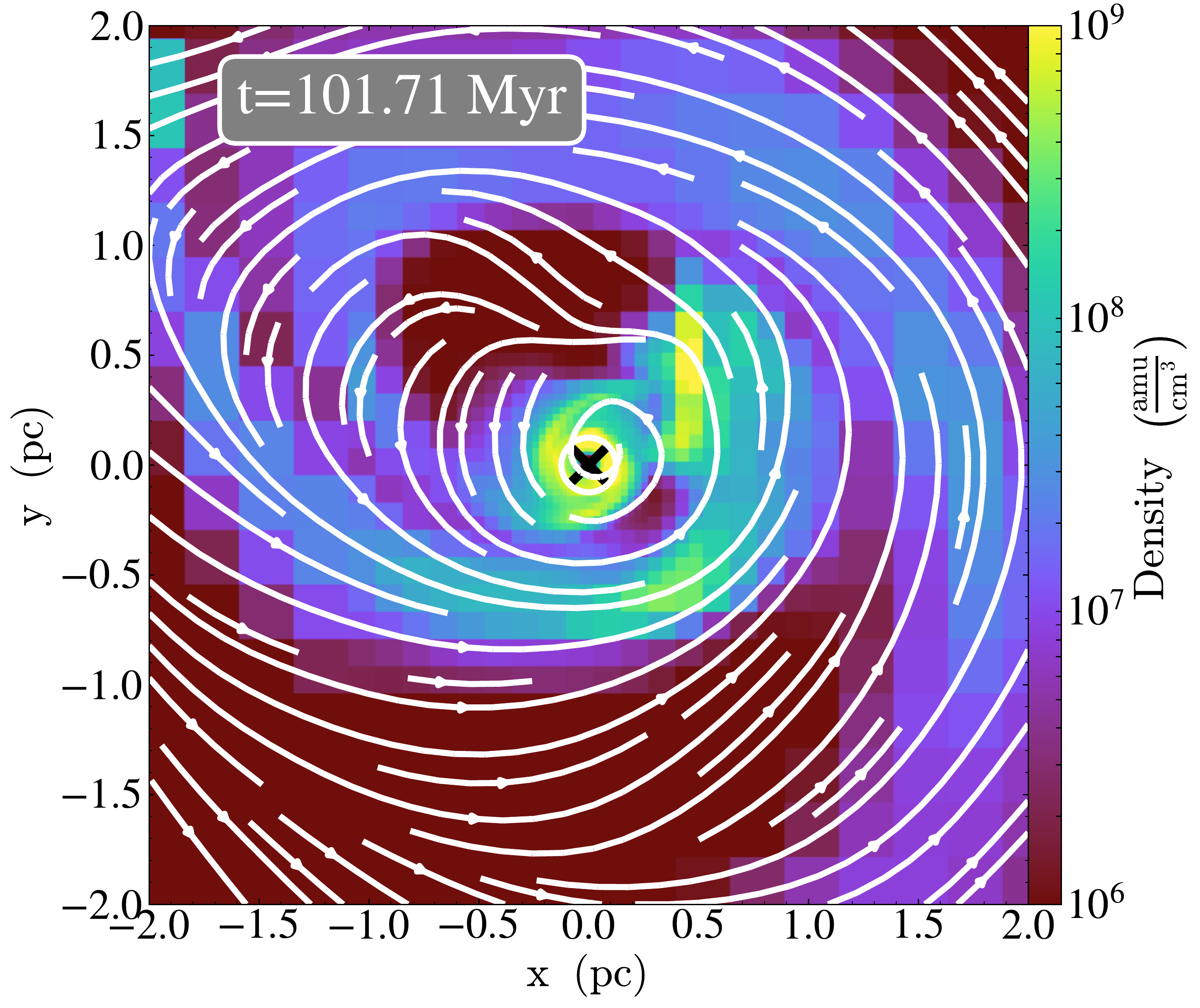}\\
		\includegraphics[width=0.32\textwidth]{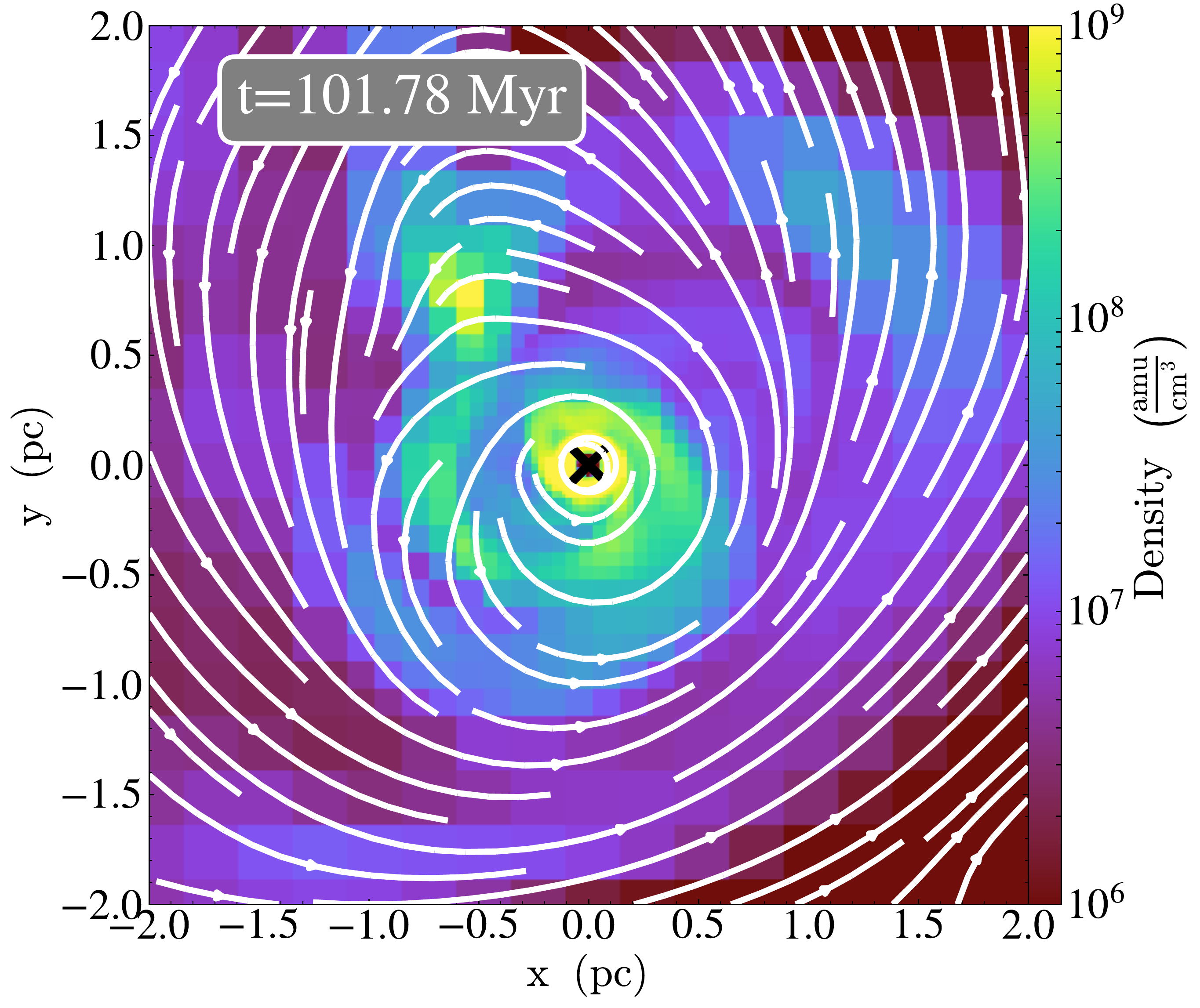}&
		\includegraphics[width=0.32\textwidth]{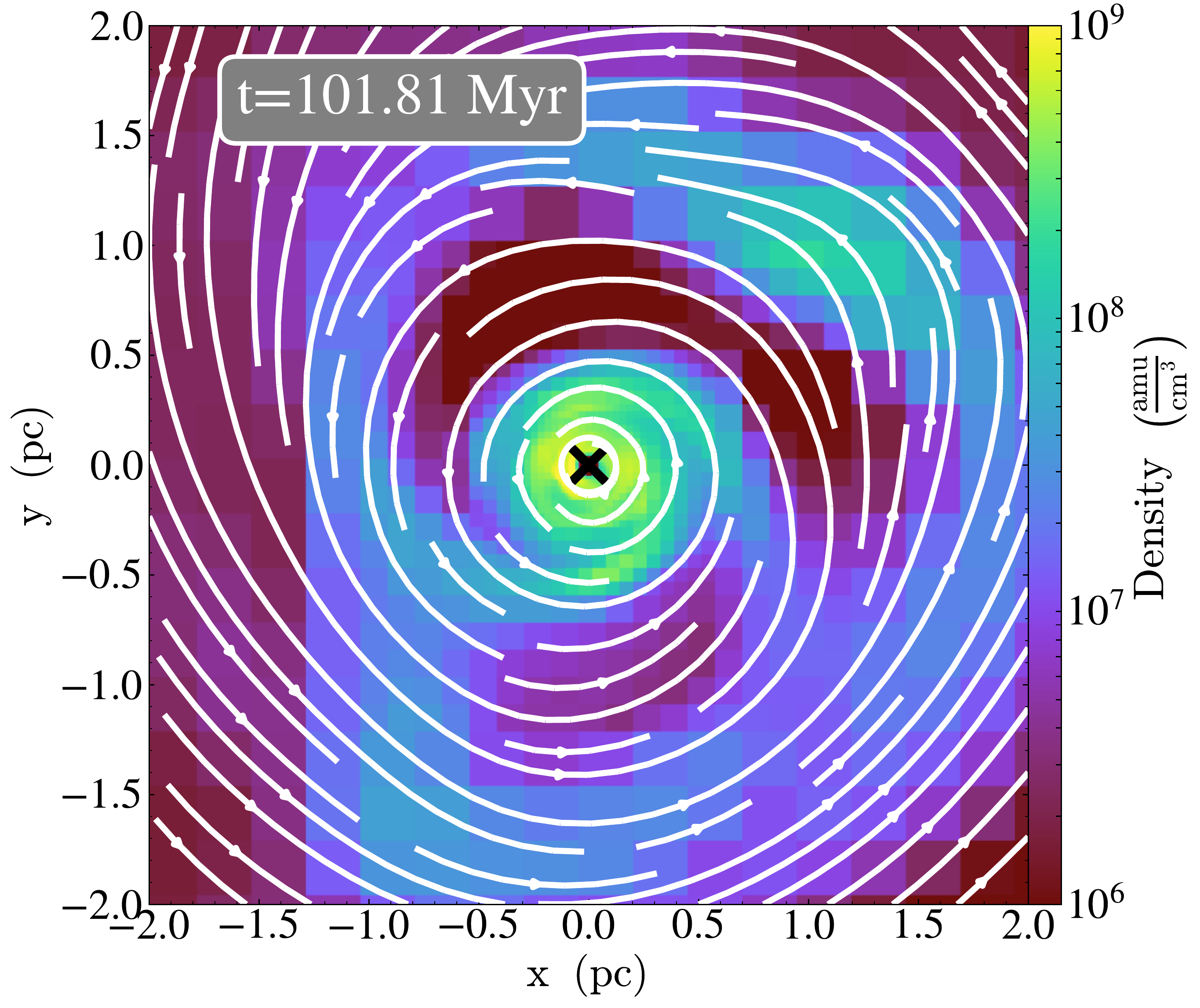}&
		\includegraphics[width=0.32\textwidth]{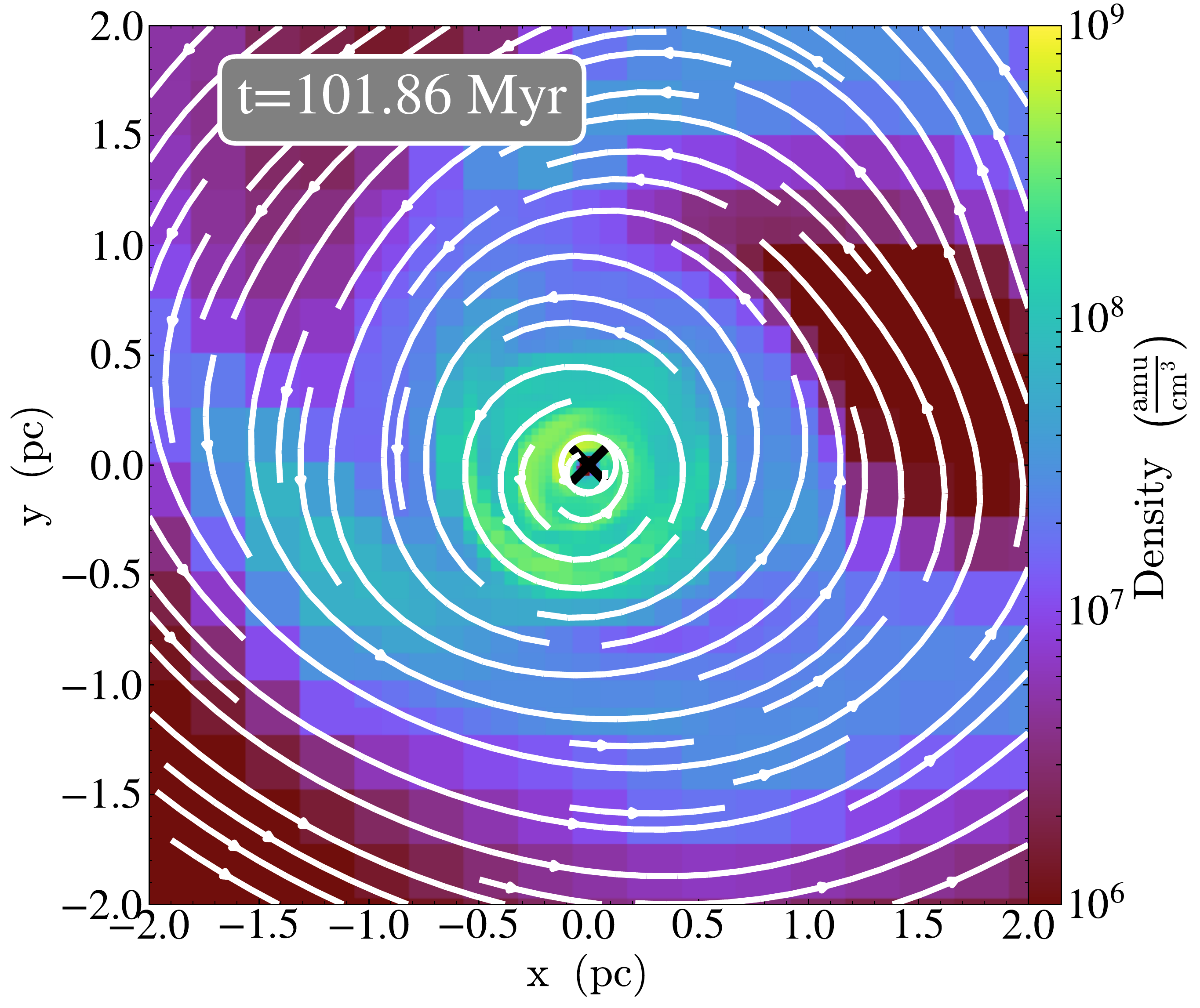}\\
	\end{tabular}
	\caption[Gas density projections of the BH environment during the accretion burst around $t=101.8 \rm \ Myr$ in R\_l26]{Gas density projections of the BH environment during the accretion burst around $t=101.8 \rm \ Myr$ in R\_l26. The accretion spike is caused by the disruption of a gas clump. The BH is marked by the black cross and streamlines are shown in white.}
	\label{fig:slices_merger}
\end{figure*}

\begin{figure}
	\centering
	\includegraphics[width=1.0\columnwidth]{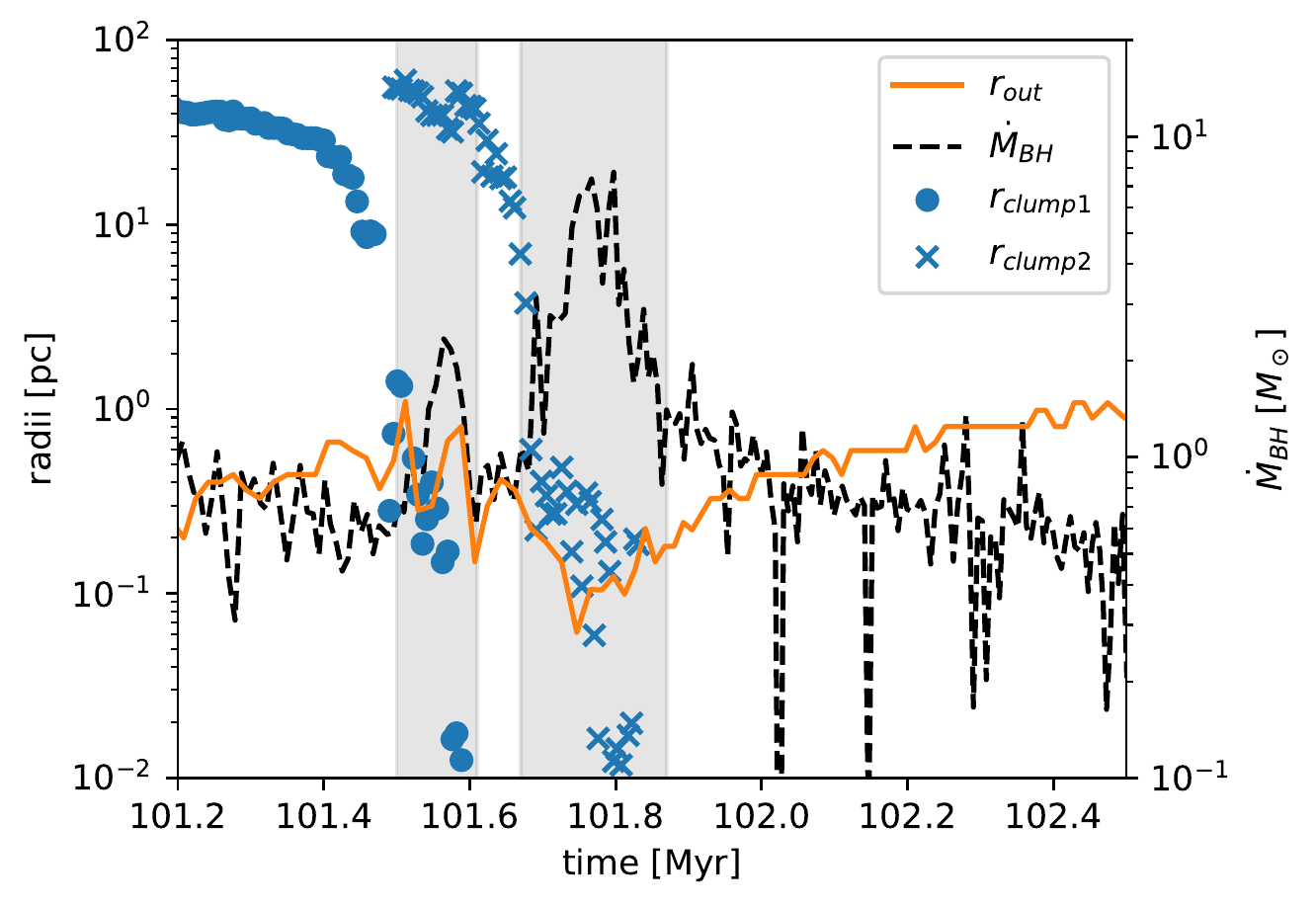}
	\caption{A detailed view of the accretion rate (dashed line) around the merger event of clump 2 shown in Figure \ref{fig:slices_merger}, overlaid with time evolutions of the outer disc edge $r_{\rm out}$ (solid line), defined to be the radius at which the Toomre parameter $Q$ falls below one, and the in-spiral of two clumps (clump 1 and 2, shown by round markers and crosses respectively.) Grey shaded regions indicate the epoch during which the BH mass accretion from each clump is calculated.}
	\label{fig:disc_instability}
\end{figure}

As can be seen in Figure \ref{fig:disc_instability}, infalling gas clumps trigger episodes of increased accretion. In the example from R\_l26 shown in this figure, two clumps reach the centre of the disc in rapid succession. The first, smaller clump, with a mass of $1.8 \times 10^5 \ \rm M_\odot$, causes a small, short-lived accretion spike that leaves the nuclear disc mostly unaffected. During this episode, the BH grows by $1.1\times 10^5 \ \rm M_\odot$, i.e. slightly less than the clump mass. By contrast, the second, more massive clump, with a mass of $8.8 \times 10^5 \ \rm M_\odot$, induces a violent disc instability that results in shrinking the nuclear disc outer radius from $r_{\rm out} = 1.2$ pc down to $r_{\rm out} = 0.4$ pc, as can be seen in Figure \ref{fig:disc_instability}, and from the density projections plotted in Figure \ref{fig:slices_merger}. The resulting redistribution of angular momentum allows gas to fall into the accretion region, boosting the accretion rate, before the disc resettles and the BH returns to smooth accretion. During this second infall event, the BH grows by a total mass of $1.3 \times 10^6 \rm \ M_\odot$, which is equal to 1.5 times the clump mass. The extra mass is drained from the nuclear disc. These accretion bursts are the dominant accretion mechanism as soon as the nuclear gas disc is resolved. Note that such a growth by clump capture is also reported in \citet{Lupi2016}, who perform numerical simulations of seed BHs in a galactic context at resolutions comparable to R\_l23. 

The differences in local gas properties and accretion duty cycles at higher resolution should have major consequences for BH feedback. 
Assuming $\dot{E}_{\rm BH} \propto \dot{M}_{\rm BH}$, R\_l20 would produce ``rapid fire'' feedback, consisting of fast bursts carrying over an order of magnitude more energy than that powered by the smoother accretion in R\_l26 during the same amount of time. Conversely, R\_l23, and particularly R\_l26 and R\_l28, would produce much more sustained weaker feedback, interspersed with a few episodes of strong feedback bursts that would interact with the dense gas of the nuclear disc. These different duty-cycles are, in turn, expected to significantly impact the effect of feedback on the BH host galaxy, particularly when combined with the effect reported in \citet{Negri2016}, i.e. that resolution impacts the ability of BHs to self-regulate as it affects the size of the gas reservoir into which feedback energy is deposited. Like accretion and dynamical friction, feedback sub-grid algorithms will have to be carefully re-examined for use in high-resolution environments of the kind presented here, and an investigation into the effect of feedback on BH accretion is therefore postponed to future work.

One advantage of the BH zoom algorithm is its relatively low computational cost, as can be seen in Table \ref{tab:fixed_simulations}. The main limitation for increasing $l_{\rm zoom}$ is the shrinking time step, set (among other criteria) by the sound-crossing time or the flow crossing time of the smallest cell in the simulation. Adding an extra level of resolution at identical gas conditions would halve the time step. Simulations with higher $l_{\rm zoom}$ also have a higher sound speed in the smallest cells near the BH, as well as a higher flow velocity into the BH potential, further decreasing the time step. Some of the extra cost can be mitigated by sub-cycling and by distributing the smallest cells over a larger number of processors, but ultimately the number of extra resolution levels that can be added is limited by the necessity of achieving a reasonable simulation real run time. 

In conclusion, this section demonstrated the difficulty of extrapolating mass accretion histories of BHs using lower-resolution simulations. Already in the highly idealised example presented in this work, the accretion onto the BH is regulated by the rotational support of a pc size nuclear disc and disc-instabilities triggered by dense, sub-pc size gas clumps. While extrapolation to even smaller scales is difficult, the results presented here suggest that BH accretion models in simulations with only $\sim$ pc (or worse) resolution need to be treated with caution, as both accretion pattern and magnitude are still strongly resolution dependent on these scales. We discuss this issue in more detail in the next section. 

\section{Discussion}
\label{sec:discussion}

\subsection{Sub-grid algorithms for accretion}
Based on the results present in Section \ref{sec:mseed} and \ref{sec:resolution}, none of the modified Bondi-schemes introduced in the introduction to this paper appear to be an appropriate choice to capture the accretion behaviour revealed by zooming in on the BH.  While BHL based schemes have an explicit dependence on the BH mass squared, the results presented here show that for light BHs in particular, the accretion rate mostly depends on the collapsing gas cloud feeding it, not the mass of the seed BH itself.

Figure \ref{fig:bondiplot_fixed} clearly shows that lower resolution simulations systematically result in the presence of lower density gas in the vicinity of the BH. Despite this, we do not advocate compensating with a boost factor, as suggested by \citet{Booth2009}, for two reasons. Firstly, despite a lower gas density, R\_l20 {\em over-accretes} in comparison to R\_l26. Secondly, adding a boost factor to the BHL formula is inconsequential once accretion has transitioned to SLA, as the mass accreted by the BH is no longer determined by the value calculated by the BHL accretion rate in Equation \ref{eq:BHL}, boosted or otherwise.

Work presented here showed the main limiting factor to accretion is the rotational support of the gas on pc and sub-pc scales. This favours accretion models that explicitly take the rotation of the gas into account. \mbox{\citet{Krumholz2004}} and \mbox{\citet{Curtis2016}} suggest models in which the BHL accretion rate should be scaled down by a factor reflecting the vorticity of the gas. Alternatively, \citet{Angles-Alcazar2015} proposes to use gravitational torques on $\sim 100 \rm \ pc$ scales to calculate the accretion rate onto the BH, which seems an even more appropriate choice, as it is based on a global gas measure and independent of BH mass. However, we note that the strong rotational support of gas found here could be due to the coherent rotation added to the halo as part of the initial conditions. Based on current knowledge, there is no guarantee that high-redshift mini-halos hosting early BHs also are rotation dominated, particularly once both stellar and BH feedback have been added. \citet{Angles-Alcazar2017} find that when using the torque-based model in a galactic context, BHs do not accrete efficiently until the galactic bulge forms at $z \sim 2$. This could be tentative evidence that early SMBHs, observed at $z>6$, accrete primarily through chaotic accretion rather than smooth accretion mediated by a galactic gas disc. To investigate this matter further, it will be necessary to conduct a high-resolution study of the kind presented in this work, but in an explicit cosmological context to 
capture the properties of the galactic gas flow during the first billion years of the Universe.

\subsection{From nuclear disc to ISCO}
\begin{figure}
	\centering
	\includegraphics[width=1.0\columnwidth]{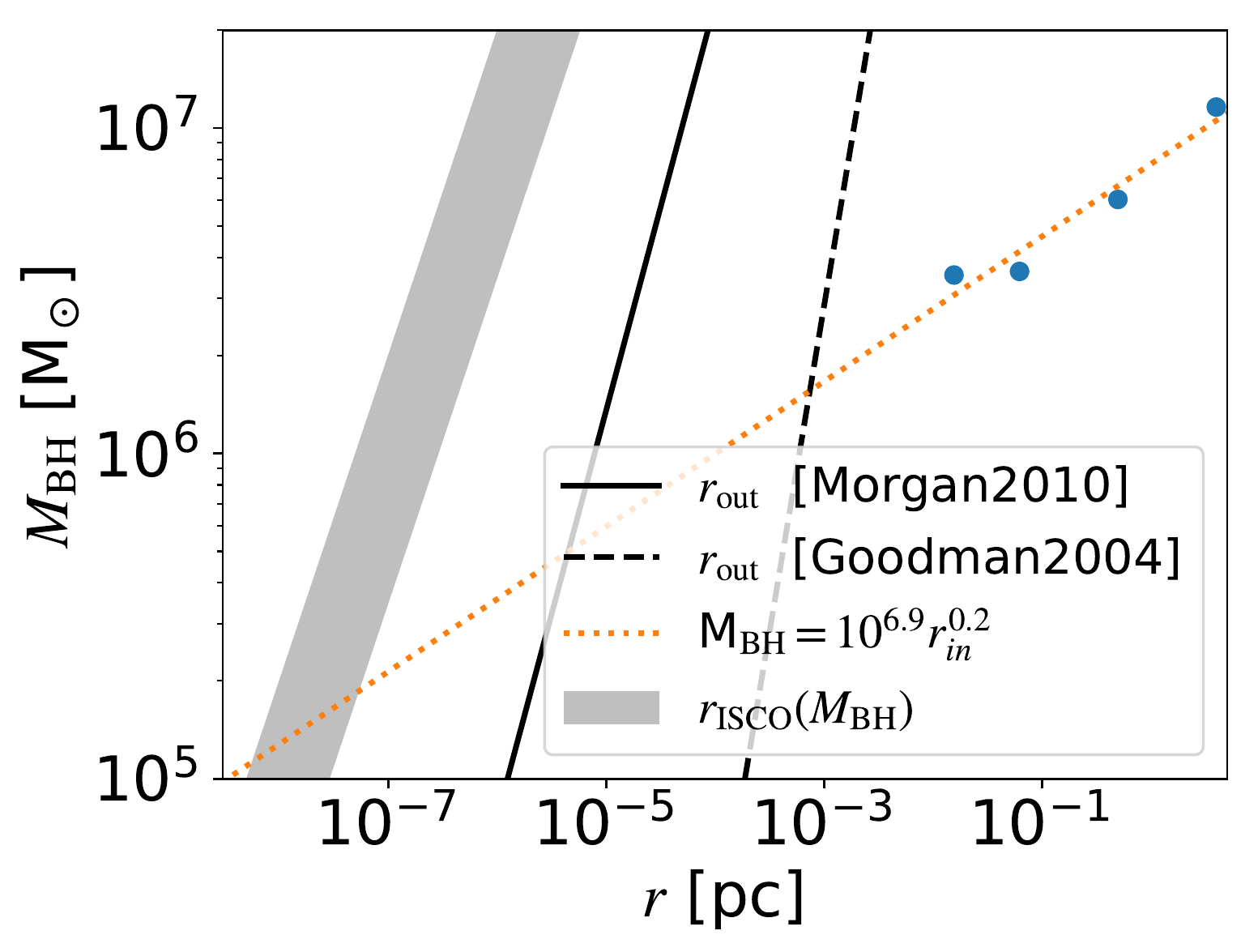}
	\caption{Dependence of the BH mass at $t=105 \rm \ Myr$  on the inner radius of the nuclear disc, $r_{\rm in}$. Each simulation is shown as a round marker, and a power-law fit is shown as the dotted line. The innermost stable orbit, $r_{\rm ISCO}$, for a range of BH spin parameters is shown as the shaded region. Also plotted is the observed extent of the accretion disc according to Equation 4 in \citet{Morgan2010} (solid line) and Equation 27 in \citet{Goodman2004} (dashed line).}
	\label{fig:BHmass_convergence}
\end{figure}

Spanning a region between 10$^{-2}$ and 1 pc, which corresponds to $10^5-10^7 \rm \ R_{SS}$ for the $\sim 10^6 \rm \ M_\odot$ BH in R\_l28, the nuclear disc found in our simulations presents an inner radius, $r_{\rm in}$, which still is very distant from the BH R$_{\rm SS}$, and an outer edge, 
$r_{\rm out}$, located much further than both observations and theoretical considerations based on a thin disk model suggest. Indeed, \mbox{\citet{Shlosman1987}} argue that Keplerian discs around massive BHs are expected to self-gravitate beyond $10^3 \rm \ R_{SS}$, which is corroborated by the more detailed analysis in \citet{Goodman2004}, who predict a self-gravity radius for the disc at $6\times 10^3 \rm \ R_{SS}$ (see Equation 27 in \citet{Goodman2004}). Observationally, SED fitting from \citet{Sirko2003} reveals an outer disc radius of $10^5 \rm \ R_{SS}$ for SMBHs, and the micro-lensing analysis from \citet{Morgan2010} places it at only $160 \rm \ R_{SS}$ for a $10^6 \rm \ M_\odot$ BH (see Equation 4 in \citet{Morgan2010}). The nuclear disc found in our simulations is therefore much larger than observed accretion disc (see Figure \ref{fig:BHmass_convergence}). 

While we expect accretion discs composed of primordial gas to have a larger extent due to the significant reduction of radiative cooling channels available, forcing the disc to remain hotter and therefore better supported against fragmentation, the magnitude of the discrepancy between observations and our results strongly suggests that the disc resolved in our work is not the real accretion disc around the BH. As argued in the previous section, we expect that, as resolution increases, a nested disc structure would develop, with a thin disc eventually forming on scales below $10^5 \rm \ R_{SS}$,  embedded within the slim nuclear disk (see Section \ref{sec:nuclear_disc} for details). Bearing in mind the difficulty to predict the behaviour of this unresolved accretion disc, the power-law extrapolation we provide in Figure \ref{fig:BHmass_convergence} suggests that the BH mass has not yet converged, even for our highest resolution simulation. If the current trend, $M_{BH} = 10^{6.9} \Big ( \frac{r_{\rm in}}{\rm pc} \Big )^{0.2} \rm \ M_\odot$, continues then the BH mass in R\_l28 is overestimated by about an order of magnitude in comparison to the limit obtained when $r_{\rm in} \rightarrow r_{\rm ISCO}$. 

This resolution dependence of the BH mass is a consequence of the two assumptions made by SLA: all gas entering the accretion region is accreted by the BH, and the timescale between gas entering the accretion region and entering the event horizon of the BH is short. The former holds for the simulations presented here, particularly R\_l23 - R\_l28 where the BH dominates the gravitational potential at the edge of the accretion region. The question of accretion timescale is more difficult, as comparing R\_l20 to R\_l23 - R\_l28 shows that in the presence of rotation, accretion can be significantly delayed on the scales resolved here. Therefore, even if all gas will eventually be accreted, one would expect a significant time delay, which could also have important consequences for BH feedback. One solution to account for this delay could be to scale down the accretion rate onto the BH by a factor reflecting the vorticity of the gas in, or just around, the accretion region. This concept would be difficult to implement within the framework of the current accretion algorithm, as SLA is an emergent phenomenon, not an explicitly calculated accretion rate. Scaling it down to compensate for vorticity is therefore not guaranteed to produce the desired effect. Indeed, one of the main justifications for using the BH zoom technique is precisely that it is able to account for the vorticity of the gas self-consistently, rather than through yet another sub-grid scheme. 

Exploring an alternative approach, an explicit flux-based scheme to calculate gas accretion onto sink particles was introduced in \textsc{ramses} by \citet{Bleuler2014}, based on previous work by \citet{Gong2013}. The idea is to compute accretion directly from hydrodynamics, using the Riemann fluxes at the boundaries of the cell where the BH resides. The authors compare this scheme to a standard BHL scheme and note that it performs better at high resolution, whereas the BHL scheme is more accurate at low resolution. Compared to the algorithm we presented here, the flux scheme of \citet{Bleuler2014} has the advantage that it produces a smooth transition of the flow into the accretion region and thereby avoids artificially low densities within this region and the associated pressure forces at the accretion edge boundary. On the other hand, our scheme automatically transitions from BHL to SLA, without the need to introduce an explicit transition criterion which necessarily depends on the value of  $R^{\rm A}$ and can therefore be difficult to define in a galactic context, as discussed in Section \ref{sec:fixes}.  

As mentioned in Section \ref{sec:nuclear_disc}, it is unlikely that the nuclear disc structure will continue to be self similar as $\Delta x_{\rm min} \rightarrow r_{\rm ISCO}$, as the gas is expected to become optically thick and radiatively efficient at cooling. However, the gas properties measured at the edge of the accretion region could provide input parameters for an accretion-disc based sub-grid model, like the ones presented in \citet{Power2011} and \citet{Dubois2012a}, particularly as the size of the accretion region for R\_l28 approaches the expected boundary of real SMBH accretion discs\footnote{The accretion disc around a $M_{\rm BH} = 10^9 \rm \ M_\odot$ BH has an estimated maximum extent of  $\sim 10^{-3} \rm \ pc$, using Equation 4 in \citet{Morgan2010}. This is only an order of magnitude below the inner disc edge in R\_l28, which has an accretion region of size $r_{\rm in} = 10^{-2} \rm \ pc$.}. In such models, gas is removed from the grid and added to a sub-grid ``accretion disc'', from which the accretion rate onto the BH itself is calculated, taking the viscous timescale on which angular momentum is transferred explicitly into account. This would also provide an opportunity to include physical processes not included in the original simulations, such as magnetic fields. How much more accurate such a scheme would be at capturing the accretion duty cycle of a BH on the relevant length scales remains an open question at this stage.

\subsection{Other refinement schemes}
The work presented in this paper is similar in spirit to that of \citet{Curtis2015}, who present a super-Lagrangian refinement algorithm for the BH vicinity implemented in the moving-mesh code AREPO. One significant difference between the two works is that \citet{Curtis2015} define a maximum radius, set to the gravitational smoothing length associated with the BH, within which resolution is increased as one approaches the BH. In BH zoom, a minimum radius is defined instead, within which the maximum resolution is maintained. This high resolution region is then surrounded by regions of progressively lower resolution until the resolution of cells adaptively refined using the standard quasi-Lagrangian refinement criterion is reached. Maintaining a fixed grid around the BH enforces that accretion onto it only proceeds from cells at the highest resolution. This is of key importance as convergence tests presented in Section \ref{sec:convergence_nrefine} have shown that accreting from a region refined at mixed resolution can lead to divergent mass accretion histories (see C\_l20n2, which has $r_{\rm refine} < 2 \Delta x_{\rm min}$, in Section \ref{sec:convergence_nrefine}). 

Another important difference is that in \citet{Curtis2015}, the smallest cell is of the order of the BH Bondi radius, $R^{\rm B}$. In the simulations presented here, the Bondi radius is consistently well resolved for BHs of comparable seed mass. Accretion in \citet{Curtis2015} therefore continues to rely on the BHL accretion rate, and uses a resolution at which the BHL accretion rate based on local quantities can differ by as much as a factor 5 from the analytic value at infinity. The authors combine BHL accretion with a non-isotropic accretion scheme where gas is only accreted from cells with a temperature below a threshold temperature, $T_{\rm cold}$, and when the fraction of cold gas within the softening length of the BH exceeds 25\%. This is designed to mimic accretion via a cold, dense disc expected to dominate BH accretion in a galactic environment. We employ no such sub-grid model as the nuclear disc feeding the 'true' BH accretion disk can be resolved self-consistently using the BH zoom technique (see Section \ref{sec:resolution}). \New{However, contrary to the work presented here,  \citet{Curtis2015} include AGN feedback. Based on their results, it is possible that in the presence of such a feedback our simple accretion algorithm could be unsatisfying, and that the size of the zoom region would have to be significantly larger to cope with the rapidly changing gas properties in the BH vicinity, making the simulations more computationally expensive.}

The issue of keeping the BH attached to the local gas flow is addressed in \citet{Curtis2015} by calculating  separate dynamical and sub-grid masses for each BH. The former is used for calculations of the gravitational potential whereas the latter is the `physical' mass of the BH, equivalent to $M_{\rm BH}$ in the work presented here. The advantage of such a dynamical mass model in comparison to the drag force model used in our work is that it does not rely on measuring local relative velocities and therefore avoids the associated difficulties. However, as stated in \citet{Curtis2015}, using the dynamical mass for local gravity calculations introduces significant changes to the central potential, and has a lasting impact on local gas dynamics, when the BH mass exceeds the local cell masses. At the resolutions studied in this paper, the BH mass frequently exceeds the local cell mass even for simulations where dynamical friction is unresolved. The ratio $M_{\rm BH}/M_{\rm cell} \propto M_{\rm BH}/(\rho_{\rm cell} \Delta x_{\rm zoom}^3)$ (important for the dynamical mass model) depends on the local density $\rho_{\rm cell}$ whereas the ratio $R^{\rm A}/\Delta x_{\rm zoom} \propto M_{\rm BH}/( \Delta x_{\rm zoom} v^2)$ (important for the drag force model) depends on the relative velocity between BH and gas, $v$, instead. A supersonically moving BH in a low density environment can therefore dominate the local potential but still have under-resolved dynamical friction. One such example is the BH in  D\_l26\_small in Section \ref{sec:dynamical_friction} that fails to settle in the disc despite its mass exceeding local cell masses by more than an order of magnitude. 

Finally, we note that another super-Lagrangian refinement scheme for BHs in \textsc{ramses} has been proposed by \citet{Lupi2014a}, in which the BH host cell is forced to always be at the highest resolution available in the simulation. Their scheme thus differs from the work presented here in that the BH environment is only resolved as well as the densest gas, not better, as with BH zoom. We independently implemented the refinement scheme of \citet{Lupi2014a} in the fiducial runs without BH zoom to avoid local cells de-refining when the density in the accretion region drops as one enters the SLA regime. It is obviously much more costly computationally to reach a given resolution in the BH vicinity with such a scheme than with BH zoom, as one needs to refine the whole galactic environment at the same resolution, which makes running extremely high resolution simulations such as those we presented in this work impossible in practice.  

\subsection{BH zoom and angular momentum}
\New{As can be seen in Figure \ref{fig:bondiplot_fixed}, the resolution in the BH zoom region has no discernible influence on the amplitude of the BH initial mass jump, with all four simulation BHs gaining $\approx 5 \times 10^5 \rm \ M_\odot$ during the cloud initial collapse phase. Naively one would expect the BH in the better resolved simulations to have a smaller initial mass jump than the ones in less resolved simulations, as some of the gas should remain on the grid further away from the accretion region. This does not seem to be the case here as a similar mass of gas free-falls onto the BH in all four simulations. This could be a consequence of the fact that cartesian grid codes, such as the one used here, struggle with angular momentum conservation when the grid structure changes dramatically. If angular momentum is poorly conserved during the zoom triggering period, even gas at comparatively large radii would lose angular momentum and be accreted onto the BH, boosting the effective seed mass. Adding refinement more slowly could mitigate this effect but it would also mean that either accretion onto the BH has to be suspended for several dynamical times, or that early accretion onto the BH occurs at less than full resolution, which in itself would lead the BH to over-accrete in comparison to full resolution. Bearing in mind the cartesian grid caveat, the density slices in Figure \ref{fig:multiplot_fixes} clearly show a disc forms during the initial refinement period, {\em before} the target resolution is reached and the BH starts to accrete. They also show that this disc survives the collapse: there is no sign, at any stage, of an excessive amount of low angular momentum gas collapsing directly onto the BH. This suggest that while the angular momentum of the gas could be somewhat under-estimated, it is not catastrophically so, and the resolution-independent amplitude of the BH initial growth burst is very likely an attribute of the particular host cloud, rather than a numerical issue.}

\New{Once the BH has formed and the BH zoom refinement completed, the grids around the BH remain 'static' despite the BH moving around the galaxy, behaving more like a set of nested grids than like an adaptively refined mesh. From that moment on, angular momentum is conserved with the standard accuracy of a fixed cartesian grid \citep[which compares quite well with smooth particle hydrodynamics, see e.g.][]{Commercon2008} and plays a deciding role in the gas dynamics within the zoom region (see Section \ref{sec:nuclear_disc} and the rotationally supported Keplerian disc forming around the black hole). We therefore conclude that overall, the level of numerically induced loss of angular momentum in our simulations is comparable with what is expected given our simulation technique and resolution.}

\section{Conclusions}
\label{sec:conclusions}

The origin of SMBHs is a complex problem that involves gas flows and gravitational torques over many orders of magnitude in length scale, from the Mpc scales of the cosmic web to the au scales of the event horizon. This paper studied the early evolution of BHs, fed by collapsing gas clouds using a ``black hole zoom'' refinement algorithm, a super-Lagrangian scheme for the AMR code \textsc{ramses} in which the BH is surrounded by a spherical region consistently refined at high resolution. Simulations presented in this paper used BH zoom to improve the spatial resolution in the vicinity of the BH by two orders of magnitude. 

Accurately capturing the BH orbit within the galaxy is of uttermost importance in the early evolution stages, when the accretion history of the BH is set by the internal structure of the self-gravitating gas cloud from which it formed. More specifically, we showed that dynamical friction plays a crucial role in studies of early BH accretion. For initially sufficiently massive BHs, dynamical friction is self-consistently resolved. For lighter seeds, which have $R^{\rm A}_\bullet < \Delta x_{\rm zoom}$, a sub-grid model is required to properly capture the BH dynamical evolution. However, reliably measuring local gas properties becomes increasingly difficult when resolution increases, as small changes in the movement of the BH lead to large fluctuations of local quantities. \New{The relative velocity between the BH and local gas is particularly vulnerable to being over-estimated. As a subgrid algorithm is only ever as reliable as its input parameters, we propose to use the simplest possible sub-grid recipes in high resolution environments, and caution the reader to carefully question the assumptions that go into each one (see Appendix \ref{sec:convergence_levelmax} for details).}

Nevertheless, we demonstrated that robust results for the BH mass accretion history, largely independent of the (somewhat arbitrary and resolution dependent) choice of BH seed mass can be achieved using two simple sub-grid models that require a minimum number of input parameters. Firstly, we show that a BHL based accretion algorithm automatically transitions to supply limited accretion (SLA) at sufficiently high resolution. During SLA, the accretion rate onto the BH is determined by the mass flux into the accretion region, an inherently local measure that will converge to the correct (Newtonian) solution as the size of the accretion region approaches the size of the BH event horizon. Secondly, a drag force model with constant maximum magnitude is needed to include {\em unresolved} dynamical friction. Such a procedure allows light BHs to remain reliably attached to collapsing local gas clouds, \New{and makes the long-term mass evolution history of the BH largely independent of seed mass}. As already shown in \citet{Beckmann2018a}, sub-grid drag force prescriptions should be switched off when $R^{\rm A}_\bullet > 0.2 \Delta x_{\rm zoom}$, as they can unphysically accelerate the BH once the gravitational wake becomes resolved.

As for the impact of resolving the black hole environment on accretion, work presented in this paper showed that, \New{in the absence of any kind of feedback,} the mass accretion history of a BH is driven by the internal structure of the clouds from which it forms and accretes. Resolving the gravitational zone of influence of the BH is therefore a necessary but not sufficient condition to capture a BH's accretion history. 

Indeed, when we compared the evolution of BHs in simulations with identical initial conditions, but where the BH local environment was resolved at $0.99 \ \rm pc$, $0.16 \ \rm pc$, $0.01\ \rm pc$ and $0.002 \ \rm pc$ respectively, the gas properties differed so notably that the mass of the BHs in the simulations never converged, despite the fact that all had a well resolved Bondi radius, with $R^{\rm B} / \Delta x_{\rm min} > 10^{3}$ at all times. In the least resolved simulation, the BH accretes continuously but chaotically, as gas clumps are disrupted and accreted as soon as they fall into the accretion region. At higher resolution, by contrast, the simulations resolve more and more of a slim nuclear gas disc that smoothly feeds the BH. Once such a nuclear disc has formed, mass growth proceeds mostly through accretion bursts, triggered when in-falling massive gas clumps accrete onto the disc and cause a disc instability which feed the BH. These accretion episodes occur much more rarely but last much longer than the oscillations in the accretion rate measured in the lowest resolution simulation. Resolving the nuclear disc around the BH therefore changes the black hole accretion duty cycle from chaotic to episodic.

Using an admittedly simplified numerical experiment set-up, we established that the accretion behaviour of the BH is highly non-linear and dominated by small scale features in the immediate vicinity of the BH. Moreover, resolving the characteristic scale length of the BH gravitational zone of influence only provides an upper limit to gas accretion, as internal structure of the gas clouds feeding the BH and rotational support on sub-pc scales have long-term consequences on BH accretion histories. In particular, we have drawn attention on the potential existence and importance of a nuclear disc feeding the true BH accretion disc on yet smaller scales. We caution that this could be due to the lack of feedback in the simulations presented here, as well as the coherent rotation added to the DM halo as part of the initial conditions. Better understanding the early accretion history of SMBHs will require performing high resolution simulations of the kind we presented in an explicit cosmological context, which include both stellar and BH feedback. 


\section*{Acknowledgements}
We would like to thank Harley Katz, Yohan Dubois and Marta Volonteri for useful discussion, and the anonymous referee for constructive comments. The research of RSB is supported by Science and Technologies Facilities Council (STFC) and by the Centre National de la Recherche Scientifique (CNRS) on grant ANR-16-CE31-0011, and the research of AS and JD at Oxford is supported by Adrian Beecroft. This work is part of the Horizon-UK project, which used the DiRAC Complexity system, operated by the University of Leicester IT Services, which forms part of the STFC DiRAC HPC Facility (\mbox{\url{http://www.dirac.ac.uk}}). This equipment is funded by BIS National E-Infrastructure capital grant ST/K000373/1 and  STFC DiRAC Operations grant ST/K0003259/1. DiRAC is part of the National E-Infrastructure. All visualisations in this work were produced using the yt-project \citep{Turk2010}.




\bibliographystyle{mnras}
\bibliography{references} 



\appendix
\section{Measuring local gas properties in collapsing gas clouds}
\label{sec:convergence_levelmax}

\begin{figure}
	\centering
	\includegraphics[width=\columnwidth]{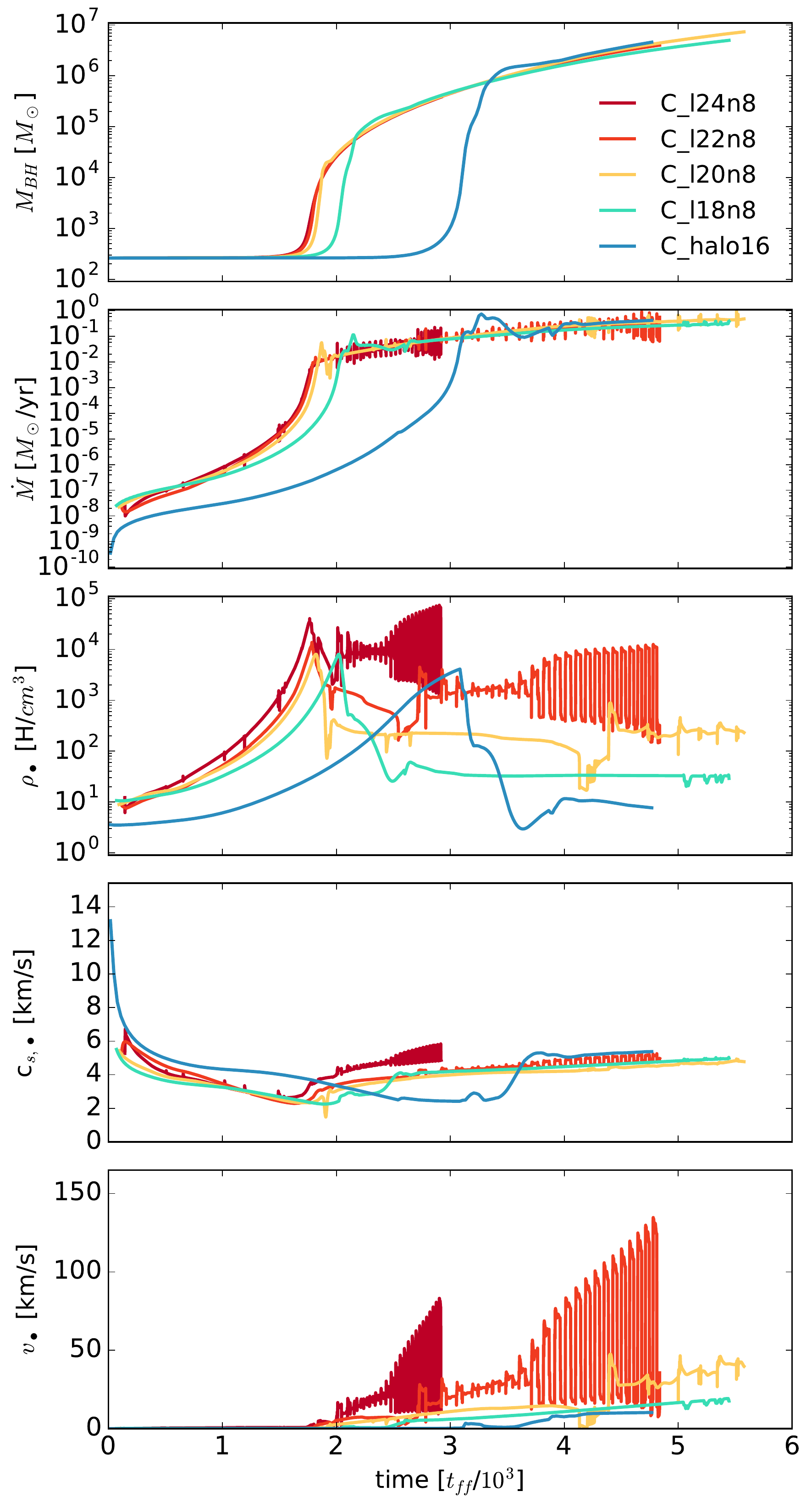}
	\caption[Time evolution of gas and BH properties, for simulations with different $\Delta x_{\rm zoom}$]{Time evolution of gas and BH properties, for simulations with different $\Delta x_{\rm zoom}$ but otherwise identical in all aspects (see Table \ref{tab:convergence} for details).}
	\label{fig:convergence_levelmax}
\end{figure}

Even in the highest resolution simulations of galaxies currently available, the smallest resolution extends orders of magnitude beyond the physical size of the BH, whose event horizon is best measured in au. Explicit sub-grid schemes therefore remain indispensable to handle unresolved physics, as is demonstrated for dynamical friction in Section \ref{sec:mseed}, even in sub-pc scale resolution simulations. However, these sub-grid schemes require input parameters, which in turn requires local or non-local gas properties to be measured. The spherically collapsing test case from Section \ref{sec:zoom-within-zoom} is used to demonstrate some of the pitfalls of measuring local gas properties, such as density, as well as less local measures, such as relative velocity, within a resolved collapsing gas cloud. In order to investigate the impact of resolution in the BH vicinity on gas properties as measured by the BH, this section presents a comparison of BH zoom simulations with a range of maximum refinement levels, i.e. $16 \leq l_{\rm zoom} \leq 20$ (see Table \ref{tab:convergence} for details). 

Figure \ref{fig:convergence_levelmax} shows that at all resolutions, the BH transitions to the SLA regime. As the accretion rate onto the BH thus becomes driven by the gas mass flux into the accretion region, which is approximately independent of radius for the free-falling gas studied here, BH accretion rates and BH masses converge for all resolutions probed. Simulations with higher $l_{\rm zoom}$ transition earlier because dense gas increases the accretion rate during the initial BHL phase, allowing the BH to build up mass faster. Simulations with higher $l_{\rm zoom}$ also have smaller accretion regions, which lead to smaller transition masses as $M_{\rm SLA} \propto \sqrt{V_\bullet (l_{\rm zoom})}$. 

Figure \ref{fig:convergence_levelmax} shows that the time-averaged value of the local gas density, $\rho_\bullet$, is resolution dependent. This has two reasons, one physical and one numerical. Physically, higher resolution simulations resolve more of the central density peak of the collapsing cloud, thus embedding the BH in denser gas. Numerically, SLA removes exactly 75\% of the mass contained in a gas cell during accretion. The residual mass, $M_{\rm res}$, left in the accretion region at the end of an accretion step, adjusts until the inflowing mass between accretion steps, $M_{\rm inflow}$, becomes exactly equal to 75 \% of the total mass from which the BH accretes, i.e. until $M_{\rm inflow} = 3 M_{\rm res}$. The process is self-balancing, as surplus gas is deposited in, or removed from, $M_{\rm res}$ until it holds. For the free-falling gas in this test case, $M_{\rm inflow}$ is constant and independent of the radius through which it is measured, and therefore approximately resolution-independent, but $M_{\rm res}$ depends on the size of the accretion region. Therefore, to maintain $M_{\rm res} = 1/3 M_{\rm inflow} = {\rm constant}$, simulations with smaller accretion regions must achieve higher densities, as seen in Figure \ref{fig:convergence_levelmax}. 

From visual inspection, the BHs in the simulations presented in this section remain located at the bottom of the potential well, with no obvious movement relative to the box. Their relative velocity is therefore approximately zero compared to its host cloud. By contrast, as can be seen in Figure \ref{fig:convergence_levelmax}, the local measured relative velocity, $v_\bullet$, is non-negligible at all resolutions, with more resolved simulations exhibiting a higher relative velocity. In fact, a comparison to the local sound speed, $c_\bullet$, would suggest that the BH is moving super-sonically relative to the local gas flow, a conclusion which is not supported by global gas flow patterns and BH movement. All simulations with $l_{\rm zoom} \geq 20$ also develop an oscillatory behaviour that appears in the local density, $\rho_\bullet$, the relative velocity, $v_\bullet$, and to a lesser extent in the sound speed, $c_\bullet$. Higher resolution simulations get unsettled earlier and oscillate with shorter periods. 

\begin{figure}
	\centering
	\includegraphics[width=\columnwidth]{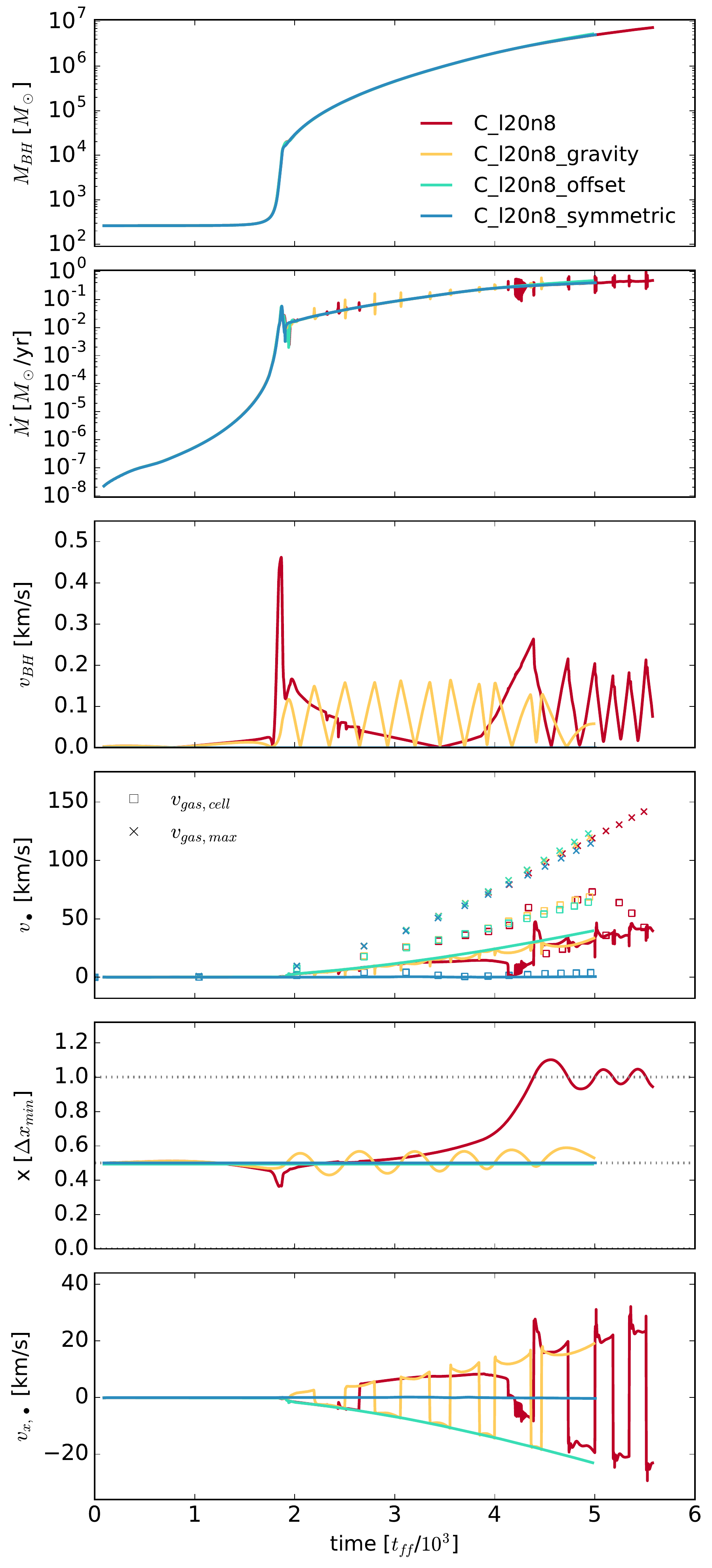}
	\caption[Time evolution of absolute and relative velocities, as measured in the vicinity of the sink particle, for a variety of setups]{Time evolution of various quantities associated with the BH including mass, $M_{\rm BH}$, and accretion rate, $\dot{M}_{\rm BH}$. $v_{\rm BH}$ is the absolute velocity of the BH in the frame of the box, $v_\bullet$ is the magnitude of the relative velocity between gas and BH as measured by the cloud particles and $v_{x,\bullet}$ is its x-component. Square markers denote the relative velocity between BH and its host cell, $v_{\rm gas,cell}$, and crosses show the maximum relative velocity within the region probed by the cloud particles $v_{\rm gas,max}$. $x$ is the x-coordinate of the BH position and dotted grey lines on the fifth panel denote the cell boundaries. Details regarding the simulations' set-up can be found in the text.}
	\label{fig:convergence_velocities}
\end{figure}

To investigate the origin of these oscillations and the high $v_\bullet$, the original simulation, C\_l20n8, where momentum is conserved during accretion and the BH moves under gravity, is compared to two simulations where the BH position is held fixed. In C\_l20n8\_symmetric, the BH is located exactly at the cell centre, and the cloud particles are spaced using $\Delta x_{\rm cloud} = \Delta x_{\rm zoom}/2.02$ to avoid ambiguity in determining a cloud particle host cell. In C\_l20n8\_offset, the BH is also fixed but its position is shifted in all three dimensions by $  - 0.01 \times \Delta x_{\rm zoom}$ relative to the cell centre. In this simulation the cloud particles remain spaced at $\Delta x_{\rm zoom} / 2.0$ and now preferentially probe cells to the bottom left of the BH in the frame of the box. In C\_l20n8\_gravity the BH is only accelerated by gravity, i.e. momentum is not conserved during accretion. For all simulations, the black hole is seeded at the local minimum of the analytic gravitational potential.

\begin{figure*}
	\centering
	\setlength{\tabcolsep}{0pt}
	\begin{tabular}{lcr}
		\includegraphics[width=0.3\textwidth]{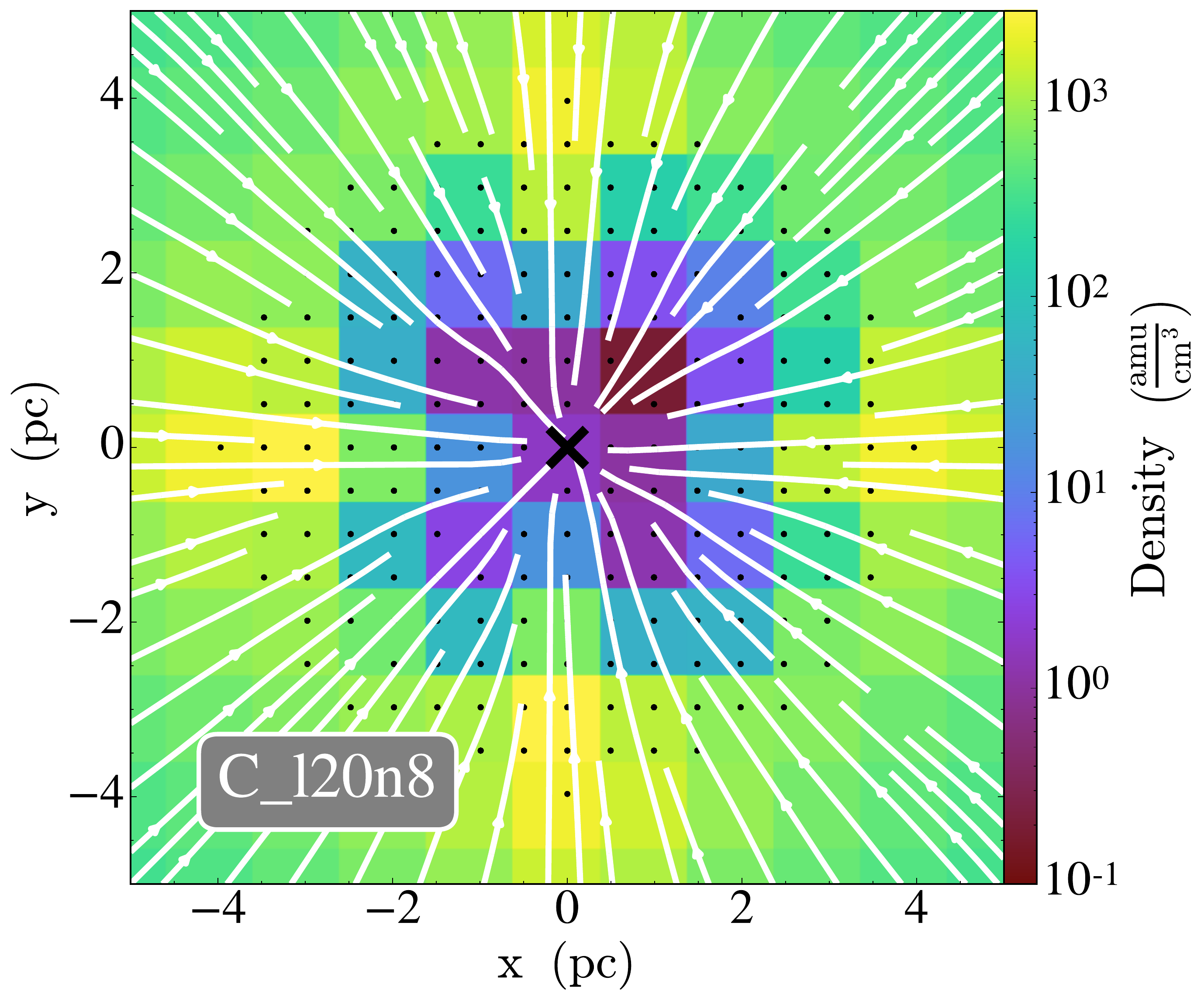} &
		\includegraphics[width=0.3\textwidth]{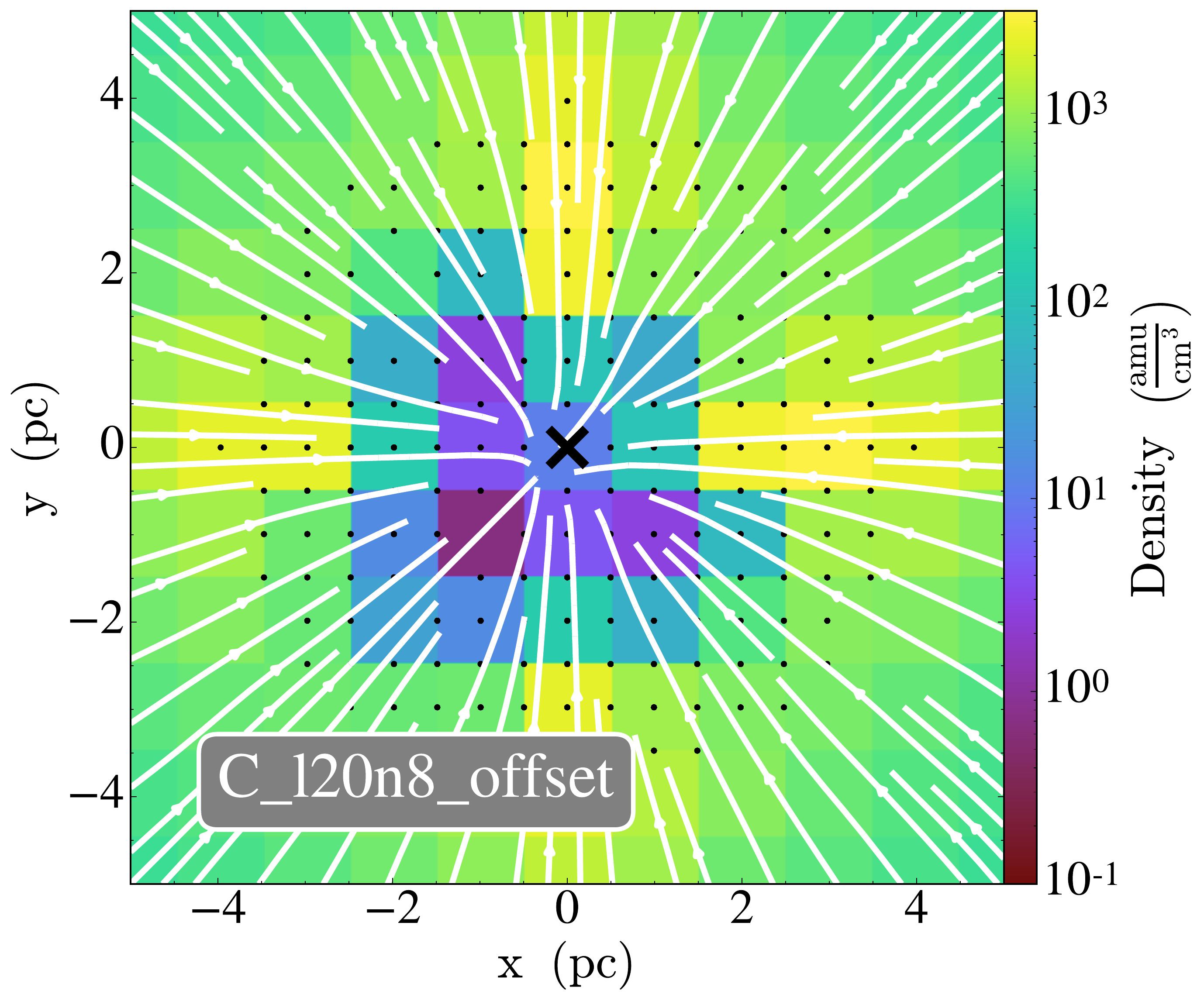} &
		\includegraphics[width=0.3\textwidth]{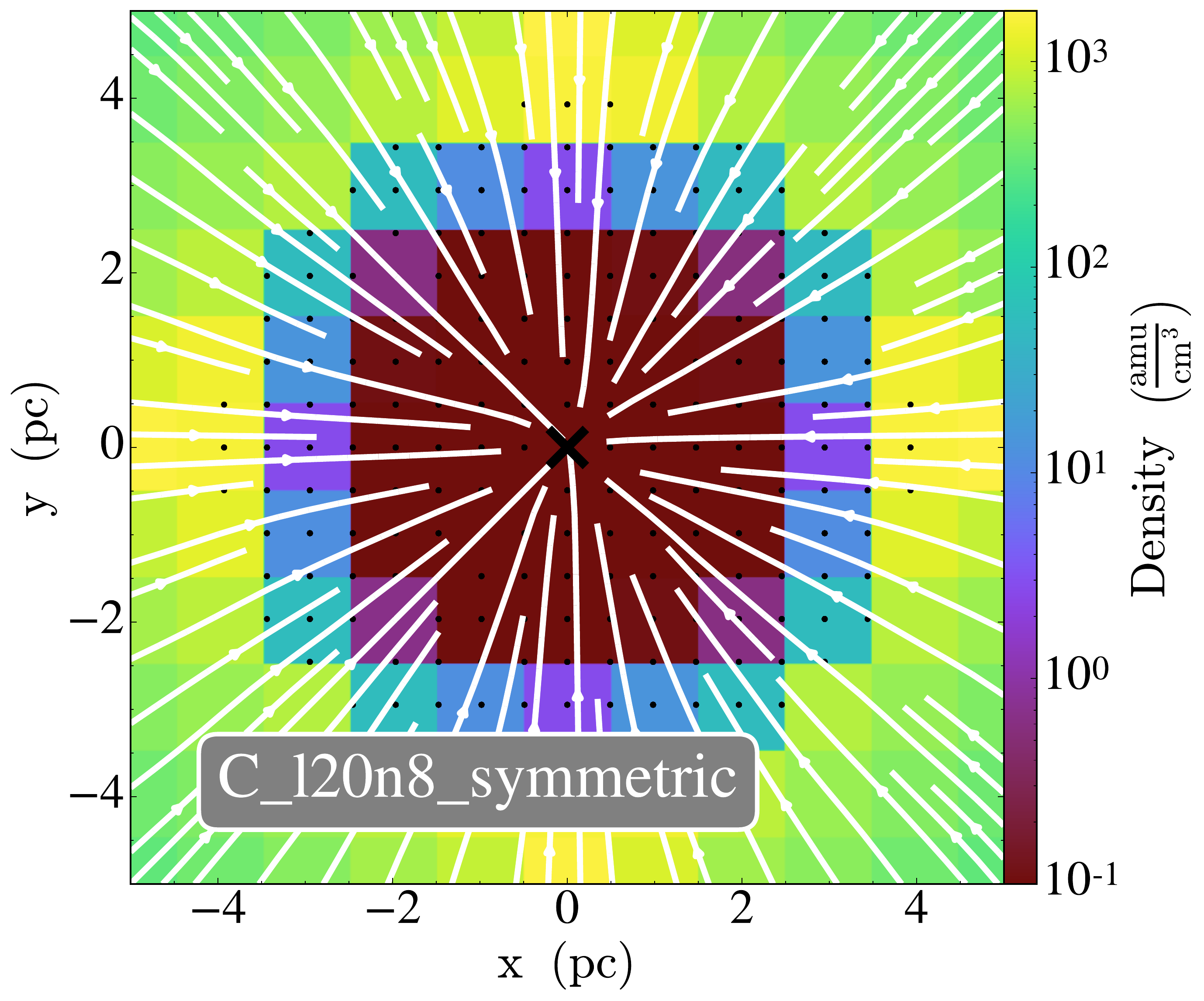} \\
		\includegraphics[width=0.3\textwidth]{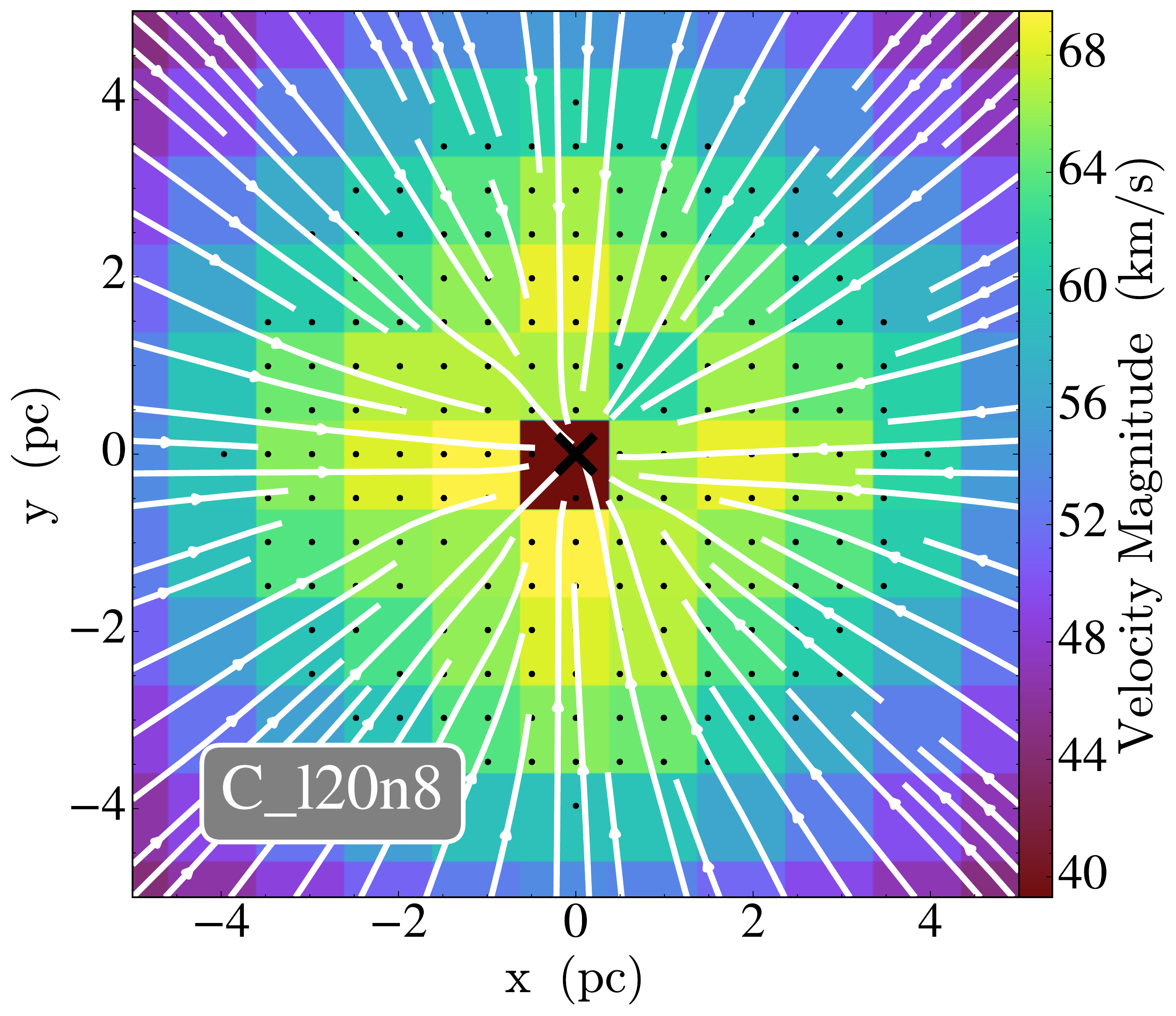} &
		\includegraphics[width=0.3\textwidth]{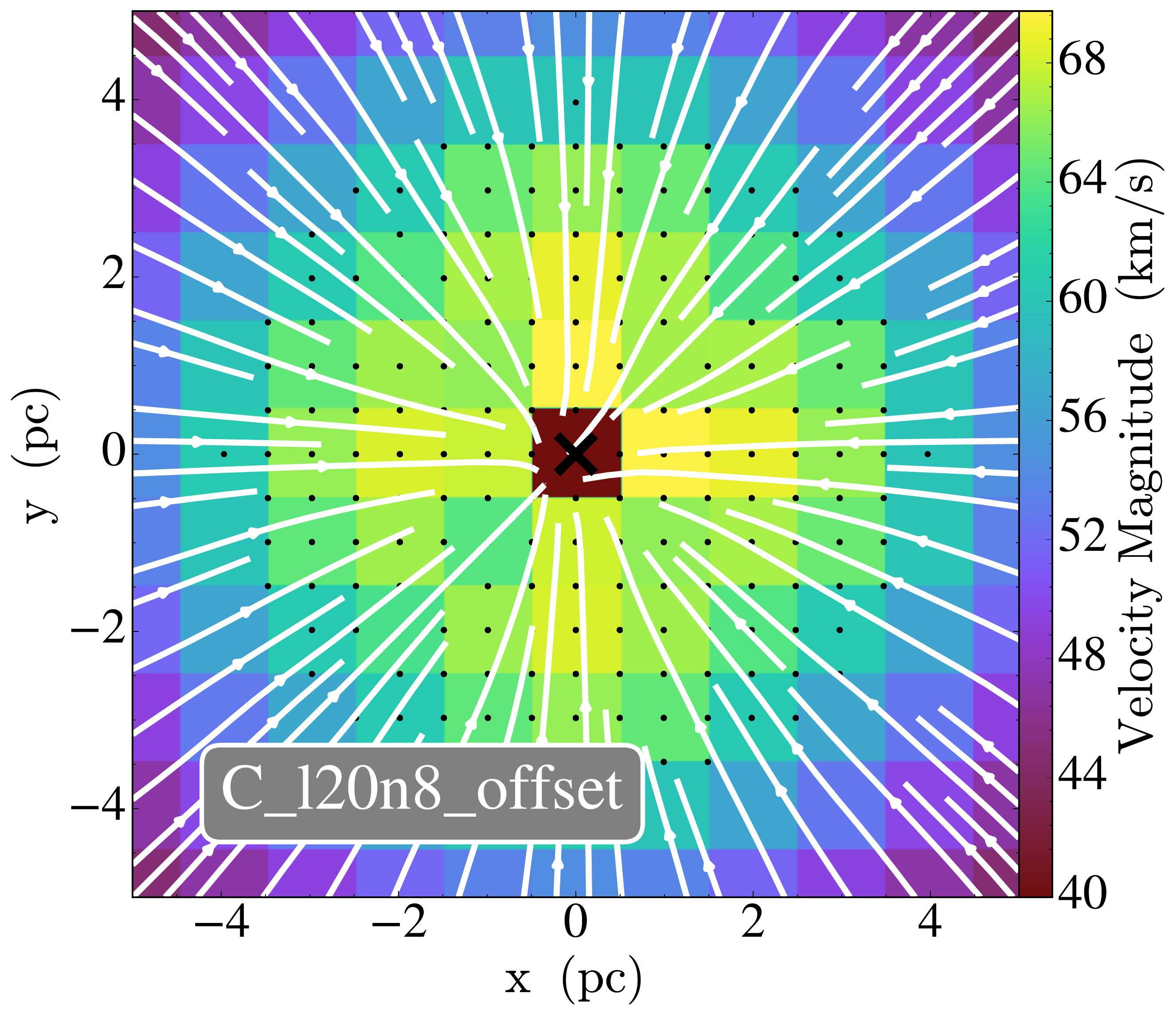} &
		\includegraphics[width=0.3\textwidth]{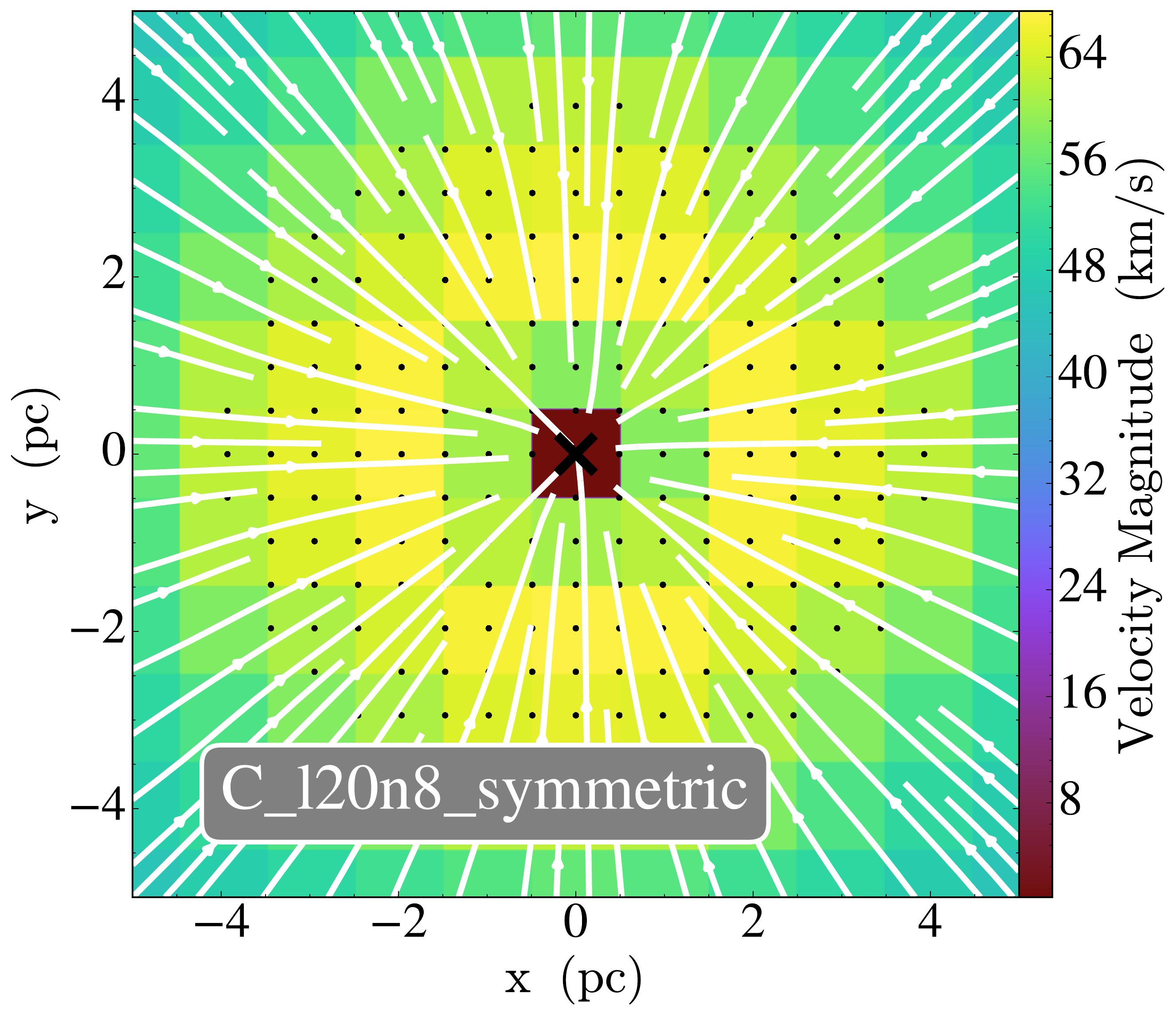} \\
	\end{tabular}
	\caption[Gas density and velocity magnitude slices in the accretion region of the BH for three different set-ups]{Gas density and velocity magnitude slices in the accretion region of the BH for C\_l20n8 (left), C\_l20n8\_offset (center) and C\_l20n8\_symmetric (right) show that only cells from which mass is removed during accretion are affected by the lack of symmetry. Streamlines are annotated in white, and the location of the BH is marked by a black cross. Cloud particle positions are shown as black dots.}
	\label{fig:velocity_slices}
\end{figure*}

Figure \ref{fig:convergence_velocities} shows that, as expected, accretion rates as well as BH masses converge for all four simulations. As can be seen in Figure \ref{fig:convergence_velocities}, $v_\bullet$ is significantly lower than the corresponding inflow velocity (crosses) and equal to or lower than the corresponding velocity of the BH host cell (squares) in all simulations, but higher than the velocity of the BH relative to the box $v_{\rm BH}$. For both for C\_l20n8 and the deliberately asymmetric simulation C\_l20n8\_offset the density distribution at the edge of the accretion region is not symmetric (see top panel, Figure \ref{fig:velocity_slices}), and the central gas cell has a non-negligible velocity relative to the box (bottom panel). By contrast, the perfectly symmetric case, C\_l20n8\_symmetric, has both a spherically symmetric density distribution and a noticeably lower velocity in the host cell ($0.01 \rm \ km\,s^{-1}$ compared to $\sim \rm \ 40 km\,s^{-1}$). The relative velocity as measured by the BH, $v_\bullet$, therefore simply reflects the fact that fluxes at the bottom of the potential well cancel imperfectly. The vector addition over the cloud particles successfully smoothes out some of the local discretisation in the gas, but a residual velocity remains. Higher resolution simulations have higher relative velocities because the gas is more strongly accelerated by the deeper potential well.

\begin{figure*}
	\centering
	\setlength{\tabcolsep}{0pt}
	\begin{tabular}{lcr}
		\includegraphics[width=0.3\textwidth]{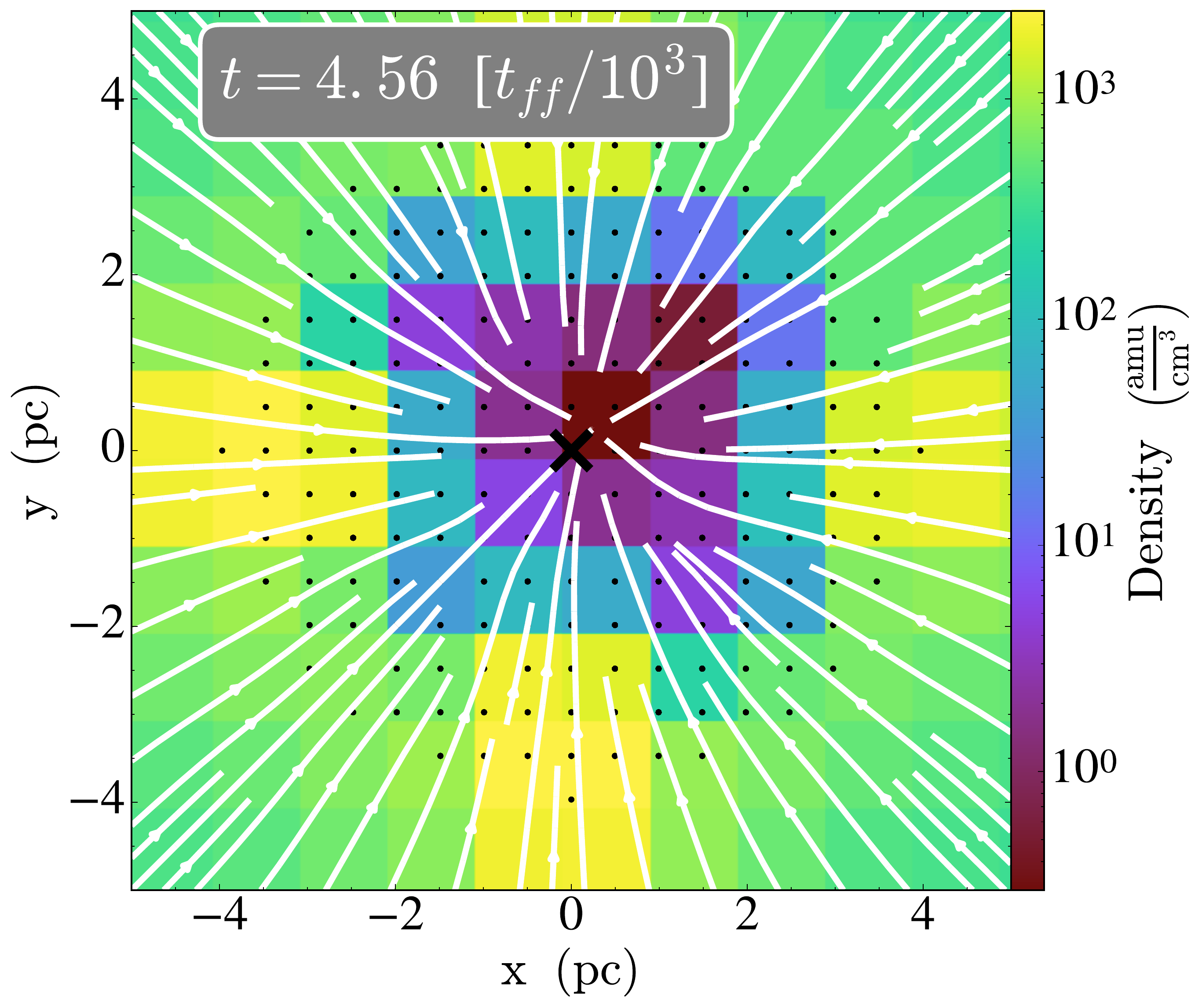} &
		\includegraphics[width=0.3\textwidth]{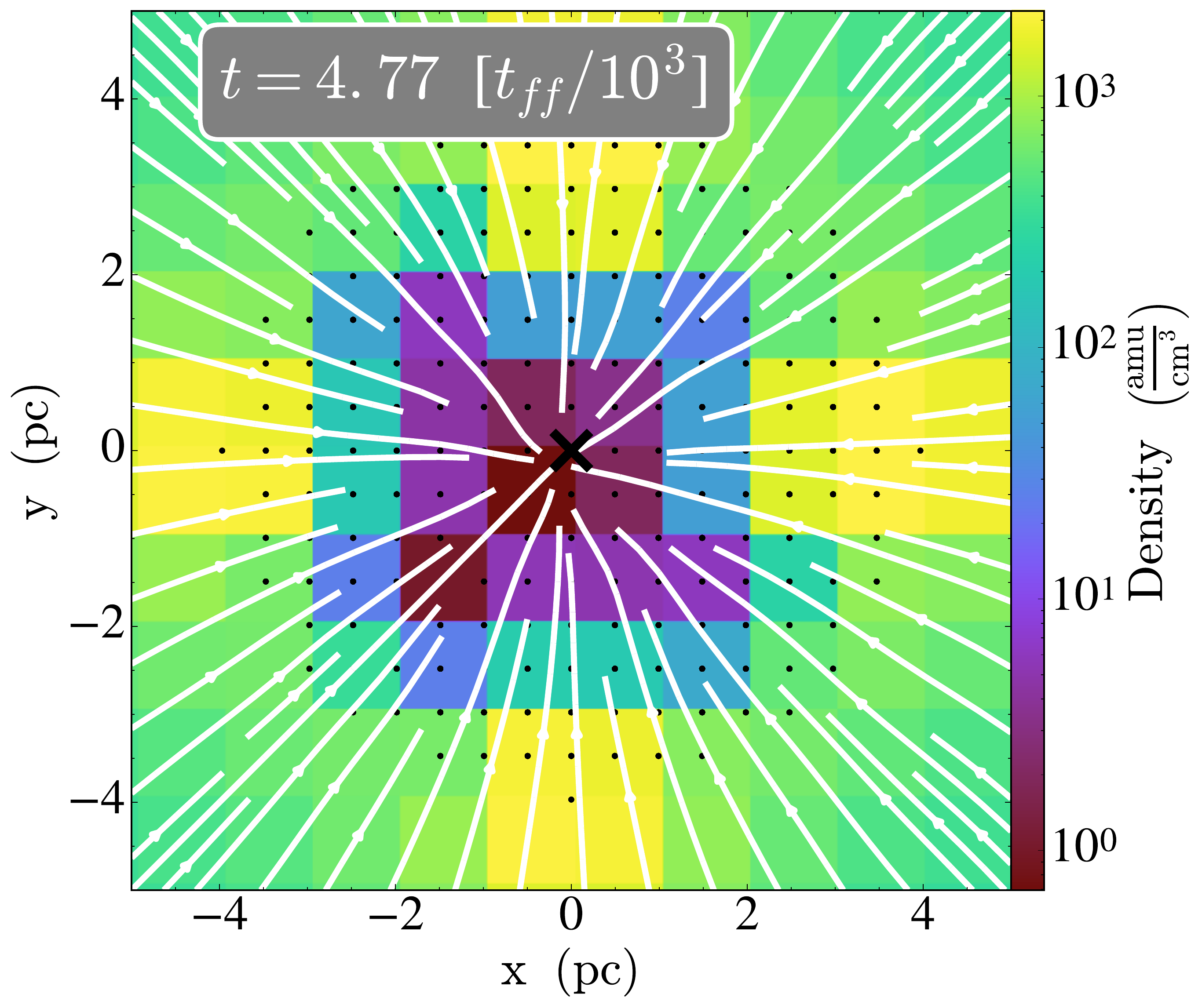} &
		\includegraphics[width=0.3\textwidth]{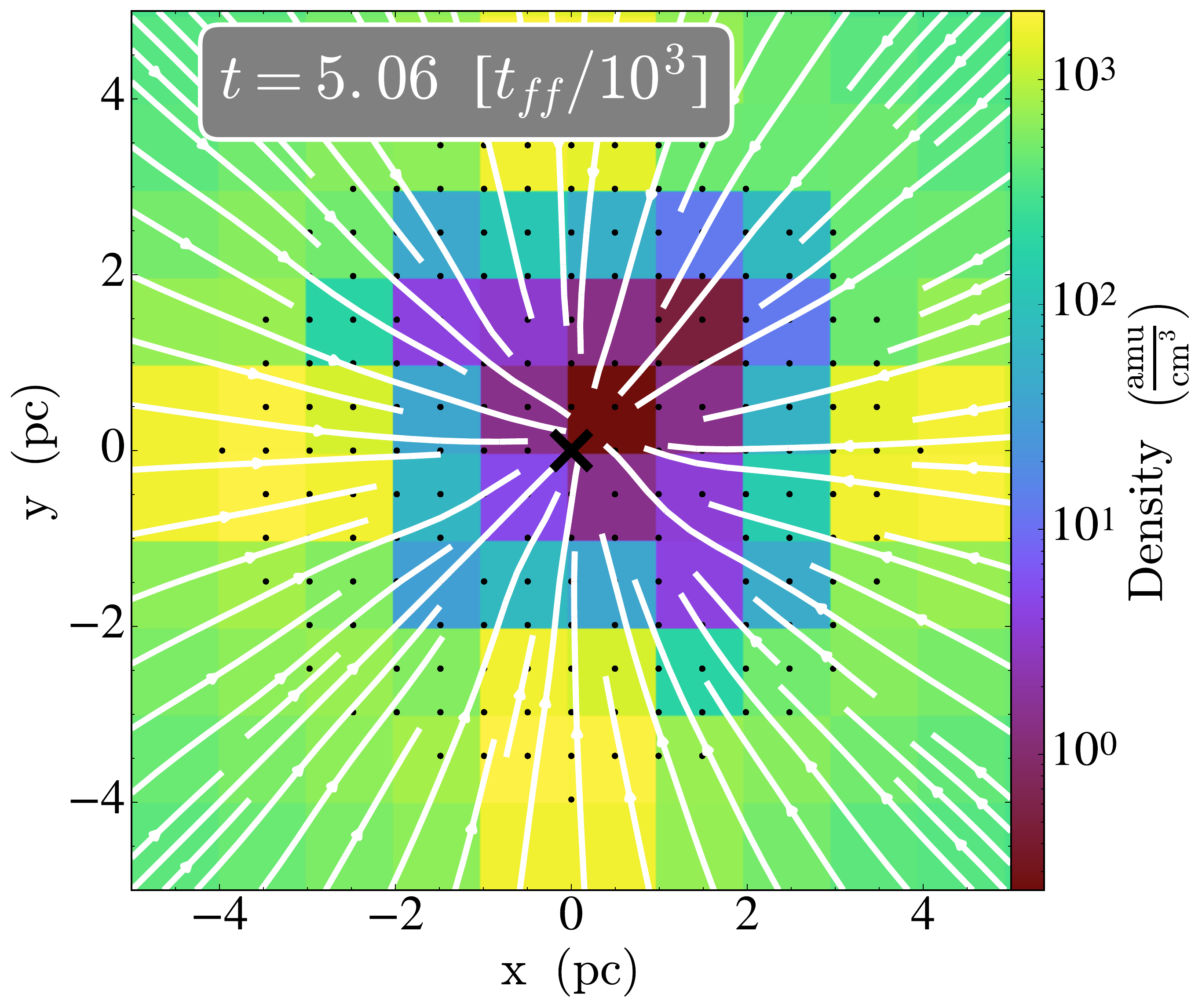} \\
	\end{tabular}
	\caption[Gas density and velocity magnitude slices in the accretion region of the BH at three points in time show the flipping symmetry as the BH passes across a cell boundary]{Gas density and velocity magnitude slices in the accretion region of the BH at three points in time show the flipping symmetry as the BH passes across a cell boundary. Streamlines are annotated in white, and the location of the BH is marked by a black cross. Cloud particles are shown as black dots.}
	\label{fig:slices_fluctuations}
\end{figure*}

When the sink particle is moving only due to gravity, in C\_l20n8\_gravity, $v_\bullet$ has the same magnitude as in the fixed C\_l20n8\_offset, but its direction flips as the BH traces out a small circle within the host cell (see bottom two panels of Figure \ref{fig:convergence_velocities}). As there is no momentum transferred during accretion and the BH initial position coincides with the bottom of the analytic potential, the BH must be accelerated by the gravitational force exerted by the local gas distribution. Figure \ref{fig:slices_fluctuations} shows that after the transition to SLA at $t=2 \ t_{\rm ff}/10^3$, the BH is surrounded by an irregular shell of leftover gas, which indeed imparts a small net gravitational pull on it.  

If the BH is also accelerated by momentum conservation during accretion (C\_p20n8) then even though the relative velocity at  $t =1.9 \ t_{\rm ff}/10^3$ is comparatively small at $v_\bullet = 2.3 \rm \ km\, s^{-1}$, the large amount of mass accreted ($M_{\rm acc,tot} > 10^5 \ \rm M_\odot$) compared to the initial BH mass ($M_{\rm init} = 260 \ \rm M_\odot$) means that accreted momentum dominates and the BH receives a velocity kick during the transition to SLA that dislodges it from its position at the cell centre (see third and fifth panel of Figure \ref{fig:convergence_velocities}). When entering a new cell, gas is removed preferentially from a different set of cells, changing the symmetry of the local density distribution (see Figure \ref{fig:slices_fluctuations}). The local symmetries in the density distribution flip, the BH reverses direction and passes back into the original cell, the symmetries flip again, and the oscillations continue. Note that this issue is purely numerical and only involves cells within the accretion region, outside of which the gas properties remain as spherically symmetric as possible on a Cartesian grid.

This movement of the BH causes all local mass weighted quantities to vary by several orders of magnitude on short timescales, including the accretion rate, but the cumulative effect on the mass evolution of the BH is negligible as the accretion algorithm is self-balancing. Any mass not accreted at a given time-step remains on the grid until accreted.  As can be seen in the top panel of Figure \ref{fig:convergence_levelmax}, the mass of C\_l24n8 is indistinguishable from the mass of stable simulations such as C\_l18n8 after more than $2 \ t_{\rm ff}/10^3 $. These types of numerical oscillatory features are expected to be less prominent for massive BH in a more realistic astrophysical context. Indeed, simulations involving turbulence, other compact objects and angular momentum are unlikely to produce anything close to a perfectly symmetric collapse for extended periods of time, and local asymmetries in the density distribution around the BH are thus more likely to be caused by the gas flow itself than by BH accretion.

We conclude that both the oscillations and the high relative velocities are numerical artefacts due to a local loss of spherical symmetry in and around the accretion region of the BH (see for example the right hand panel of Figure \ref{fig:flux}), caused by small motions of the BH with respect to the cartesian grid. The local measures of gas properties, when conducted inside collapsing clouds, therefore become less reliable with increasing resolution. This is particularly true for non-local measures such as the relative velocity, and is compounded by the fact that many sub-grid models take the local measure as a proxy for a global one. BHL accretion models, for example, assume quantities are measured far from the BH, not deep in its gravitational potential (see \citet{Beckmann2018a} for details). A potential solution would be to measure gas properties on larger scales (e.g. as in the torque-driven accretion model by \citet{Angles-Alcazar2015}). However, as discussed further in Section \ref{sec:discussion}, these types of models make assumptions about gas dynamics on pc scales, which are unsupported by our current understanding of conditions in high-redshift mini-halos. As the local structure around the object under investigation becomes better resolved, local gas properties should be treated with an increasing degree of caution and sub-grid models based on a minimum number of input parameters should be employed. SLA satisfies these requirements, as it only relies on the gas mass flux into the accretion region of the BH, an inherently local measure that will converge to the correct Newtonian solution as the size of the accretion region approaches the physical extent of the BH.
   
Note that the simplicity of the setup studied in this section might contribute to the BH mass convergence. As both the gas and the BH are gravitationally bound at the bottom of the potential well, the point at which mass is transferred from the grid to the sink particle has no effect on the long term evolution. In a more complex environment, the finite lifetime of the cloud feeding the BH could become a concern. From Figure \ref{fig:convergence_levelmax}, the transition to SLA happens around $t \simeq 1.7 \  t_{\rm ff}/10^3$, but it is significantly delayed for C\_l18n8 and C\_halo16. In other words, should the cloud feeding the BH be disrupted at $t \simeq 1.9 \  t_{\rm ff}/10^3$, before the transition to SLA for C\_l18n8 and C\_halo16, the difference in BH mass at that point would have led to very different later evolution histories for these two simulations. 


\bsp	
\label{lastpage}
\end{document}